\documentclass[12pt]{article}
\pdfoutput=1
\usepackage{putex}
\usepackage{graphicx}
\usepackage{epstopdf}
\usepackage{enumerate}
\usepackage[nosort]{cite}
\usepackage{tensor}
\usepackage{slashed}
\usepackage{feynmf}
\usepackage{hyperref}
\usepackage{float}
\usepackage[bottom]{footmisc}
\usepackage[utf8]{inputenc}
\usepackage{amsmath}
\usepackage[usenames,dvipsnames,svgnames,table]{xcolor}
\usepackage[bf,hang,footnotesize]{caption}
\usepackage{bbold}

\numberwithin{equation}{section}
\hypersetup{
	pdfstartview={FitH},    
	pdftitle={Carrollian Partition Functions and the Flat Limit of AdS},    
	pdfauthor={Kraus, Myers},     
	colorlinks=true,       
	linkcolor=blue,          
	citecolor=blue!59!black! ,        
	filecolor=magenta,      
	urlcolor=blue!75!black,           
	linktoc=all
}

\makeatletter
\g@addto@macro\bfseries{\boldmath}
\makeatother

\numberwithin{equation}{section}

\newcommand{\eq}[2]{\begin{align}\label{#1}#2\end{align}}

\newcommand {\be} {\begin {equation}}
\newcommand {\ee} {\end {equation}}

\newcommand{\p}{\partial}

\newcommand\rt{{\rightarrow}}
\def\eps{\epsilon}

\newcommand{\rf}[1]{(\ref{#1})}
\newcommand{\rff}[1]{\ref{#1}}

\newcommand{\zb}{\overline{z}}

\newcommand{\veps}{\varepsilon}

\newcommand{\arrow}{\rightarrow}

\definecolor{greenC}{rgb}{0.0, 0.38, 0.18}

\newcommand{\R}{\mathbb{R}}

\newcommand{\phib}{\bar{\phi}}

\newcommand{\Ab}{\overline{A}}

\newcommand{\ab}{\overline{a}}

\newcommand{\xh}{\hat{x}}

\newcommand{\dr}{d}

\newcommand{\scrI}{\mathcal I}
\newcommand{\pv}{\vec{p}}
\newcommand{\normord}[1]{\mathopen{:}#1\mathclose{:}}

\newcommand{\Ic}{{\cal I}}
\newcommand{\Gc}{{\cal A}}
\newcommand{\Gct}{{\tilde{\cal A}}}

\newcommand{\ph}{\hat{p}}
\newcommand{\bb}{\overline{b}}
\newcommand{\cb}{\overline{c}}
\newcommand{\db}{\overline{d}}
\newcommand{\Ah}{\hat{A}}
\newcommand{\Abh}{\hat{\overline{A}}}
\newcommand{\Phib}{\overline{\Phi}}

\newcommand{\nabh}{\hat{\nabla}}

\newcommand{\D}{\mathcal{D}}

\newcommand{\pa}{\overset{\leftrightarrow}{\partial} }
\newcommand{\Gt}{\tilde{G}}
\newcommand{\epsh}{\hat{\varepsilon}}
\newcommand{\etah}{\hat{\eta}}
\newcommand{\qv}{\vec{q}}
\newcommand{\qh}{\hat{q}}
\newcommand{\yh}{\hat{y}}

\begin{document}

\institution{UCLA}{ \quad\quad\quad\quad\quad\quad\quad\ ~ \, $^{1}$Mani L. Bhaumik Institute for Theoretical Physics
		\cr Department of Physics \& Astronomy,\,University of California,\,Los Angeles,\,CA\,90095,\,USA}

\title{ Carrollian Partition Functions \hspace{3in} and the \hspace{6in} Flat Limit of AdS}

\authors{Per Kraus$^{1}$, Richard M. Myers$^{1}$}
	
\abstract{   The formulation of the S-matrix as a path integral with specified asymptotic boundary conditions naturally leads to the realization of a Carrollian partition function defined on the boundary of Minkowski space. This partition function, specified at past and future null infinity in the case of massless particles, generates Carrollian correlation functions that encode the S-matrix. We explore this connection, including the realization of symmetries, soft theorems arising from large gauge transformations, and the correspondence with standard momentum space amplitudes. This framework is also well-suited for embedding the Minkowski space S-matrix into the AdS/CFT duality in the large radius limit. In particular, we identify the AdS and Carrollian partition functions through a simple map between their respective asymptotic data, establishing a direct correspondence between the actions of symmetries on both sides. Our approach thus provides a coherent framework that ties together various topics extensively studied in recent and past literature.
 }
	
	\date{}
	
	\maketitle
	\setcounter{tocdepth}{2}
	\begingroup
	\hypersetup{linkcolor=black}
	\tableofcontents
	\endgroup
	

\section{Introduction}

A defining feature of holography is that it allows for the computation of observables via some sort of dimensionally reduced description, the best understood example being AdS/CFT.  Compared to AdS/CFT, our understanding of a holographic description of flat space is primitive.  In (asymptotically) flat space, a natural observable to focus on is the S-matrix, and for massless particles  the natural location for a putative holographic dual is null infinity, $\scrI$, or a subspace thereof, though more abstract possibilities may also be entertained.  A large body of work  has arisen in recent years trying to develop this picture.

Most of the recent progress on flat space holography has been from a bottom-up perspective, where there are presently two major approaches in the literature. Celestial holography\footnote{The literature on celestial holography is by now extensive and we direct the reader to \cite{Strominger:2017zoo,Pasterski:2021rjz,Raclariu:2021zjz,Pasterski:2021raf, McLoughlin:2022ljp,Donnay:2023mrd} for a selection of reviews and further references. } proposes to formulate S-matrix elements in four dimensions as correlators in some putative CFT$_2$ on the celestial sphere. This framework, in the absence of gravity, uses the isomorphism between the Lorentz group and the conformal group in two fewer dimensions to match the basic symmetries on both sides of the duality.  The other proposal, which will be our primary interest in this work, is Carrollian holography \cite{Duval:2014uoa,Duval:2014uva,Hartong:2015xda,Hartong:2015usd,Bagchi:2016bcd,Ciambelli:2018wre,Ciambelli:2019lap,Salzer:2023jqv,Bagchi:2019clu,Hansen:2021fxi,Donnay:2022aba,Bagchi:2022emh,Donnay:2022wvx,Bagchi:2023fbj,deBoer:2023fnj,Saha:2023hsl,Nguyen:2023vfz,Nguyen:2023miw,Bagchi:2023cen,Mason:2023mti,Cotler:2024xhb}. This proposal interprets S-matrix elements as correlators in a putative Carrollian CFT$_3$ supported on $\scrI$. One approach \cite{Bagchi:2016bcd} to Carrollian CFTs is to think of them as the hyperboosted limit of a standard CFT$_3$, making them a natural candidate for a theory living on a null hypersurface. This construction uses that the bulk Lorentz group is isomorphic to the boost limit of the conformal algebra in one dimension lower.

In the presence of gauge and gravitational dynamics, the existence of large gauge transformations makes identifying the precise symmetry group of asymptotically flat spacetimes more subtle. For example, in the presence of gravity it has long been understood that the Lorentz group is extended to an infinite dimensional group found by Bondi, van der Burg, Metzner, and Sachs (BMS) \cite{Bondi:1962px,Sachs:1962zza}. But there are several consistent extensions to the BMS group and the correct one depends on the precise definition used for asymptotically flat spacetime \cite{Barnich:2009se,Barnich:2010eb,Barnich:2010ojg,Campiglia:2014yka,Campiglia:2015yka,Flanagan:2019vbl,Freidel:2021fxf}. Such subtleties in identifying the true symmetry group of asymptotically flat spacetimes are not isolated to considerations of holography. As symmetries, they must have real, measurable consequences for scattering processes. Indeed, via the discovery of the IR triangle \cite{Strominger:2017zoo}, these asymptotic symmetry groups are closely related to the soft theorems \cite{Low:1958sn,Weinberg:1965nx} that control the soft expansion of amplitudes. At tree level, one can organize these symmetries to predict an infinite tower of constraints on amplitudes \cite{Strominger:2021mtt}.  But beyond the tree approximation such statements become entangled with the usual issue of IR divergences that modify the soft expansion and make it subtle to give a precise definition of the S-matrix \cite{Kulish:1970ut,Prabhu:2022zcr}. Which symmetry implications are obeyed by the usual Feynman diagrams? Do these symmetries survive, perhaps with a deformation, to loop order?  To address such questions it is advantageous to develop a formalism that allows symmetries to be represented in a clean way.

\vspace{.2cm}
\noindent 
{{\bf{\large{ \underline{Carrollian partition function}}}}}
\vspace{.2cm}

\begin{figure}[h]
    \centering
    \includegraphics[scale=0.3]{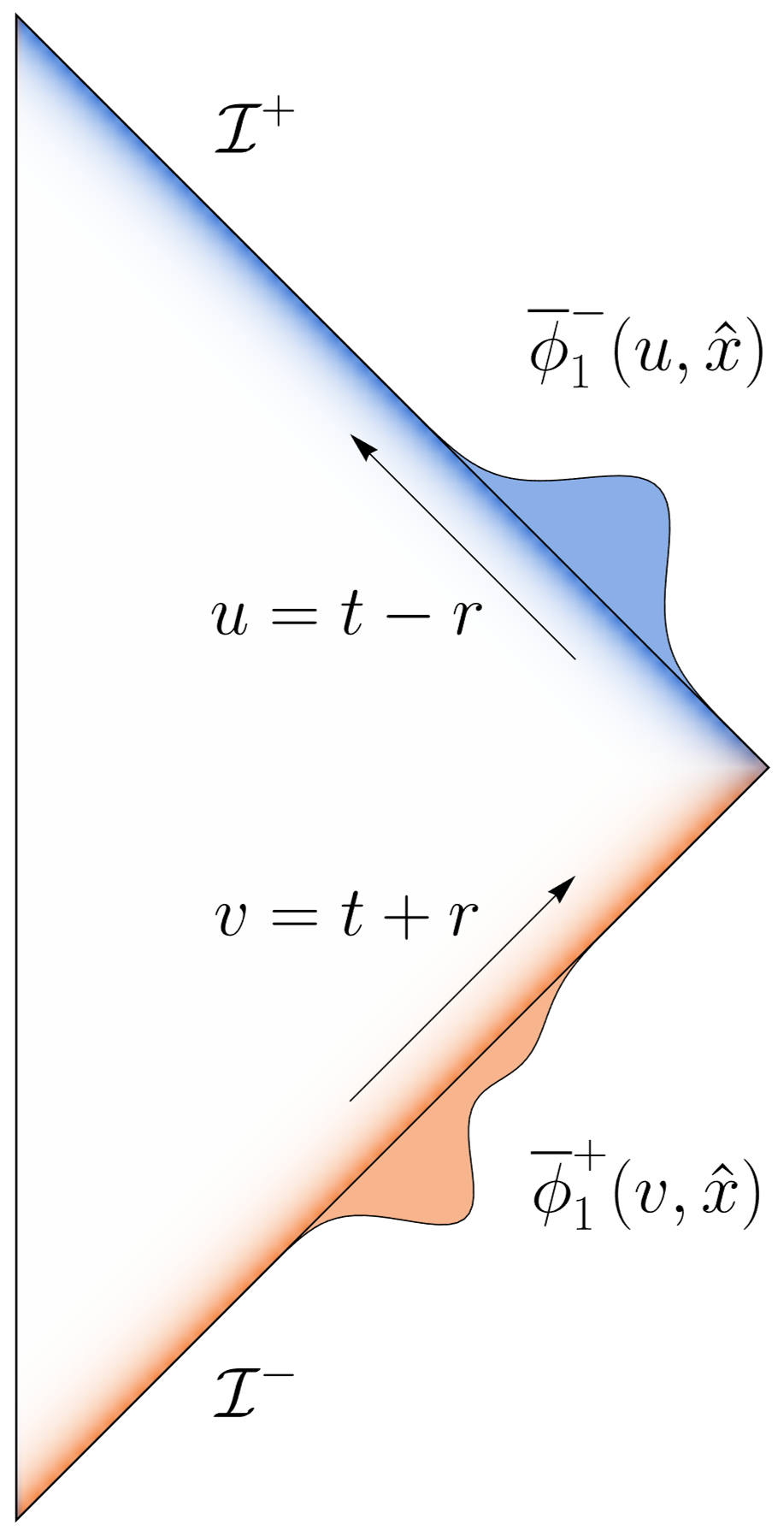}
    \caption{Minkowski Penrose diagram with asymptotic boundary conditions indicated; a pure negative (positive) frequency waveform is specified along $\scrI^+$($\scrI^-$).}
    \label{fig:triangle_with_waveform}
\end{figure}

In the recent work \cite{Kim:2023qbl}, we explored a formulation of the S-matrix in terms of a path integral with specified asymptotic boundary data, as originally proposed by Arefeva, Faddeev, and Slavnov (AFS)\footnote{In our context, AFS can also stand for asymptotically flat spacetime.} \cite{Arefeva:1974jv}. The path integral defined in this way serves as a generating functional for S-matrix elements, in close analogy with the GKP/W dictionary \cite{Gubser:1998bc, Witten:1998qj}  of AdS/CFT that defines boundary correlators in terms of the path integral over fields in the bulk with specified boundary conditions.\footnote{This is to be contrasted with the BDHM, or ``extrapolate", dictionary \cite{Banks:1998dd} where boundary correlators are obtained by taking the boundary limit of bulk correlators, a formulation that bears more similarity to LSZ reduction.  }   
The specific boundary conditions relevant for the S-matrix can be understood as fixing the positive (negative) frequency content\footnote{The realization of this formalism in Euclidean signature has recently been explored in \cite{Jain:2023fxc}. In that context, one can arrange to use Dirichlet boundary conditions instead.} of the field configurations in the path integral along $\scrI^-$ ($\scrI^+$), see figure \ref{fig:triangle_with_waveform},
\eq{int1}{
    Z[\bar\phi_1^-, \bar \phi_1^+] = \int\mathcal{D}\phi e^{iI[\phi] + iI_\text{bndy}[\phi,\bar\phi]}
}
where $(\phib_1^-,\phib_1^+)$ encode the asymptotic boundary conditions as 
\eq{int1a}{ \phi(x) \approx \begin{cases}
			{1\over r} \phib_1^-(u,\xh) + {\rm positive~frequency} &  {\rm on}~ \Ic^+\\
         {1\over r} \phib_1^+(v,\xh) + {\rm negative~frequency} &  {\rm on}~ \Ic^-
		 \end{cases} ~.}
As in AdS/CFT, the boundary term in the action  can be determined by demanding a good variational principle. In section \ref{AFSreview} we review how $Z$ may be used to generate the usual S-matrix elements. Since asymptotic symmetries are naturally defined to act on the asymptotic data of the fields, the AFS formulation is well-suited to describing such symmetries and their implications, treating asymptotic symmetries on the same footing as standard global symmetries. For example, it was shown in \cite{Kim:2023qbl} that the leading soft photon theorem follows easily from the invariance of $Z$ under large gauge transformations. In section \ref{Sec:Invariance under large gauge transformations and soft theorems} we show that this demonstration can be simplified even further to sidestep direct reference to symplectic structures.

In this work we further explore the AFS generating functional with an eye towards the Carrollian description of flat holography\footnote{The conversion of basis between the Carrollian and celestial descriptions is straightforward and has already been worked out in \cite{Donnay:2022wvx}.}, where the fixed frequency data amounts to specifying the waveforms  $\bar\phi_1^-(u, \hat x)$ and $\phib_1^+(v,\xh)$. In terms of this Carrollian data, the AFS generating functional \eqref{int1} for the S-matrix can be interpreted as a generating functional for Carrollian correlators; for example the contribution corresponding to $2\rt 2$ scattering, in the case of a real scalar, is  
\eq{int2}{   Z[\bar \phi_1^-, \bar \phi_1^+]&= {1\over (2!)^2} \int_{\Ic^-} d^3Y_1 d^3Y_2 \int_{\Ic^+} d^3Z_1 d^3Z_2   G_{2,2}(Y_1,Y_2,Z_1,Z_2) \cr
& \!\! \!\!\!\!\!\!\!\times\p_{v_1} \phib_1^+(Y_1) \p_{v_2} \phib_1^+(Y_2)\big(-\p_{u_1} \phib_1^-(Z_1)\big)   \big(-\p_{u_2} \phib_1^-(Z_2)\big)}
where $Y_i=(v_i,\xh_i)$ denote coordinates on $\Ic^-$ and $Z_i=(u_i,\xh_i)$ are coordinates on $\Ic^+$. Written in this form, the AFS generating functional is supported only on the boundary. In analogy with the AdS partition function in the presence of dual CFT sources, we refer to this object as the Carrollian partition function.  Since \eqref{int1} and \eqref{int2} are two equivalent ways of writing the same object, we use it to work out the relation between Carrollian correlators and standard momentum space amplitudes which, as proposed in previous works \cite{Banerjee:2018gce,Donnay:2022wvx,Bagchi:2023cen,Mason:2023mti}, essentially amounts to a Fourier transform in $u$ or $v$ space.  The approach adopted here nicely explains the underlying origin of this relation.

Since the partition function \eqref{int2} descends from the path integral \eqref{int1}, it's clear that the action of symmetries on our asymptotic data will be an invariance of the Carrollian partition function.  Also, since only the asymptotic transformations of fields enter, it's  clear that large gauge/diffeomorphism transformations are treated on an equal footing with more familiar global symmetries \footnote{This is strictly only true of symmetries that do not mix the fixed and unfixed frequency content. Poincar\'e transformations and all internal symmetries have this property. More general symmetries would imply that $Z$ obeys a functional differential equation.} of the Carrollian partition function. We demonstrate how this invariance can be used to deduce the properties of Carrollian correlators, first in the example of a scalar field and then in scalar QED. In both cases we obtain the action of Lorentz transformations on the asymptotic data and demand invariance of \eqref{int2} to obtain the Lorentz invariance properties of the Carrollian correlators. In the case of scalar QED, we do the same for large gauge transformations, which yields the Carrollian representation of the leading soft photon theorem. 

One can also pose the inverse question: given a relation satisfied by the Carrollian correlators, what is the transformation of the asymptotic data that would imply this property? We use the example of the tree-level subleading soft photon theorem \cite{Low:1958sn} to demonstrate how this question may be answered. While the deduced transformation  acts non-locally on the asymptotic field as was noted in \cite{Lysov:2014csa}, it has a  local action on $\p_u\bar \phi_1$, which is the only combination appearing in \eqref{int2}. As a related  example, we obtain the action of (infinitesimal) special conformal transformations on the asymptotic data in appendix \ref{Sec:Asymptotic special conformal transformations}, which also appears to act non-locally on the asymptotic field data while acting locally on $\p_u\bar \phi_1$.

\vspace{.2cm}
\noindent 
{{\bf{\large{\underline{ Flat limit of AdS}}}}}
\vspace{.2cm}

\begin{figure}[h]
    \centering
    \includegraphics[scale=0.4]{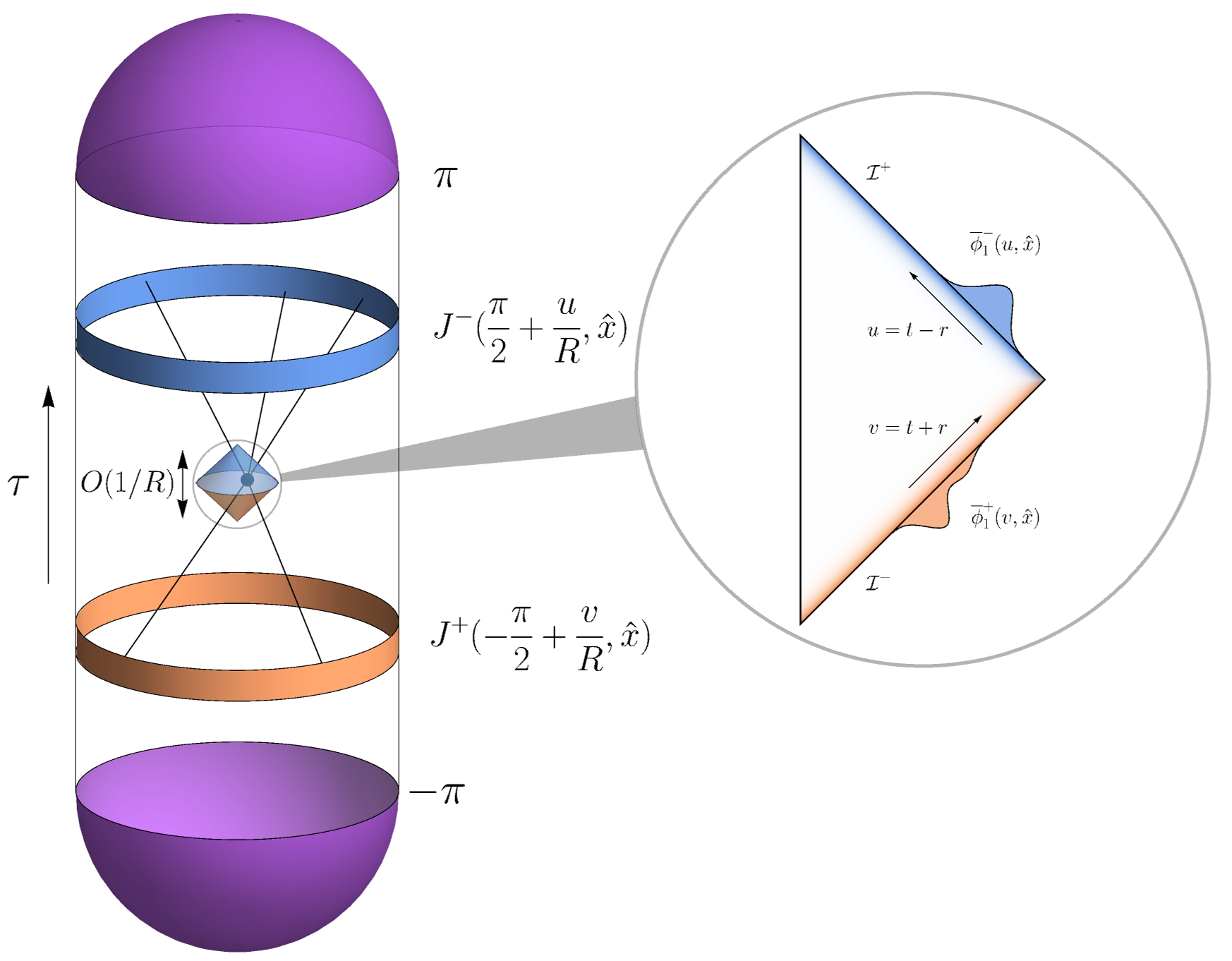}
    \caption{Flat space may be embedded in a small region in the center of Lorentzian AdS, much smaller than the curvature radius $R$. The flat space data on $\scrI^\pm$ maps to AdS Dirichlet data in thin strips about $\tau = \pm\frac{\pi}{2}$ in global AdS time. The violet caps represent a Euclidean state preparation that projects onto the vacuum.}
    \label{Lorentzian pill}
\end{figure}

Another approach to obtain a holographic understanding of the S-matrix is by attempting to import our understanding of holography in AdS/CFT by taking the flat limit of AdS. This is an idea with a history almost as old as AdS/CFT itself, with most of the important conceptual points laid out early on in \cite{Polchinski:1999ry,Susskind:1998vk}, and further developed in many works including \cite{Gary:2009ae,Gary:2009mi,Giddings:1999jq,Fitzpatrick:2011jn,Hijano:2019qmi,Hijano:2020szl,Komatsu:2020sag,Li:2021snj,Duary:2022pyv,Bagchi:2023fbj,Campoleoni:2023fug,deGioia:2024yne,Alday:2024yyj,Marotta:2024sce}. The essential idea of the limit is to consider high energy, with respect to global AdS time, particles sent into the bulk of AdS, aimed such that their interactions occur only on a scale much smaller than the AdS radius. In this special case, the interactions are insensitive to the curvature of AdS, making quantities calculated from the process agree with their flat counterparts. In practice, one constructs wavepackets emanating from  thin strips of the AdS boundary, as shown in figure \ref{Lorentzian pill}. The distance between these strips can be chosen such that, as emphasized in \cite{Gary:2009ae}, the bulk point singularity \cite{Maldacena:2015iua} of the bulk Witten diagrams occurs within a region much smaller than the AdS curvature radius $R$.

While the above arguments, and indeed most of the literature on the subject, are phrased with the BDHM dictionary in mind, it's straightforward to rephrase the setup in terms of the GKP/W dictionary. In that case, one chooses the boundary sources localized to the same strips on the AdS boundary, in global AdS time centered on $\tau=\pm\frac{\pi}{2}$. It is simplest to construct a source $J$ with the desired support by going to momentum space. For example, a positive frequency source centered about $\tau=-\frac{\pi}{2}$ may be written as
\eq{int3}{
    J^+(\tau, \hat x) = \int_0^\infty\frac{\dr\omega}{2\pi}\tilde J(\omega, \hat x)e^{-i\omega R(\tau + \frac{\pi}{2})}.
}
At large $R$ the resulting $ J^+(\tau, \hat x) $ is concentrated around $\tau = -{\pi\over 2}$.
Then under the assumption that no particle interactions occur outside the flat region in the center of AdS, we just need to follow this data into the flat region by computing\footnote{This prescription is valid for particles that become massless in the flat region, corresponding to $\Delta \sim O(R^0)$ on the boundary. A similar prescription for massive particles, having $\Delta \sim O(R)$, may be found in \cite{Hijano:2019qmi,Hijano:2020szl}. See \cite{Komatsu:2020sag} for a recent discussion of additional subtleties that arise in the massive limit.} $\phi(x) = \int\dr^dx'\sqrt{-h}K(x;x')J^+(x')$, where $K(x;x')$ is the usual bulk-boundary propagator, in the limit where the bulk point $x$ is within the flat region and at large $R$.  There is a corresponding expression for a negative frequency source $J^-$ localized near $\tau= {\pi \over 2}$. This approach shares some features with the HKLL reconstruction used in \cite{Hijano:2019qmi,Hijano:2020szl} to find a map between the creation/annihilation operators of particles in flat space and normalizable modes in AdS. The flat limit has recently been approached from a Carrollian perspective in \cite{Bagchi:2023fbj,Bagchi:2023cen,Alday:2024yyj} and the celestial perspective in \cite{deGioia:2022fcn,Raclariu:2021zjz}. See \cite{Li:2021snj} and references therein for a recent survey of other approaches.
 
We  compute the bulk-boundary propagator integral along with the analogous expression for a bulk gauge field. The result is most simply written in Carroll form, and provides a map between the AdS data $J$ on the strip and the Carroll data that appears in the Carrollian partition function \eqref{int2}.   Under the assumption that interactions are effectively confined to the flat region in the large $R$ limit, we deduce that the AdS and AFS generating functionals are equal in this limit.  In the case of a massless bulk scalar this relation reads
\eq{int5}{
    Z_\text{AdS}[J^+(\tau,\xh),J^-(\tau,\xh)] = Z_\text{AFS}[\phib_1^-(u,\xh), \phib_1^+(v,\xh)]~.
}
with the data mapping
\eq{int5a}{\phib_1^-(u,\xh) = -R^2\p_u J^-\Big(\frac{\pi}{2} + \frac{u}{R}, \hat x\Big) ~,\quad \phib_1^+(v,\xh) =R^2 \p_v J^+\Big(-\frac{\pi}{2} + \frac{v}{R}, \hat x\Big) ~.}
The appearance of the $u$ and $v$ derivatives is dictated by symmetry, and the factors of $R$ by dimensional analysis.   The above formulas  provide a very simple and direct embedding of flat space physics in AdS.  This includes both the extraction of S-matrix elements and, because the dictionary is written in position space, the incorporation of large gauge transformations (which are singular in Fourier space).

The dictionary in the case of gauge fields is even simpler.  On the flat space side, the AFS data consists of the leading $r^0$ asymptotics of the angular components of the gauge field, whose positive and negative frequency components on $\Ic^-$ and $\Ic^+$ we denote as $\Abh_{A,0}^\pm$.  The AdS boundary sources for the gauge field are written as  $a_A^\pm(\tau,\xh)$.  The corresponding partition functions are then equal under the map
\eq{int5b}{ \hat{\overline A}_{A,0}^-(u,\xh) &= a^-_A\Big(\frac{\pi}{2} + \frac{u}{R}, \hat x\Big)~,\quad \hat{\overline A}_{A,0}^+(v,\xh) = a^+_A\Big(-\frac{\pi}{2} + \frac{v}{R}, \hat x\Big)~.   }
Furthermore, by working out the extension of a large transformation in flat space to the AdS boundary, 
the invariance of  $Z_{\rm AdS}$ under gauge transformations of the boundary sources  maps to the invariance of $Z_{\rm AFS}$ under large gauge transformations,  which implies the leading soft photon theorem, a result obtained from a slightly different perspective in
 \cite{Hijano:2020szl}.  
 For massless scalar QED, the general relation between the Carrollian and AdS partition can be summarized as
 $$ \boxed{Z_{\rm AdS}[J^+(\tau,\xh),J^-(\tau,\xh),a_A^+(\tau,\xh),a_A^-(\tau,\xh)] =  Z_{\rm AFS} [\phib_1^+(v,\xh),\phib_1^-(u,\xh),\Ab^+_{A,0}(v,\xh),\Ab^-_{A,0}(u,\xh)]} $$
 under the data mapping displayed above. 

When thinking about a hypothetical Carrollian theory holographically dual to Minkowski space a natural question is how degrees of freedom on $\Ic^-$ and $\Ic^+$ are related.  When embedded in AdS, these flat space boundaries map to the thin strips on the AdS boundary located at $\tau = \pm {\pi\over 2}.$   In this CFT picture the degrees of freedom in the two strips are related by Hamiltonian evolution over the $\Delta \tau = \pi$ interval of time separating them.  This suggests that there is no simple local way to couple the degrees of freedom on $\Ic^-$ to those on $\Ic^+$, since this coupling would have to encode the highly nontrivial CFT time evolution.

As summarized above, the main purpose of this work is to use the path integral formulation of the flat space S-matrix and AdS partition functions to arrive at simple and coherent derivations of the Carrollian dictionary for flat space amplitudes and their recovery from the AdS/CFT correspondence, including the incorporation of large gauge transformations.   The results so obtained are consistent with previous results, and in particular resonate particularly strongly with the relatively recent works \cite{Hijano:2020szl,Donnay:2022wvx,Bagchi:2023fbj}, which in turn build on the literature noted above.

The organization of the paper is as follows.   In section \rff{AFSreview} we briefly review the AFS approach to the S-matrix.   The AFS path integral expressed in terms of boundary data serves as a generating functional of the S-matrix that may be thought of as a Carrollian partition function; this interpretation is developed in section  \rff{Carroll-sec} in the case of scalar field theory.  In section \rff{ScalarQED} this is extended to scalar QED, followed by a discussion of obtaining soft photon theorems from invariance under large gauge transformations in section \rff{Sec:Invariance under large gauge transformations and soft theorems}.      We turn to general comments on the  flat limit of AdS in \rff{Flat space limit of AdS}, and then proceed to  work out  the relation between Carrollian and AdS partition functions for scalar fields (section  \rff{Flat limit scalar fields}) and scalar QED (section \rff{Flat limit gauge fields}).  The relationship between spacetime and gauge symmetries acting on the flat space boundary versus AdS boundary is discussed in section \rff{Flat space vs AdS symmetries}.   A series of appendices contain  supplementary material on the action of special conformal transformations on the flat space boundary data, on the relationship between flat space and AdS boundary terms, and on the derivation of the massless spin-1 bulk-boundary propagator.

\section{Review of AFS approach to the S-matrix}
\label{AFSreview}

The S-matrix in asymptotically flat spacetime may be computed by evaluating a path integral with specified asymptotic boundary conditions \cite{Arefeva:1974jv}.   We focus here on theories of massless particles, for which the boundary conditions are imposed at past and future null infinity, $\Ic^-$ and $\Ic^+$ respectively, though the extension to massive particles is straightforward.   Roughly speaking, we specify the positive(negative) frequency part  of fields at $\Ic^- (\Ic^+)$.  The ``roughly speaking" clause refers to the fact that in theories with gauge symmetry we also need to specify the zero frequency Goldstone modes, as we review in due course.

We begin by reviewing the simplest case, that of a massless real scalar field, following \cite{Kim:2023qbl}.  Working in $(-,+,+,+)$ signature the  action is
\eq{a1}{ I[\phi,\phib] & = \int\! d^4x \left( \frac{1}{2} \phi \nabla^2 \phi -V(\phi) \right) + I_\text{bndy}[\phi,\phib]  }
where $\phib$ encodes the boundary conditions, as will be  specified momentarily along
with the  corresponding boundary terms.   To specify boundary conditions we write the Minkowski metric as
\eq{a2}{ ds^2
&= -du^2 -2dudr +r^2\dr\Omega^2\cr
&= -dv^2 +2dvdr +r^2\dr\Omega^2}
with
\eq{a3}{ u=t-r~,\quad v=t+r~.}
Points on the unit $S^2$ will be denoted  here by  $\xh$, and later by the  complex coordinates $(z,\zb)$. 
We are interested in scattering solutions for which $\phi$ behaves as a free field at large $r$, falling off as $\phi \approx {1\over r} \phi_1 $  at fixed $v$ (on $\Ic^-$) or $u$ (on $\Ic^+$).    For the S-matrix the appropriate boundary conditions involve fixing the positive frequency part of  $\phi_1$ on $\Ic^-$ and the negative frequency part on $\Ic^+$.  Here a ``positive frequency" function is one whose Fourier expansion in time contains terms $e^{-i\omega t}$ with $\omega\equiv p^0>0$.   We thus write the asymptotics as
\eq{a3a}{ \phi(x) \approx \begin{cases}
			{1\over r} \phib_1^-(u,\xh) + {\rm positive~frequency} &  {\rm on}~ \Ic^+\\
         {1\over r} \phib_1^+(v,\xh) + {\rm negative~frequency} &  {\rm on}~ \Ic^-
		 \end{cases} }
so that the boundary conditions are encoded in the functions $\phib_1^-(u,\xh),\phib_1^+(v,\xh)$, with the superscripts denoting the frequency content, as shown in figure \ref{fig:triangle_with_waveform}.
It is useful to package this boundary data into a free field $\phib(x)$, obeying $\nabla^2 \phib(x)=0$, whose asymptotics also take the form \rf{a3a}, and which can be decomposed into positive and negative frequency parts,
\eq{a3b}{ \phib(x) = \phib^+(x) + \phib^-(x)~.}
The free field $\phib(x)$ contains the same information as the pair $\big( \phib_1^-(u,\xh),\phib_1^+(v,\xh) \big) $.

The boundary term in \rf{a1} may now be deduced by demanding that the variation of $I[\phi,\phib] $ with respect to $\phi$ takes the form $\delta I = \int\! d^4x ({\rm eqs~of~motion}) \delta \phi$ under variations that respect the boundary conditions, i.e. for which $\delta \phib_1^-|_{\scrI^+} = \delta \phib_1^+|_{\scrI^-}=0$.   A suitable choice is
\eq{a3c}{ I_\text{bndy}[\phi,\phib] =   (\phib^-,\phi)_{\Ic^+} - (\phib^+,\phi)_{\Ic^-}~,}
with
\eq{a3d}{  (\phib^-,\phi)_{\Ic^+} & = {1\over 2} \lim_{r\rt \infty}  \int \! dud^2\xh r^2 ( \phib^-\p_u \phi - \p_u \phib^- \phi ) =  {1\over 2}\int_{\Ic^+} du d^2 \xh  ( \phib_1^-\p_u \phi_1 - \p_u \phib_1^- \phi_1  ),\cr
 (\phib^+,\phi)_{\Ic^-} & = {1\over 2}\lim_{r\rt \infty}  \int \! dvd^2 \xh  r^2 ( \phib^+\p_u \phi - \p_u \phib^+ \phi ) = {1\over 2} \int_{\Ic^-} dv d^2 \xh  ( \phib_1^+\p_v \phi_1 - \p_v \phib_1^+ \phi_1  ),   }
where $\phi = \frac{1}{r}\phi_1 + (\text{subleading in }r)$.

The basic object of interest is the path integral viewed as a functional of the boundary data,
\eq{a3e}{ Z[\phib] = \int_{\bar\phi}{\cal D}\phi e^{iI[\phi,\phib]}~. }
$Z[\phib]$ is a generating functional for the S-matrix.  To make this explicit it is convenient to introduce mode expansions.  The boundary-condition-encoding  free field $\phib(x)$ admits the mode expansion
\eq{a4}{ \phib(x) = \int\frac{\dr^3p}{(2\pi)^3}\frac{1}{2\omega_p}\left( b(\vec p)e^{ip\cdot x} +  b^\dag(\vec p)e^{-ip\cdot x}\right), \ \ \ \ p^0 = \omega_p = |\vec{p}|~. }
The large $r$ asymptotics are obtained from the following  formula obtained by saddle point approximation of the integral at large $r$ (see e.g. \cite{Strominger:2017zoo}),
\eq{a4a}{ \int {d^2 \hat{p} \over (2\pi)^2}\omega_p f(\hat{p}) e^{ip\cdot x} = \begin{cases}  -{i\over 2\pi r} f(\xh) e^{-i\omega u} \quad {\rm on~} \Ic^+ \\
  {i\over 2\pi r} f(-\xh) e^{-i\omega v} \quad {\rm on~} \Ic^-  \end{cases}  }
where $d^2 \hat{p}$ is the standard measure on the unit sphere in $\vec{p}$ space. 
This gives 
\eq{a5}{ \phib(x) \approx \left\{ \begin{array}{cc}
   {-i\over 8\pi^2 r} \int_0^\infty \! d\omega  \left( b(\omega \xh) e^{-i\omega u} -  b^\dagger(\omega \xh) e^{i\omega u}  \right) & {\rm on}~ \scrI^+   \\  & \\
   {i\over 8\pi^2 r} \int_0^\infty \! d\omega \left( b(-\omega \xh) e^{-i\omega v} -  b^\dagger(-\omega \xh) e^{i\omega v} \right) & {\rm on}~ \scrI^-
\end{array} \right.  }
The boundary conditions  therefore may be expressed in terms of the modes $\big(b(\pv),b^\dagger(\pv)\big)$  via
\eq{a6}{ \phib_1^-(u,\xh) & = {i\over 8\pi^2} \int_0^\infty \! d\omega b^\dagger(\omega \xh) e^{i\omega u}  \cr
 \phib_1^+(v,\xh) & = {i\over 8\pi^2} \int_0^\infty \! d\omega b(-\omega \xh) e^{-i\omega v}~.}
 To compute $Z[\phib]$ in perturbation theory one can expand around the free field solution $\phib$ by writing  $\phi = \phib + \phi_G$ and then perform the path integral over $\phi_G$ order by order in the coupling.  One then arrives at the standard diagrammatic expansion, where the use of the Feynman propagator for internal $\phi_G$ lines is dictated by the boundary conditions.   Related methods have been used to study tree level scattering in curved space in \cite{Adamo:2017nia,Adamo:2021rfq,Gonzo:2022tjm}.

In terms of this mode data, S-matrix elements are obtained as
\eq{a7}{
\langle q_1,\ldots, q_M|\hat S|p_1,\ldots,p_N\rangle =\left[\prod^N_{k \text{ in}}\left( {2\omega_{p_k}(2\pi)^3}\frac{\delta}{\delta b_k(\vec p_k)} \right)\prod^M_{\ell \text{ out}}\left( {2\omega_{q_\ell}}{(2\pi)^3}\frac{\delta}{\delta b^\dag_\ell(\vec q_\ell)} \right)Z[\bar\phi] \right]_{\bar\phi = 0}.
}
Alternatively, the S-matrix operator can be written directly by promoting $\big(b(\pv),b^\dagger(\pv)\big)$ to operators obeying
\eq{a7a}{ [ \hat b(\pv),\hat b^\dagger(\pv')] = (2\pi)^3 2\omega_{\pv} \delta^{(3)}(\pv-\pv')}
as
\eq{a8}{ \hat{S} =  \normord{ e^{-iI_\text{bndy}[\hat{\phib},\hat{\phib}]} Z[\hat{\phib}] }   }
where the prefactor sets $\hat{S}=1$ in the case of a free field in Minkowski space.\footnote{For a free field in a curved asymptotically flat spacetime there is in general nontrivial scattering and particle creation, corresponding to the fact that prefactor does not cancel the path integral in such cases; see \cite{Kim:2023qbl}.}    These statements may be derived either by starting from the coherent state representation of the path integral, or by verifying perturbative equivalence with the LSZ prescription for the S-matrix.  See \cite{Arefeva:1974jv,Balian:1976vq,Jevicki:1987ax,Kim:2023qbl} for more details.

It is also straightforward to combine \eqref{a4} and \eqref{a6} to compute the bulk-boundary propagator for the Carrollian data. That is, an arbitrary free solution obeying the boundary conditions \eqref{a3a} may be written\footnote{These expressions assume suitable falloff conditions in $u'$ or $v'$ for the data on scri.}
\eq{aa1}{
    \phi(x) = \int_{\scrI^+}\dr^3x' K_{\scrI^+}(x; x')\bar\phi^-_1(x') + \int_{\scrI^-}\dr^3x' K_{\scrI^-}(x;x')\bar\phi_1^+(x')
}
where $x' = (u',\hat x')$ or $x' = (v', \hat x')$ with
\eq{aa2}{
    K_{\scrI^+}(x;x') &= \frac{i}{(2\pi)^2}\frac{1}{(u' + q(\hat x')\cdot x - i\epsilon)^2}\cr
    K_{\scrI^-}(x;x') &= \frac{i}{(2\pi)^2}\frac{1}{(v' + q(-\hat x')\cdot x + i\epsilon)^2}.
}
To keep this expression compact we have defined $q^\mu(\hat x') = (1, \hat x')$. These expressions are closely related to the Kirchoff-d'Adhemar formulas studied in \cite{Donnay:2022wvx}. We make use of \eqref{aa2} in sections \ref{Witten diagrams for flat space} and \ref{position space flat limit}.

In analogy with the bulk-boundary propagator of AdS, these expressions may also be related to the usual Feynman bulk-bulk propagator, $G_F(x-y) = \langle 0 |T\phi(x) \phi(y) |0\rangle$, which in position space takes the form
\eq{aaa1}{
    G_F(x) = \frac{1}{4\pi^2} \frac{1}{x^2 + i\epsilon},
}
and obeys $\nabla^2 G_F(x) =i\delta^{(4)}(x)$.
By direct comparison with \rf{aa2} we have
\eq{aaa2}{
    K_{\scrI^+}(x; x') =  2i\lim_{r'\arrow \infty}r' \p_{u'}G_F(x; x')
}
where the limit should be taken at fixed $u'$.  Similarly, 
\eq{aaa2z}{
    K_{\scrI^-}(x; x') =  -2i\lim_{r'\arrow \infty}r' \p_{v'}G_F(x; x')
}
with the limit taken at fixed $v'$.   These relations can also be derived  by a standard Green's function argument: for any solution $\phi$ of the wave equation with asymptotics \eqref{a6},
\eq{aaa3}{
    \phi(x) &= \int\dr^4 x' \delta(x - x') \phi(x')\cr
    &= i\int_{\scrI^+}\dr^3x' r^2(\p_{u'} G_F\phi^- - G_F\p_{u'}\phi^-) - i\int_{\scrI^-}\dr^3 x' r^2(\p_{v'}G_F\phi^+ - G_F\p_{v'}\phi^+)\cr
    &= \int_{\scrI^+}\dr^3x'(2ir\p_{u'}G_F)\phi^-_1 + \int_{\scrI^-}\dr^3x' (-2ir\p_{v'}G_F)\phi^+_1~.
}

\section{Structure of the Carrollian partition function}
\label{Carroll-sec}

\subsection{Definition of the partition function}

The generating functional $Z[\phib]$ is properly thought of as depending on the functions  $\phib_1^+(v,\xh)$ and $\phib^-_1(u,\xh)$, defined on $\Ic^-$ and $\Ic^+$ respectively. It follows that $Z[\bar\phi]$ can be viewed as the generating function for ``correlators'' supported on $\scrI^\pm$.     For example, a particular term in $Z$, namely one that turns out to describe $2\rt 2$ scattering, takes the form\footnote{This is equivalent to \eqref{int2}, with a less condensed notation.}
\eq{b1}{ \!\!\!Z^{\rm bndy}_{2,2}[ \phib] &= {1\over (2!)^2} \int_{\Ic^-} \!dv_1 d^2\xh_1 dv_2 d^2\xh_2 \int_{\Ic^+} \!du_3 d^2\xh_3 du_4 d^2\xh_4   G_{2,2}(v_1,\xh_1;v_2,\xh_2;u_3,\xh_3;u_4,\xh_4) \cr
& \!\! \!\!\!\!\!\!\!\times\p_{v_1} \phib_1^+(v_1,\xh_1) \p_{v_2} \phib_1^+(v_2,\xh_2)\big(-\p_{u_3} \phib_1^-(u_3,\xh_3)\big)   \big(-\p_{u_4} \phib_1^-(u_4,\xh_4)\big)~. }
 As discussed below, the  correlator $G_{2,2}$ is essentially the Fourier transform with respect to energy of the standard momentum space  on-shell $2\rt 2 $  amplitude; the $u$ and $v$ derivatives appearing in \rf{b1} are there to make this connection appear in the most direct fashion, as is the convention that each $u$ derivative is accompanied by a minus sign.\footnote{In fact, each $u$ derivative originally appears in the combination $\phib(u,\xh) {1\over 2} \pa_u G(u,\xh)$.  For present purposes it is convenient to integrate by parts so that the $u$ derivative acts purely on $\phib$.  The relative minus sign between $u$ and $v$ derivatives is due to the fact that $\p_u$ and $\p_v$ are outward and inward pointing normals respectively.} Note that we take the $\phi_1$ functions to die off at infinity sufficiently rapidly that we can freely integrate by parts in $u$ or $v$ and neglect any boundary terms.    $G_{2,2} $ can be assumed to be purely positive frequency with respect to its $u$ arguments, and purely negative frequency with respect to its $v$ arguments, since the integrations against the $\phib$ project out the opposite frequency components. 

 It is sometimes convenient to rewrite \rf{b1} in more compact notation.  We write the terms  at $n$th order in $\phib$ as
\eq{b2}{ Z^{\rm bndy}_n[\phib] = {1\over n!} \int_{\Ic} G_n(u_I,\xh_I) \prod_{i=1}^n \big(-\p_{u_i} \phib_1(u_i,\xh_i)\big)~.}
Here we are using the notation $u_I = \{u_1,u_2, \ldots, u_n \}$ and similarly for $\xh_I$. 
To recover the previous form \rf{b1} for $n=4$, $\p_u \phib_1(u,\xh)$ should first be replaced by $  \p_u  \phib_1^-(u,\xh)-\p_v \phib_1^+(v,\xh)$.  We then expand out the product and keep the term corresponding to \rf{b1}, while also replacing $(u_1,u_2)$ by $(v_1,v_2)$ in both the integration measure and in $G_4$.  When appropriate we may also make explicit the dependence on fields living on $\Ic^+$ versus $\Ic^-$.

\subsection{Relation to conventional amplitudes}
\label{conamps}

We now address the question of how the correlators appearing in the expansion of $Z[\phib]$ relate to conventional amplitudes computed via Feynman diagrams in momentum space.  To that end, let $\Gct_n(p_1,\ldots, p_n)$ denote the off-shell, momentum space, amputated, $n$-point scalar amplitude computed via Feynman diagrams, including the momentum conserving delta function.  We define off-shell position space amplitudes via Fourier transform,\footnote{With this sign convention for the Fourier transform, $\tilde {\mathcal{A}}$ is defined with all external momenta outgoing.}
\eq{b3}{ \Gc_n(x_1,\ldots x_n) = \int \left( \prod_{i=1}^n {d^4p_i \over (2\pi)^4}  \right)\Gct_n(p_1,\ldots p_n) e^{i\sum_i p_i \cdot x_i}~.}
From this we define
\eq{b4}{ Z^{\rm bulk}_n[\phib] = {1\over n!} \int\! \left( \prod_{i=1}^n d^4x_i\right) \Gc_n(x_1,\ldots x_n) \phib(x_1 ) \ldots \phib(x_n)~,}
where $\phib$ is a free field with mode expansion \rf{a4}. As the notation indicates, this is in fact the same object as appears in \rf{b2}, but written in the ``bulk description".  To see this, note that we should think of \rf{b4} as defining the generating functional for the S-matrix computed via the LSZ prescription, where the free field $\phib(x)$ supplies the external on-shell wavefunctions that project the amplitude to its on-shell part.  The equivalence of the two expressions for $Z_n$  is the same as the equivalence between the LSZ and AFS prescriptions for the S-matrix, which is known to be true (at least within perturbation theory) \cite{Arefeva:1974jv,Balian:1976vq,Jevicki:1987ax,Kim:2023qbl}.

We now wish to relate $\Gc$ and $G$  by equating \rf{b2} and  \rf{b4}.  We begin by writing
 \eq{b5}{ \phib(x) = \int {d^4p \over( 2\pi)^4} 2\pi \delta(p^2) B(p^0,\pv) e^{ip\cdot x}   }
 where
 \eq{b6}{ B(p^0,\pv) = \left\{
\begin{array}{cc}
  b(\pv) & p^0>0     \\
  b^\dagger(-\pv) & p^0<0
\end{array}
  \right.  }
It is straightforward to plug \rf{b3} and \rf{b5} into \rf{b4} and perform the spatial integrals to obtain
\eq{b6a}{ Z^{\rm bulk}_n[\phib] = {1\over n!} \int \left( \prod_{i=1}^n {d^4p_i\over (2\pi)^4}  2\pi \delta(p_i^2)B(p^0_i,\pv_i)\right) \Gct(-p_1,\ldots, -p_n)  }
Doing the $p^0$ integrals using the delta functions generates $2^n$ terms,
\eq{b6b}{ Z^{\rm bulk}_{n}[\phib] = {1\over n!}  \sum_{\eta_i = \pm} \int \left( \prod_{i=1}^n{d^3p_i\over (2\pi)^3} {1\over 2 \omega_{p_i}} B(\eta_i \omega_{p_i},\pv_i)\right) \Gct(-\eta_1 \omega_{p_1},- \pv_1 ;\ldots; -\eta_n \omega_{p_n},-\pv_n )  }
with $\omega_{p_i} = |\pv_i|$.  We then use \rf{b6} and change integration variables so that the argument of $b^\dagger$ is given according to $b^\dagger(\pv)$. Finally, we convert to spherical coordinates in momentum space.   Explicitly, for  the $n=2 $ case we  have 
\eq{b6c}{Z^{\rm bulk}_n[\phib] 
& = {1\over 2} \int {d^2\ph_1 \over (2\pi)^2} \int_{0}^\infty {d\omega_{p_1} \over 2\pi} {\omega_{p_1}\over 2} \int {d^2\ph_2\over (2\pi)^2 }\int_{0}^\infty {d\omega_{p_2} \over 2\pi} {\omega_{p_2}\over 2} \cr
& \quad\quad\quad\quad\quad\quad\quad\times \Big[ b(\omega_{p_1}\ph_1)b(\omega_{p_2}\ph_2) \Gct( -\omega_{p_1},-\omega_{p_1}\ph_1;-\omega_{p_2},-\omega_{p_2}\ph_2) \cr
& \quad\quad\quad\quad\quad\quad\quad\quad + b^\dagger(\omega_{p_1}\ph_1)b(\omega_{p_2}\ph_2) \Gct( \omega_{p_1},\omega_{p_1}\ph_1;-\omega_{p_2},-\omega_{p_2}\ph_2)  \cr
& \quad\quad\quad\quad\quad\quad\quad\quad  +b(\omega_{p_1}\ph_1)b^\dagger(\omega_{p_2}\ph_2)\Gct( -\omega_{p_1},-\omega_{p_1}\ph_1;\omega_{p_2},\omega_{p_2}\ph_2)\cr
&\quad\quad\quad\quad\quad\quad\quad\quad  +b^\dagger(\omega_{p_1}\ph_1)b^\dagger(\omega_{p_2}\ph_2)\Gct( \omega_{p_1},\omega_{p_1}\ph_1;\omega_{p_2},\omega_{p_2}\ph_2)    \Big] }
For present purposes it will be sufficient to be more schematic and  explicitly display dependence on a single field variable only, so that we write
\eq{b6d}{ Z^{\rm bulk}[\phib] =  \int {d^2\ph \over (2\pi)^2} \int_{0}^\infty {d\omega_p \over 2\pi} {\omega_{p}\over 2} \Big[ b(\omega_p \ph)\Gct(-\omega_p,-\omega_p \ph) + b^\dagger(\omega_p \ph)\Gct(\omega_p,\omega_p \ph) \Big] }

We now turn to the boundary partition function.  In $Z^{\rm bndy}_n$ there are once again $2^n$ terms corresponding to taking each $\phib$  to live on either $\Ic^+$ or $\Ic^-$. As in the bulk case, we can adopt a schematic notation and display only a single field insertion.  This gives
\eq{b7}{ Z^{\rm bndy}[\phib] & = \Big[   \int_{\Ic^-} d^2\xh dv \p_v \phib^+_1(v,\xh) G(v,\xh)+ \int_{\Ic^+} d^2\xh du \big(-\p_u \phib^-_1(u,\xh)\big) G(u,\xh) \Big] }
Plugging in \rf{a6} yields 
\eq{b8}{Z_{\rm bndy}[\phib] & = \Bigg[  {1\over 2\pi} \int_{\Ic^-} d^2 \xh \int_0^\infty {d\omega \over 2\pi} { \omega \over 2} b(\omega\xh) \Gt(-\omega,-\xh)+ {1\over 2\pi} \int_{\Ic^+} d^2 \xh \int_0^\infty {d\omega \over 2\pi} { \omega \over 2} b^\dagger(\omega\xh) \Gt(\omega,\xh)   \Bigg]   }
with
\eq{b9}{ \Gt(-\omega,-\xh) &= \int_{-\infty}^\infty \! dv G(v,-\xh) e^{-i\omega v}~, \cr
\Gt(\omega,\xh) &= \int_{-\infty}^\infty \! du G(u,\xh) e^{i\omega u}~. }
Equating this with \rf{b6d} we deduce
\eq{b10}{ \Gt(-\omega,-\xh) & = {1\over 2\pi} \Gct(-\omega,-\omega \xh)~, \cr
\Gt(\omega,\xh) & =  {1\over 2\pi}\Gct(\omega,\omega \xh) }
Note that the second argument in $\Gct$ originally involved a momentum variable $\hat{p}$, but now involves a position variable $\xh$;  this is just a change of labeling,  where we note that both variables are unit vectors that define a point on $S^2$.
Here $\omega>0$ in both relations, but we may combine them to write
\eq{b11}{ \Gt(\omega,\xh) =  {1\over 2\pi}\Gct(\omega,|\omega|\xh)}
which then holds for either sign of $\omega$.   Restoring all the arguments, the full relation is 
\eq{b12}{ \Gt(\omega_1,\xh_1 ; \ldots ; \omega_n, \xh_n) = {1\over (2\pi)^n}  \Gct(\omega_1,|\omega_1|\xh_1; \ldots ; \omega_n,|\omega_n|\xh_n) }
The sign choices for the frequencies determine which boundary component the corresponding argument refers to:  $\omega_i<0$ means that the $i$th argument lives on $\Ic^-$, while  $\omega_i>0$ means that the $i$th argument lives on $\Ic^+$.  

The position space boundary amplitudes are recovered by  Fourier transform
\eq{b13}{ &  G(v_1,\xh_1; \ldots ;v_m,\xh_m; u_{m+1},\xh_{m+1}; \ldots ; u_n, \xh_n) \cr
& = \left(\prod_{i=1}^m \int_{-\infty}^0 {d\omega_i \over 2\pi}  \right)  \left(\prod_{j=m+1}^n \int^{\infty}_0 {d\omega_j \over 2\pi}  \right)  \Gt(\omega_1,\xh_1 ; \ldots ; \omega_n, \xh_n)e^{-i \sum_{i=1}^m \omega_i v_i - i\sum_{j=m+1}^n \omega_j u_j   }  \cr
& = {1\over (2\pi)^n} \left(\prod_{i=1}^m \int_{-\infty}^0 {d\omega_i \over 2\pi}  \right)  \left(\prod_{j=m+1}^n \int^{\infty}_0 {d\omega_j \over 2\pi}  \right)  \Gct(\omega_1,|\omega_1|\xh_1); \ldots ; \omega_n,|\omega_n|\xh_n) e^{-i \sum_{i=1}^m \omega_i v_i - i\sum_{j=m+1}^n \omega_j u_j} \cr    }
Thus the Carrollian boundary amplitudes are given by the frequency Fourier transform of the on-shell momentum space amplitudes, as proposed in \cite{Banerjee:2018gce,Donnay:2022aba,Donnay:2022wvx,Bagchi:2022emh,Bagchi:2023cen}, and with explicit examples worked out in \cite{Bagchi:2023cen,Salzer:2023jqv,Nguyen:2023miw,Mason:2023mti}.  
The derivation given here makes the origin of this relation particularly transparent. 

Since the amplitude  contains an overall energy conserving delta function,\\$\Gct(\omega_1,|\omega_1|\xh_1); \ldots ; \omega_n,|\omega_n|\xh_n) \propto \delta(\omega_1 + \ldots +\omega_n)$, it follows that any Carrollian correlator with all legs on the same boundary component vanishes.   On the other hand, this conclusion need not hold if Minkowski space were replaced by a time-dependent asymptotically flat background.

\subsection{Witten diagrams for flat space}
\label{Witten diagrams for flat space}

Given the Carrollian bulk-boundary propagators \eqref{aa2}, one can construct the flat space analog of AdS Witten diagrams by replacing each external leg in a bulk diagram by a bulk-boundary propagator terminating somewhere along $\scrI$. What object do such diagrams compute?

Consider again \eqref{b4}, but now using \eqref{aa1}:
\eq{bb1}{
    Z_n^\text{bulk} = \frac{1}{n!}\int_{\scrI^+}\left( \int_{\mathcal{M}_4}\mathcal{A}_n(x_1,\ldots,x_n)K_{\scrI^+}(x_1; x'_1)\cdots K_{\scrI^+}(x_n; x'_n) \right) \bar\phi^-_1(x_1')\cdots \bar\phi^-_1(x_n').
}
Here we have written the expression for the case where all boundary points are on $\scrI^+$. Though this term is technically zero by energy conservation, we make this choice to save writing. The same expression holds for any combination of points on $\scrI^\pm$.

Since $\mathcal{A}_n$ is given by the sum of amputated diagrams, it's clear that the coefficient of the data $\bar \phi_1^-$ in \eqref{bb1} is, by definition, the sum over Witten diagrams. Comparing with our earlier definition \eqref{b2} of Carrollian correlators, we see that the sum of Witten diagrams computes a  differentiated Carrollian correlator.  Explicitly, 
\eq{bb1z}{  \p_{u_1'}  \ldots  \p_{u_n'}G_n(x'_1, \ldots ,x'_n) = \int_{{\cal M}_4}  \left(\prod_{i=1}^n d^4x_i  \right)  {\cal A}_n(x_1,\ldots,x_n) K_{\scrI^+}(x_1; x'_1) \ldots K_{\scrI^+}(x_n; x'_n)~.      }
In the case where any of the external lines are instead on $\scrI^-$, we should apply the replacement $\p_{u'} \arrow -\p_{v'}$.

\subsection{Carrollian two-point function}

The two-point function is a special case, and is worth spelling out in more detail. It may be computed in the free theory, where we assume that the wavefunction renormalization factor has been set to unity by a field redefinition if necessary.  The on-shell  bulk action reduces to a sum of boundary terms given by \rf{a3c}.  After an integration by parts we have
\eq{b4ca}{ I_{bndy} &= (\phib^-,\phi)_{\Ic^+} - (\phib^+,\phi)_{\Ic_{-} }\cr
& = \int_{\Ic^+}\! d^3x \phib^-_1(x) \p_u \phi_1(x) - \int_{\Ic^-}\! d^3x \phib^+_1(x) \p_{v} \phi_1(x)~.  }
In the first term only the positive frequency part of $\p_u \phi_1(x)$ survives the $u$ integration, and this part is sourced entirely by the fixed data on $\Ic^-$, so for this term we can write $\phi(x) = \int_{\Ic^-}\! d^3x' K_{\Ic^-}(x,x') \phib_1^+(x')$ and extract the leading $1/r$ coefficient.   Similarly, in the second term we can write $\phi(x) =$  $ \int_{\Ic^+}\! d^3x'  K_{\Ic^+}(x,x') \phib_1^-(x')$.  The action can then be written as 
\eq{b4caa}{ I_{bndy} & = \int_{\Ic^+}\! d^3x \int_{\Ic^-}\! d^3x' \p_u K^{(1)}_{\Ic^-}(x,x') \phib_1^-(x) \phib_1^+(x') - \int_{\Ic^-}\! d^3x \int_{\Ic^+}\! d^3x' \p_v K^{(1)}_{\Ic^+}(x,x') \phib_1^+(x) \phib_1^-(x')  }
where we use the notation $K^{(1)}_{\Ic^\pm}(x,x') = \lim_{r\rt \infty} [rK_{\Ic^\pm}(x,x')] $, the factor of $r$ coming from the asymptotic relation $\phib_1 \sim r \phib$.
The expressions \rf{aa2}  yield the near boundary asymptotics\footnote{These expressions may be obtained by noting that in \rf{aa2} we have $1/r^2$ falloff except when $\xh+\xh'=0$.  This signals the presence of $\delta^2(\xh+\xh') {1\over r}$, whose prefactor may be obtained by integrating over $\xh'$.  This is the same strategy as employed to deduce the boundary behavior of the AdS bulk-boundary propagator \cite{Witten:1998qj}.  }
\eq{b4cb}{  K_{\Ic^+}(x,x')   \approx   \left( {i\over 2\pi}  {1\over u'-v-i\eps}\right)  \delta^2(\xh+\xh') {1\over r} + \ldots \cr
 K_{\Ic^-}(x,x')   \approx  - \left( {i\over 2\pi}  {1\over v'-u+i\eps}\right)  \delta^2(\xh+\xh') {1\over r} + \ldots  }
where the delta function is defined by $\int\! d^2 \xh'  \delta^2(\xh+\xh')=1 $.  The two lines in \rf{b4cb} are equivalent since they only differ by the order in which we take points to the boundary.   Inserting these expressions into \rf{b4caa} yields
\eq{b4cc}{ I_{bndy} & =  -{i\over \pi} \int_{\Ic^+}\! d^3x \int_{\Ic^-}\! d^3x' {1\over (u-v'-i\eps)^2} \delta^2(\xh+\xh') \phib_1^-(x) \phib_1^+(x')~.  }
On the other hand, the definition of the Carrollian two-point function is 
\eq{b4cd}{ Z_2 & = iI_{bndy} = \int_{\Ic^+}\! d^3x \int_{\Ic^-}\! d^3x' \p_u (-\p_{v'}) G_2(x,x')  \phib_1^-(x) \phib_1^+(x')~.  }
Comparing yields 
\eq{b4ce}{ \p_u (-\p_{v'})  G_2(x,x') =   {1\over \pi} {1\over (u-v'-i\eps)^2} \delta^2(\xh+\xh') ~.  }
Integrating to get $G_2$ requires boundary conditions, which are not fixed by our general considerations.   Finally, we note that the two-point functions with both points on the same component of $\Ic$ vanish, as noted earlier in the general case.

\subsection{Bulk Poincar\'e transformations}

The Carrollian partition function is based on the Poincar\'e invariant bulk action and so inherits that symmetry.  The symmetry acts on the data appearing in the partition function, which consists of the asymptotic field data.   The invariance of the partition function under transformations of this data implies invariance properties of the boundary amplitudes, which are of course to  be understood as the Ward identities associated with Poincar\'e symmetry.

We will focus on Lorentz transformations, commenting on translations at the end.   Formulas look the simplest if we recast Lorentz transformations $x^\mu \rt x_\Lambda^\mu = \Lambda^\mu_{~\nu} x^\nu$ in SL(2,C) language; see e.g.  \cite{Oblak:2015qia}.  Define
\eq{b14}{ X =
\left(
\begin{array}{cc}
  x^0-x^3  &  x^1+ix^2    \\
  x^1- ix^2 &    x^0 +x^3
\end{array}
\right)
 }
and let $S$ be an element of SL(2,C),
\eq{b15}{ S = \left(
\begin{array}{cc}
  a  &  b   \\
  c &   d
\end{array}
\right)~,\quad ad-bc=1
 }
A Lorentz transformation then acts as
\eq{b16}{ X~ \rt  ~ X_\Lambda = S X S^\dagger~,\quad  S \in SL(2,C) }
This looks rather messy when written in terms of the $x^\mu$, but
since we are interested in the transformation of the asymptotic field data we only need the asymptotic transformation formulas for the coordinates.      Let's consider the transformations near $\Ic^+$, as usual defined as taking $r \rt \infty$ at fixed $u=t-r$.   Further, it is convenient to use complex coordinates on the sphere, defined via stereographic projection.  We then have coordinates $(r,u,z,\zb)$ defined as 
\eq{b17}{
\vec{x}^2=r^2, \quad t=u+r, \quad x^1+i x^2=\frac{2 r z}{1+z \bar{z}}, \quad x^3=r \frac{1-z \bar{z}}{1+z \bar{z}} 
}
with inverse
\eq{b18}{
r^2=\vec{x}^2, \quad u=t-r, \quad z=\frac{x^1+i x^2}{x^3+r}, \quad \zb=\frac{x^1-i x^2}{x^3+r} . }
The metric on the unit $S^2$ is 
\eq{b19}{ d\Omega^2 & = d\theta^2 + \sin^2 \theta d\phi^2 \cr& = 2\gamma_{z\zb}dzd\zb  }
with
\eq{b20}{\gamma_{z \bar{z}}=\frac{2}{(1+z \bar{z})^2}~.}
The full metric is then 
\eq{b21}{ ds^2 & = -dt^2 + d\vec{x}^2 \cr
& =-d u^2-2 d u d r+2 r^2 \gamma_{z \bar{z}} d z d \bar{z} }
To obtain the formulas on $\Ic^-$ simply replace $u\rt -v$ in the above\footnote{Note that the sphere coordinates on $\scrI^\pm$ used here are {\em not} related by an antipodal map.}.  At large $r$ we then have 
\eq{b22}{ & r~\rt ~ r_\Lambda=   {r \over  f_\Lambda(z,\zb) } + O(r^0)\cr
& u ~\rt ~u_\Lambda=  f_\Lambda(z,\zb)  u +O(r^{-1}) \cr
& z~\rt ~ z_\Lambda= {az+b\over cz +d} +O(r^{-1}) \cr
& \zb~\rt ~ \zb_\Lambda= {\ab\zb+\bb\over \cb\zb +\db} +O(r^{-1}) }
Here we have defined 
\eq{b23}{  f_\Lambda(z,\zb) =  \frac{1+z \bar{z}}{|a z+b|^2+|c z+d|^2}~, }
which is the Weyl factor picked up by the unit sphere metric under the M\"obius transformation $z\rt z_\Lambda$,
\eq{b24}{  d\Omega^2 \rt  d\Omega_\Lambda^2=  \big( f_\Lambda(z,\zb) \big)^2 d\Omega^2~. }

For a general Poincar\'e transformation we just replace the $u$ transformation in  \rf{b22}   by $u\rt  f_\Lambda(z,\zb) \big( u+ \alpha(z,\zb) \big)$  where $\alpha(z,\zb)$ is any linear combinations of the spherical harmonics $Y_\ell^m$ with $\ell=0,1$.  To see this, consider an infinitesimal translation by $b^\mu$. This acts by
\eq{b24a1}{
    \delta r = \frac{\vec b\cdot \vec x}{r},\ \ \ \delta u = b^0 - \delta r,\ \ \ \delta z = \frac{b^1 + i b^2}{x^3 + r} - (b^3 + \delta r)\frac{z}{x^3 + r}.
}
It's clear that the above only act at subleading order in the large $r$ expansion, with the exception of $\delta u$,
\eq{b24a2}{
    \delta u = b^t - \vec b\cdot \hat x = b^t - b^1 \frac{z + \overline z}{1 + z \overline z} - b^2 \frac{-i(z - \overline z)}{1 + z \overline z} - b^3 \frac{1 - z\overline z}{1 + z\overline z}.
}
In terms of stereographic coordinates the $\ell = 0,1$ harmonics $Y_\ell^m$ are
\eq{b24a3}{
    Y_0^0(z, \overline z) &= \frac{1}{\sqrt{4\pi}},\\
    Y^{-1}_1(z, \overline z) = \sqrt{\frac{3}{2\pi}}\frac{\overline z}{1 + z \overline z},\ \ \ Y^0_1(z,\overline z) &= \sqrt{\frac{3}{4\pi}}\frac{1 - z\overline z}{1 + z \overline z},\ \ \ Y^1_1(z, \overline z) = -\sqrt{\frac{3}{2\pi}}\frac{z}{1 + z \overline z},
}
hence the asymptotic transformation of $u$ is a linear combination of these functions as claimed. Furthermore, the asymptotic action of translations integrates up trivially, making our statements valid for finite translations as well.

We will also need momentum space versions of these formulas.   We label the components of a null momentum $q^\mu$ by its energy $q^0 = \omega_q$ and the angle on the sphere to which it points, described in complex coordinates by $(z_q,\zb_q)$.  The Cartesian components of  $q^\mu$ are then 
\eq{b25}{ q^\mu& = (q^0,q^1,q^2,q^3) \cr 
& = \frac{\omega_q}{1+z_q \bar{z}_q} \big(1+z_q \bar{z}_q, z_q+\bar{z}_q,-i(z_q-\bar{z}_q), 1-z_q \bar{z}_q\big)
  }
The Lorentz transformation $q^\mu \rt q^\mu_\Lambda=  \Lambda^\mu_{~\nu} q^\nu$ then acts as
\eq{b26}{  z_{ q_\Lambda} = {az_q+b\over cz_q+d},\ \ \ 
 \zb_{q_\Lambda} = {\ab\zb_q+\bb\over \cb\zb_q+\db},\ \ \ 
\omega_{q_\Lambda} = { \omega_q \over f_\Lambda(z_q,\zb_q) }. }

\subsection{Lorentz invariance of the Carrollian partition function}

Turning now to the transformation of the field variables, the scalar field transforms as
\eq{b27}{ \phib(x)~ \rt ~\phib_\Lambda(x) = \phib(\Lambda x)~. }
Recalling the definition of $\phib_1$ as $\phib(x) \approx {1\over r} \phib_1(u,z,\zb)$, as well as the transformation of $r$ given in \rf{b22}, we have that
\eq{b28}{ \phib_1(u,z,\zb)~ \rt ~ \phib_{1\Lambda} (u,z,\zb) = {r\over r_\Lambda} \phib_1(u_\Lambda,z_\Lambda,\zb_\Lambda) = f_\Lambda(z,\zb)  \phib_1(u_\Lambda,z_\Lambda,\zb_\Lambda)~.}
Lorentz invariance of the partition function implies that the boundary correlators must obey
\eq{b29}{  G_n(Y_1,\ldots,Y_n) =   \left( \prod_{i=1}^n  f_\Lambda(z_i,\zb_i) \right)   G_n(Y^\Lambda_1,\ldots,Y^\Lambda_n)~.  }
with $Y_i=(u_i,z_i,\zb_i).$
To derive \rf{b29} we can start from the Lorentz invariance of the standard momentum space amplitudes and then use the results in section  \ref{conamps}  to convert this to a transformation law for the boundary correlators, the main point being the transformation from frequency space to $u$ space. 
In general, if a momentum space function obeys
\eq{b32}{ \tilde{F}(\omega,z,\zb) = \tilde{F}(\omega_\Lambda,z_\Lambda,\zb_\Lambda) }
then in $u$-space, defined by
\eq{b33}{ F(u,z,\zb) = \int \! \frac{d\omega}{2\pi} \tilde{F}(\omega,z,z) e^{-i\omega u}, }
it obeys 
\eq{b34}{ F(u,z,\zb)  = f_\Lambda(z,\zb) F(u_\Lambda,z_\Lambda,\zb_\Lambda)   }
from which \rf{b29} follows.

Alternatively, we can verify  Lorentz invariance directly from the Carrollian partition function. Using \rf{b28} and \rf{b29} gives
\eq{b30}{ Z_n[\phib] & ~\rt~  Z_n[\phib_\Lambda ]\cr
& = {1\over n!} \int_{\Ic} G_n(u_i,z_i,\zb_i) \prod_{i=1}^n \big(-\p_{u_i} \phib_{1\Lambda}(u_i,z_i,\zb_i)\big)\cr
& =  {1\over n!} \int_{\Ic} G_n(u_{\Lambda i},z_{\Lambda i},\zb_{\Lambda i} ) \prod_{i=1}^n \dr u_i d^2\xh_i \big( f_\Lambda(z_i,\zb_i) \big)^2 \big(-\p_{u_i} \phib_{1\Lambda}(u_{\Lambda i},z_{\Lambda i},\zb_{\Lambda i} \big)~.   }
We now use \rf{b24} to deduce
\eq{b31}{     du_i d^2\xh_i   \big( f_\Lambda(z_i,\zb_i) \big)^2 \p_{u_i} = du_{\Lambda i} d^2 \xh_{\Lambda i} \p_{u_{\Lambda i}}   }
so that a change of integration variable in \rf{b30} gives $Z_n[\phib] = Z_n[\phib_\Lambda] $.

\section{Scalar QED}
\label{ScalarQED}

 We now consider massless scalar QED,  following the treatment in \cite{Kim:2023qbl}.  The action is 

\eq{c1}{ I  &=  \int\! d^4x  \left( {1\over 2} A^\mu \nabla^2 A_\mu +{1\over 2}\phi^* D^2 \phi + {1\over 2} (D^2 \phi)^*  \phi \right)+I_\text{bndy}~.  }
where UV counterterm and ghost actions are suppressed.  We work in Lorenz gauge, $\nabla^\mu A_\mu=0$.    The covariant derivative for a charge $Q$ scalar is
\eq{c2}{ D_\mu \phi = (\p_\mu -i eQ A_\mu)\phi}
The scalar field is assumed to obey the same asymptotics as in \rf{a3a}.  The scalar mode expansions are generalized in the obvious way; e.g. in \rf{a4} $b^\dagger(\pv)$ is replaced by $c^\dagger(\pv)$, the creation operator for anti-particles.

In Lorenz gauge the asymptotic data associated with the gauge field consists of the order $r^0$ behavior of the angular components $A_A$.  The timelike and longitudinal components are fixed in terms of these by the Maxwell equations and gauge condition.  For the purposes of studying large gauge transformations and their associated soft theorems it's important to isolate the Goldstone mode by writing \cite{He:2014cra}
\eq{c2a}{ A_\mu =  \Ah_\mu + \p_\mu \Phi. }
The hatted part $\Ah_\mu$  admits a Fourier expansion and asymptotically can be separated into positive and negative frequency parts. It thus obeys boundary conditions analogous to those of the scalar field,
\eq{c3}{ \Ah_A(x) \approx \begin{cases}
		 \Abh_{A,0}^-(u,\xh) + {\rm positive~frequency} &  {\rm on}~ \Ic^+\\
         \Abh_{A,0}^+(v,\xh) + {\rm negative~frequency} &  {\rm on}~ \Ic^+
		 \end{cases} }
where the remaining components can be determined through the Maxwell equations and the Lorenz gauge condition.
The Goldstone mode $\Phi$ obeys $\nabla^2 \Phi = 0$ by virtue of the Lorenz gauge condition. We allow $\Phi$ to have $r^0$ leading behavior, and the Laplace equation then imposes 
\eq{c4}{ \Phi(x) \approx \left\{\begin{array}{cc}
  \Phib_0(\xh) & {\rm on~}\scrI^+  \\
  \Phib_0(-\xh) & {\rm on~}\scrI^- 
\end{array}  \right.  }
where $-\xh$ denotes the sphere point antipodal to $\xh$.   Our boundary data for the gauge field  thus consists of the functions $\big( \Abh_{A,0}^-(u,\xh), \Abh_{A,0}^+(v,\xh),\Phib_0(\xh)\big) $.  As with the scalar field, it is convenient to adopt a compact notation and express the boundary data in terms of a single function $\Ab_{A,0}(u,\xh)$.   When evaluated on $\Ic^+$ this stands for $\Abh^-_{A,0}(u,\xh)+\p_A \Phib_0(\xh)$, while on $\Ic^-$ it stands for $\Abh^+_{A,0}(v,\xh)+\p_A \Phib_0(\xh')$.

Given these boundary conditions we can write down suitable  boundary terms in the action,
\eq{c4a}{ I_\text{bndy} &= (\Ab^{A-}_0,A_{A})_{\Ic^+} + (\phib^{*-}_1,\phi)_{\Ic^+} + (\phib^-_1,\phi^*)_{\Ic^+} \cr
&  \quad -(\Ab^{A+}_0,A_{A})_{\Ic^-}  - (\phib^{*+}_1,\phi)_{\Ic^-} - (\phib_1^+,\phi^*)_{\Ic^-}~,}
where we use the notation \rf{a3d} and where $A$ indices are raised using $\gamma^{AB}$ obeying $\gamma^{AB}\gamma_{BC}=\delta^A_C$. We are also assuming standard falloffs of fields  at infinity; see e.g. \cite{Campiglia:2015qka}.

The mode expansion of the non-Goldstone part of the boundary-condition-encoding free gauge field takes the form 
\eq{c5}{ \Abh_\nu(x)= \sum_{\alpha= \pm} \int \frac{d^3 q}{(2 \pi)^3} \frac{1}{2 \omega}\left[\varepsilon_\nu^{* \alpha}(\vec{q}) a_\alpha(\vec{q}) e^{i q \cdot x}+\varepsilon_\nu^\alpha(\vec{q}) a_\alpha(\vec{q})^{\dagger} e^{-i q \cdot x}\right]  }
where $\alpha =\pm$ denotes the two photon helicity states.   A convenient choice of polarization vectors is
\eq{c6}{ \varepsilon^{+\mu}(\vec{q})=\frac{1}{\sqrt{2}}(\bar{z}, 1,-i,-\bar{z}), \quad \varepsilon^{-\mu}(\vec{q})=\frac{1}{\sqrt{2}}(z, 1, i,-z)~, }
which obey $ q_\mu \varepsilon^{ \pm \mu}(\vec{q})=0$ and $ \varepsilon_\alpha^\mu \varepsilon_{\beta \mu}^*=\delta_{\alpha \beta} $, and we write $q^\mu$ as in \rf{b25}.  The asymptotic data is best expressed in terms of $A_A= \p_A x^\mu A_\mu$ where $A=(z,\zb)$, and where the stereographic coordinates on the sphere are given as 
\eq{c6a}{x^1+i x^2=\frac{2 r z}{1+z \bar{z}}, \quad x^3=r \frac{1-z \bar{z}}{1+z \bar{z}} .
}
Using this we have on $\Ic^+$
\eq{c6b}{ \Abh_z(x) & \approx  {-i\over 8\pi^2 }  \int_0^\infty \! d\omega  \left( \epsh^{*+}_z(\omega \xh) a_+(\omega \xh) e^{-i\omega u} - \epsh^-_z(\omega \xh) a_-^\dagger(\omega \xh) e^{i\omega u}  \right) \cr
   \Abh_{\zb}(x) & \approx  {-i\over 8\pi^2 }  \int_0^\infty \! d\omega  \left( \epsh^{*-}_{\zb}(\omega \xh) a_-(\omega \xh) e^{-i\omega u} - \epsh^+_{\zb}(\omega \xh) a_+^\dagger(\omega \xh) e^{i\omega u}  \right)  }
where we have defined 
\eq{c6c}{ \epsh_A(\omega \xh) = {1\over r} \p_A x^\mu \veps_\mu(\omega \xh)~,\quad A = z,\zb }
given explicitly by
\eq{c6d}{ & \epsh^+_z(\omega \xh)=0~,\quad \epsh^+_{\zb}(\omega \xh) = {\sqrt{2} \over 1+z\zb} \cr
& \epsh^-_z(\omega \xh)= {\sqrt{2} \over 1+z\zb} ~,\quad \epsh^-_{\zb}(\omega \xh) = 0 }
and which obey
\eq{c6da}{ \epsh^A_\alpha \epsh^*_{\beta A} = \delta_{\alpha\beta},}
where $A$ indices are raised using the inverse unit sphere metric $\gamma^{AB}$.

On $\Ic^-$ we have the asymptotics 
\eq{c6e}{  \Abh_z(x) & \approx  {i\over 8\pi^2 }  \int_0^\infty \! d\omega  \left( \etah^{*-}_z(-\omega \xh) a_-(-\omega \xh) e^{-i\omega v} - \etah^+_z(-\omega \xh) a_+^\dagger(-\omega \xh) e^{i\omega v}  \right) \cr
   \Abh_{\zb}(x) & \approx  {i\over 8\pi^2 }  \int_0^\infty \! d\omega  \left( \etah^{*+}_{\zb}(-\omega \xh) a_+(-\omega \xh) e^{-i\omega v} - \etah^-_{\zb}(-\omega \xh) a_-^\dagger(-\omega \xh) e^{i\omega v}  \right)  }
where now
\eq{c6f}{ \etah^\alpha_A(-\omega \xh)  = {1\over r} \p_A x^\mu \veps^\alpha_\mu(-\omega \xh)}
Note that $\etah^\alpha_A(-\omega \xh)  \neq  \left[{1\over r} \p_A x^\mu \veps^\alpha_\mu\right]\big|_{-\omega \xh} $ because $x^\mu$ and $-\omega \xh$ lie at antipodal points on the sphere.    Expressions for $\etah_\mu(-\omega \xh)$ are given by replacing $z\rt -1/\zb$ and $ \zb \rt -1/z$ in \rf{c6}, which acts as the antipodal map.  The expressions for $ \etah_A$ therefore look different from $\epsh_A$.  To restore the symmetry between expressions on $\Ic^+$ and $\Ic^-$  one can introduce new complex coordinates $(z',\zb')$  on $\Ic^-$, related to the original ones by the antipodal map, i.e. $z'=-1/\zb$.   This gives
\eq{c6fa}{ & \etah^+_{z'}(-\omega \xh)=0~,\quad \etah^+_{\zb'}(\omega \xh) =- {\sqrt{2} \over 1+z'\zb'} \cr
& \etah^-_{z'}(-\omega \xh)=- {\sqrt{2} \over 1+z'\zb'} ~,\quad \etah^-_{\zb'}(-\omega \xh) = 0 }
which just differ from \rf{c6d} by an overall sign coming from the $x^i \rt -x^i$ antipodal action.

The boundary generating functional takes a form analogous to the pure scalar case, where the data now consists of 
\eq{c6g}{ &\phib_1^-(u,\xh)~,\quad  \phib^{*-}_1(u,\xh)~,\quad \Ab^-_{A,0}(u,\xh) \quad {\rm on~} \Ic^+ \cr
&\phib_1^+(v,\xh)~,\quad  \phib^{*+}_1(v,\xh)~,\quad \Ab^+_{A,0}(v,\xh) \quad {\rm on~} \Ic^-~, }
along with the Goldstone mode $\Phi(\xh)$. 
Note that $\phib^{*-}$ refers to the negative frequency part of $\phib^*_1$ (as opposed to the complex conjugate of $\phib_1^-$). 

To illustrate the general structure of the partition function including the gauge field we write out a particular cubic term,
\eq{c7}{ Z_{1,1,1}^{\rm bndy} &= \int_{\Ic^-} d^2 \xh_1 dv_1 \int_{\Ic^+} d^2 \xh_2 du_2\int_{\Ic^+} d^2 \xh_3 du_3\cr&  \p_{v_1} \phib_1^{*+}(v,\xh_1) \big(-\p_{u_2} \phib^-_1(u_2,\xh_2)\big) \Ab_{A,0}(u_3,\xh_3) {1\over 2} \pa_{u_3} G^A(v_1,\xh_1;u_2,\xh_2;u_3,\xh_3)~.  }
where $A\pa B = A \p B - \p A B$.  Several comments are in order.  First, the Goldstone mode is included via $\Ab_{A,0}(u,\xh) = \Abh^-_{A,0} (u,\xh)+ \p_A \Phi(\xh)$.  Second, unlike for the scalar field, for the gauge field we cannot freely integrate by parts since the Goldstone mode implies that  the vector potential cannot be assumed to go to zero at the boundaries of $\Ic$. The choice of ${1\over 2} \pa_{u_3} $ thus represents a particular choice, which as we'll see below is what is needed in order that invariance under large gauge transformations yields the leading soft photon theorem with the correct coefficient.  By carefully considering asymptotic falloffs and boundary terms it should be possible to arrive at this prescription from first principles, but we will not do so here\footnote{It seems likely that this issue is the analog of a subtle factor of $1/2$ in the canonical formulation \cite{He:2014cra}.}.  Finally, as for the scalars, if the gauge field appears on $\Ic^-$ we should replace $\pa_u \rt - \pa_v$.

More general terms may be written schematically as
\eq{c8}{ Z_{m,n,p}[\phib^*,\phib,\Ab] &= {1\over m! n! p!} \int_{\Ic}  \left(\prod_{i=1}^m \p_{u_i} \phib^*_1(u_i,\xh_i) \right) \left(\prod_{j=1}^n \p_{u_j} \phib_1(u_j,\xh_j) \right) \left(\prod_{k=1}^p  \Ab_{A_{k},0}(u_k,\xh_k)  \right)  \cr
& \quad\quad \times {1\over 2} \pa_{u_{k_1} }\ldots {1\over 2}\pa_{u_{k_p}}G_{m,n,p}^{A_1\ldots A_p}(u_i,\xh_i;u_j,\xh_j;u_k,\xh_k) }
This schematic form incorporates field insertions on both $\Ic^+$ and $\Ic^-$ according to the rules given above.

\subsection{Relation to conventional amplitudes}

We proceed using the same strategy as in the pure scalar case, starting with the bulk action, which takes the form  
\eq{c9}{ Z_{m,n,p}^{\rm bulk} &={1\over m! n!p!} \int \left( \prod_{i=1}^m d^4x_i \right)  \left( \prod_{j=m+1}^{m+n} d^4x_j \right)   \left( \prod_{k=1}^p d^4y_k \right) \Gc^{\mu_1 \ldots \mu_p}(x_1,\ldots , x_{m+n};y_1 ,\ldots y_p)  \cr& \quad \times \phib^*(x_1) \ldots \phib^*(x_m) \phib(x_{m+1})\ldots \phib(x_{m+n}) \Ab_{\mu_1}(y_1)\ldots   \Ab_{\mu_p}(y_p)}
where $ \Gc^{\mu_1 \ldots \mu_p}(x_1,\ldots , x_{m+n};y_1 ,\ldots y_p) $ is the Fourier transform of the polarization stripped, off-shell, amputated correlator with $m$ external $\phi$ lines, $n$ external $\phi^*$ lines, and $p$ external $A_\mu$ lines.   The next step is to plug in the mode expansions and perform the spatial integrals.   The gauge field analog of \rf{b5} is 
\eq{c10}{\Ab_\mu(y) & =  \int {d^4q \over( 2\pi)^4} 2\pi \delta(q^2)\sum_\alpha \veps^{*\alpha}_\mu(q) D_\alpha(\omega,\qv) e^{iqy}   }
 with 
\eq{c11}{  D_\alpha(\omega,\qv)= \left\{
\begin{array}{cc}
   a_\alpha(\qv)&  \omega>0    \\
 a_\alpha^\dagger(-\qv) &    \omega <0
\end{array}
 \right. }
and
\eq{c12}{  \veps_\mu^\alpha(\omega,\qv)= \left\{
\begin{array}{cc}
 { \veps}_\mu^\alpha(\qv)&  \omega>0    \\
\veps_\mu^{*\alpha}(-\qv) &    \omega <0
\end{array}
 \right. }
As in the scalar case it is sufficient to be schematic and  only display dependence on a single gauge field insertion.  The analog of \rf{b6d} works out to be 
\eq{c13}{  Z_{m,n,p}^{\rm bulk} &={1\over m! n!p!}  \int {d^2 \qh \over(2\pi)^2} \int_{0}^\infty \!{d\omega\over 2\pi} {\omega^2 \over 2\omega} \sum_\alpha \Big[  a_\alpha(\omega \qh) \veps^{*\alpha}_\mu(\omega \qh) \Gct^\mu(-\omega,-\omega\qh)   + a^\dagger_\alpha(\omega \qh) \veps^{\alpha}_\mu(\omega \qh) \Gct^\mu(\omega,\omega \qh)    \Big] }

We now turn to the partition function. For the comparison to ordinary amplitudes we can omit the Goldstone mode and hence are allowed to integrate by parts along $\Ic$.   Focusing, as above, on a single photon, we have 
\eq{c14}{  Z_{m,n,p}^{\rm bndy} &={1\over m! n!p!}\Bigg[\int_{\Ic^-} d^2 \yh dv \Big(\p_v \Ab_{z,0}(v,\yh) G^z(v,\yh)  + \p_v \Ab_{\zb,0}(v,\yh) G^{\zb} (v,\yh)     \Big) \cr
& \quad\quad\quad\quad\quad -\int_{\Ic^+} d^2 \yh du \Big( \p_u \Ab_{z,0}(u,\yh) G^z(u,\yh)  + \p_u \Ab_{\zb,0}(u,\yh) G^{\zb} (u,\yh)  \Big)    \Bigg]   }
We then insert the asymptotic expressions \rf{c6b} and \rf{c6a} which gives
\eq{c15}{  Z_{m,n,p}^{\rm bndy} &={1\over m! n!p!}\Bigg[  {1\over 8\pi^2} \int d^2 \qh \int_0^\infty \! d\omega \omega  a_\alpha(-\omega \qh)\etah^{*\alpha}_A(-\omega \qh) \Gt^A(-\omega,-\qh)\cr
&\quad\quad\quad\quad + {1\over 8\pi^2} \int d^2 \qh  \int_0^\infty \! d\omega \omega  a_\alpha^\dagger(\omega \qh)\epsh^\alpha_A(\omega\qh) \Gt^A(\omega,\qh)  \Bigg]  }
Comparing to \rf{c13} we find
\eq{c16}{\epsh^\alpha_A(\omega\qh) \Gt^A(\omega,\qh) & = {1\over 2\pi}  \veps^{\alpha}_\mu(\omega \qh) \Gct^\mu(\omega,\omega \qh)  \cr
\etah^{*\alpha}_A(-\omega \qh) \Gt^A(-\omega,-\qh ) & = {1\over 2\pi}\veps^{*\alpha}_\mu(\omega \qh) \Gct^\mu(-\omega,-\omega\qh) }
Using \rf{c6da} we solve the top line as 
\eq{c17}{\Gt^A(\omega,\qh) & = {1\over 2\pi }\left[\veps^{\alpha}_\mu(\omega \qh) \Gct^\mu(\omega,\omega \qh)    \right]  \epsh^{*A}_\alpha (\omega \qh)  }
Using \rf{c6fa} the solution of the bottom line may be written 
\eq{c18}{ \Gt^{A'}(-\omega,-\qh ) & = -{1\over 2\pi}\left[\veps^{*\alpha}_\mu(\omega \qh) \Gct^\mu(-\omega,-\omega \qh)    \right]  \epsh^{A'}_\alpha (\omega \qh)~,\quad A'=(z',\zb')~,   }
where we recall that the antipodally flipped complex coordinates are $(z'=-{1\over \zb},\zb' =- {1\over z})$.
The expressions above hold for $\omega>0$, and the sign of the frequency argument in $\Gt$ determines whether the argument of the correlator is on $\Ic^+$  (positive argument) or $\Ic^-$  (negative argument).

The $(u,v)$ space correlators are then recovered as in \rf{b13}.  For illustration we write this out explicitly for the correlator that appears in the partition function as in \rf{c7},
\eq{c19}{ G^A(v_1,\xh_1;u_2,\xh_2;u_3,\xh_3)&=  {1\over (2\pi)^3} \int_{-\infty}^0 {d\omega_1 \over 2\pi}  \int^{\infty}_0 {d\omega_2 \over 2\pi}  \int^{\infty}_0 {d\omega_3 \over 2\pi} \cr
& 
\cdot\veps^\alpha_\mu(\omega_3 \xh_3)\Gct^\mu(\omega_1,|\omega_1|\xh_1; \omega_2, |\omega_2|\xh_2; \omega_3,|\omega_3|\xh_3) e^{-i \omega_1 v_1 -i \omega_2 u_2 -i\omega_3 u_3} \epsh^{*A}_\alpha(\omega_3 \xh_3) \cr}
We have thus arrived at the dictionary between the Carrollian correlators and standard momentum space amplitudes in scalar QED.

\subsection{Lorentz invariance of the Carrollian partition function}

We first recall some facts regarding the Lorentz transformation of the vector potential $A_\mu$.  Under a Lorentz transformation described in SL(2,C) language \rf{b16} the polarization vectors \rf{c6} obey 
\eq{d5}{  \Lambda^\beta_{~\mu} \veps^{+ }_\beta(\Lambda q) &= {  cz_q+d\over \cb \zb_q+\db  } \veps^{+}_\mu(q) -   {c \over \cb \zb_q+\db} {1+z_q\zb_q\over \sqrt{2}\omega_q} q_\mu \cr
 \Lambda^\beta_{~\mu} \veps^{- }_\beta(\Lambda q) &= {  \cb \zb_q+\db \over  cz_q+d} \veps^{-}_\mu(q) -   {\cb \over c z_q+d} {1+z_q\zb_q\over \sqrt{2}\omega_q} q_\mu }
The inhomogeneous terms on the right hand side are responsible for the familiar fact that the physical vector potential only transforms like a four-vector up to a gauge transformation.   
From \rf{d5} we compute
\eq{d6}{  A_\mu(\Lambda x) \Lambda^\mu_{~\nu}  & =  \sum_{\alpha= \pm} \int \frac{d^3 q}{(2 \pi)^3} \frac{1}{2 \omega}\left[\varepsilon_\nu^{* \alpha}(\vec{q}) a_{\alpha\Lambda} (\vec{q}) e^{i q \cdot x}+\varepsilon_\nu^\alpha(\vec{q}) a_{\alpha\Lambda}(\vec{q})^{\dagger} e^{-i q \cdot x}\right] +\p_\nu \Phi_\Lambda(x)   }
where we have defined
\eq{d7}{ a_{+\Lambda}(\vec{q})^{\dagger} & =  {  cz_q+d\over \db \zb_q+\db  }  a_{+}(\Lambda\vec{ q})^{\dagger} \cr
  a_{-\Lambda}(\vec{q})^{\dagger} & =  { \cb \zb_q+\db \over  cz_q+d }  a_{-}(\Lambda\vec{ q})^{\dagger} }
and
\eq{d8}{ \Phi_\Lambda(x)  =  i  \int \frac{d^3 q}{(2 \pi)^3} \frac{1}{2 \omega_q} \left[  {\cb \over cz_q  +d} a_+(\vec{\Lambda q}) + {c \over \cb \zb_q  +\db}  a_-(\vec{\Lambda q})  \right] {1+z_q\zb_q\over \sqrt{2}\omega_q}  e^{i q \cdot  x} + {\rm H.C.}     }
To define a Lorentz transformation law for $A_\mu$ that is equivalent to a transformation of the mode operators  we need to include a compensating gauge transformation that cancels the $\p_\nu \Phi_\Lambda $ term,
\eq{d9}{ A_\nu(x) &~ \rt~  A^\Lambda_\nu(x) =   A_\mu(\Lambda x) \Lambda^\mu_{~\nu}  - \p_\nu \Phi_\Lambda.}
This transformation is then equivalent to the transformation of the modes given in  \rf{d7}.   Lorentz invariance requires that the compensating gauge transformation be an invariance of the bulk generating functional, which implies that the bulk amplitudes obey the usual transversality condition, e.g. $q_\mu \Gct^\mu(p_1,p_2,q)=0$, where $q$ denotes the photon momentum. It should be noted that although $\Phi_\Lambda$ is exotic in the sense that it is a state-dependent gauge transformation, $\p_\mu\Phi_\Lambda$ obeys the same fall-offs as $A_\mu$ and hence is a small gauge transformation.

The Lorentz transformation of the modes in \rf{d7} applied to the asymptotic expressions \rf{c7} is readily seen to imply the following Lorentz  transformation  of the gauge field boundary data
\eq{d10}{ \Ab_{z,0}(u,z,\zb)&~\rt ~ \Ab^{\Lambda}_{z,0}(u,z,\zb) = {dz_\Lambda \over dz} \Ab_{z,0}(u_\Lambda,z_\Lambda,\zb_\Lambda) = {1\over (cz+d)^2} \Ab_{z,0}(u_\Lambda,z_\Lambda,\zb_\Lambda) \cr
\Ab_{\zb,0} (u,z,\zb)&~\rt ~ \Ab^{\Lambda}_{\zb,0}(u,z,\zb) = {d\zb_\Lambda \over d\zb} \Ab_{\zb,0}(u_\Lambda,z_\Lambda,\zb_\Lambda) = {1\over (\cb\zb+\db)^2} \Ab_{\zb,0}(u_\Lambda,z_\Lambda,\zb_\Lambda)   }
Inserting this in the boundary generating functional \rf{c8} and performing some manipulations analogous to \rf{b30} shows that Lorentz invariance requires that the boundary correlators obey (recall that $(m,n,p)$ stand for the number of $\phi^*$, $\phi$, and $A_A$ insertions respectively)
\eq{d11}{&  \left(\prod_{i=1}^m f_\Lambda(z_i,\zb_i) \right) \left(\prod_{j=1}^n f_\Lambda(z_j,\zb_j) \right)  \left(\prod_{k=1}^p f_\Lambda(z_k,\zb_k) \right)^2 G_{m,n,p}^{A_1\ldots A_p}(u_{\Lambda i},z_{\Lambda i};u_{\Lambda j},z_{\Lambda j};u_{\Lambda k},z_{\Lambda k}) \cr
& \quad =   \left(\prod_{k=1}^p { d\xh^{A_k}_\Lambda \over d\xh^{B_k} } \right) G_{m,n,p}^{B_1\ldots B_p}(u_i,z_i;u_j,z_j;u_k,z_k)}
The factors of $\big( f_\Lambda(z_k,\zb_k)\big)^2 $ come from the transformation of the sphere measure (see \rf{b24}).  Only a single power of $f_\Lambda$ appears for the scalar arguments because the factor of $r$ stripped out to define $\phib_1$ introduces a compensating factor. More generally, Carrollian correlators and sources can all be assigned a weight corresponding to the power of $f_\Lambda$ that appears, and Lorentz invariance requires a combined weight that cancels that of the integration measure.

\section{Invariance under large gauge transformations and  soft theorems}
\label{Sec:Invariance under large gauge transformations and soft theorems}

We now consider a collection of scalar fields of charge $Q_i$.  To lighten notation, in this section we replace complex conjugate fields $\phi^*$ by  $\phi$ fields of the opposite sign charge.

\subsection{Soft photon theorems}

Let $\Gct_n(p_1,\ldots p_n)$ be the scalar $n$-point amplitude, and let $\Gct^\mu_{n,1}(p_1,\ldots p_n;q)$ be the same (polarization stripped) amplitude with the addition of a photon of momentum $q$. The soft photon expansion consists of the relation \cite{Low:1958sn}
\eq{d19}{ \veps_\mu(q)  \Gct^\mu_{n,1}(p_1,\ldots p_n;q) =  \big[ S^{(0)}+S^{(1)}\big] \Gct_n(p_1,\ldots p_n) +O(\omega_q) }
where 
\eq{d20}{ S^{(0)} = e\sum_{i=1}^n Q_i {p_i \cdot \veps(q) \over p_i \cdot q}, \quad S^{(1)} =-ie \sum_{i=1}^n Q_i {\veps_{\mu}(q) q_\nu J_i^{\mu\nu}  \over p_i \cdot q} }
and the angular momentum operator is 
\eq{d21}{ J_i^{\mu\nu} = i \left( p^\mu_i {\p \over \p p_{i\nu}}-p_i^\nu {\p \over \p p_{i\mu}}\right)~. }
The leading $O(\omega^{-1})$ part associated with $S^{(0)}$ is completely universal, holding for any type of matter at any loop order.   The situation is different for  the subleading $O(\omega^0)$ part associated with $S^{(1)}$.  For example, at loop level there is generically a $\ln \omega$ part which dominates \cite{Sahoo:2018lxl}. For this reason, when we discuss the subleading soft theorem we will restrict to tree level processes computed in the scalar QED theory defined above.  

\subsection{Leading soft theorem from large gauge transformations} 

Our interest here is in the  invariance of the partition function under large gauge transformations,
\eq{d12}{  \delta \phib_{1i } = ie Q_i \lambda \phib_{1i}~,\quad \delta \Ab_{A} = \p_A \lambda~,}
where  $\lambda$ obeys $\nabla^2 \lambda =0$ to preserve Lorenz gauge and hence has asymptotics as in \rf{c4},
\eq{d13}{ \lambda(x) \approx \left\{\begin{array}{cc}
  \lambda_0(\xh) & {\rm on~}\scrI^+  \\
  \lambda_0(-\xh) & {\rm on~}\scrI^- 
\end{array}  \right.  }
This invariance of the action is manifest from its bulk definition. 
The invariance of the partition function involves a cancellation between the variations of terms with $p$ photons and $p+1$ photons, the extra factor of the coupling associated to the latter compensated for by the factor of $e$ in $\delta \phib_{1i}$.   For notational simplicity we consider the $p=1$ case.  The relevant terms in the partition function are then 
\eq{d14}{ Z_n^\phi[\phib_i] & = \int_{\Ic}\left( \prod_{i=1}^n \big( -\p_{u_i} \phib_{1i}(u_i,z_i)\big)\right)G_n(u_i,z_i)\cr
Z_{n,1}^{\phi,A}[\phib_i,\Ab_0] & = \int_{\Ic} \left( \prod_{i=1}^n  \big(-\p_{u_i}\phib_{1i}(u_i,z_i)\big)\right) \Ab_{A,0}(u,z){1\over 2}\pa_u G^A_{n,1}(u_i,z_i;u,z)~,}
where we are using the condensed notation which allows for fields on both $\Ic^+$ and $\Ic^-$. 
The gauge variations of these are 
\eq{d16}{ \delta Z_n^\phi[\phib_i] & =  ie \int_{\Ic}  \left( \sum_{i=1}^n Q_i \lambda(z_i) \right) \left(\prod_{i=1}^n  \big(-\p_{u_i} \phib_{1i}(u_i,z_i)\big)\right)G_n(u_i,z_i) \cr
\delta   Z_{n,1}^{\phi,A}[\phib_i,\Ab_0] & = -{1\over 2} \int_{\Ic}\p_u \nabh^{(z)}_A G^A_{n,1}(u_i,z_i;u,z) \prod_{i=1}^n (-\p_{u_i} \phib_{1i}(u_i,z_i))\lambda(z) ~,}
where $\nabh_A$ is the covariant derivative on the unit $S^2$.   The bottom line should be thought of as a sum of two terms, corresponding to the gauge variation of  $\Ab_\mu$ on $\Ic^-$ and $\Ic^+$; these two contributions are equal \cite{He:2014cra} so we just retain  the $\Ic^+ $ contribution and remove the factor ${1\over 2}$. Cancellation of the two terms in \rf{d16} then  implies
\eq{d17}{ \sqrt{\gamma(\xh)} \nabh^{(x)}_A \int\! du \p_u G^A_{n,1}(u_i,\xh_i;u,\xh)  = ie \left(   \sum_{i=1}^{m} Q^{\rm out}_i   \delta^{(2)}(\xh-\xh^{\rm out}_i) - \sum_{i=1}^{n} Q^{\rm in}_i   \delta^{(2)}(\xh-\xh^{\rm in}_i) \right) G_n(u_i,\xh_i)  }
where we have now separated out the matter contributions on $\Ic^+$ and $\Ic^-$. 
This is readily solved using the Green's function on $S^2$, which in complex coordinates gives
\eq{d18}{ \sqrt{\gamma(z,\zb) }\int_{-\infty}^\infty\! du \p_u G^z_{n,1}(u_i,z_i;u,z) & =  { ie\over 2\pi} \left( \sum_{i=1}^m  {Q^{\rm out}_i \over \zb-\zb^{\rm out}_i}- \sum_{i=1}^n  {Q^{\rm in}_i \over \zb-\zb^{\rm in}_i} \right) G_n(u_i,z_i) \cr
 \sqrt{\gamma(z,\zb)} \int_{-\infty}^\infty\! du \p_u G^{\zb}_{n,1}(u_i,z_i;u,z)  &=  { ie\over 2\pi} \left( \sum_{i=1}^m  {Q^{\rm out}_i \over z-z^{\rm out}_i}-\sum_{i=1}^n  {Q^{\rm in}_i \over z-z^{\rm in}_i} \right) G_n(u_i,z_i)  }
To convert this to the usual soft theorem we first of all 
use the Fourier transform relation 
\eq{d19a}{ \int_{-\infty}^\infty \! du \p_u f(u) = \lim_{\omega \rt 0} \big[ -i\omega \tilde{f}(\omega) \big]}
so that the integrals on the left hand side of \rf{d18} pick out the $1/\omega$ poles. 
We then convert the boundary amplitudes to ordinary amplitudes using \rf{c17},
\eq{d20a}{ \Gt^{\zb}(p_i,q) & = {1\over 2\pi} {1+z\zb \over \sqrt{2}} \veps^+_\mu(\omega \qh) \Gct^\mu(p_i;\omega,\omega\qh)  \cr
\Gt^{z}(p_i,q) & = {1\over 2\pi} {1+z\zb \over \sqrt{2}} \veps^-_\mu(\omega \qh) \Gct^\mu(p_i;\omega,\omega\qh) }
Finally we use the identity
\eq{zs49}{ \frac{1+z \bar{z}}{\sqrt{2}}\left[\sum_{k=1}^m \frac{Q_k^{\text {out }}}{z-z_k^{\text {out }}}-\sum_{k=1}^n \frac{Q_k^{\text {in }}}{z-z_k^{\text {in }}}\right]=\left[\sum_{k=1}^m\frac{\omega Q_k^{\text {out }} p_k^{\text {out }} \cdot \veps^{+}}{p_k^{\text {out }} \cdot q}-\sum_{k=1}^n \frac{\omega Q_k^{\text {in }} p_k^{\text {in }} \cdot \veps^{+}}{p_k^{\text {in }} \cdot q}\right]}
where momenta are written as in \rf{b25}.   We likewise have an identity with $z \rt \zb$ and $\veps^+\rt \veps^-$.  Putting these facts together we arrive at the standard leading soft theorem for emission of a photon, 
\eq{d21a}{ \veps^+_\mu(\omega \qh) \Gct^\mu_{n,1}(p_i;\omega,\omega\qh)  =e\left[\sum_{k=1}^m\frac{ Q_k^{\text {out }} p_k^{\text {out }} \cdot \veps^{+}}{p_k^{\text {out }} \cdot q}-\sum_{k=1}^n \frac{ Q_k^{\text {in }} p_k^{\text {in }} \cdot \veps^{+}}{p_k^{\text {in }} \cdot q}\right] \Gct_n(p_i)   + \ldots  }
where $\ldots$ are higher order in $\omega$.  This generalizes in the obvious way to amplitudes with any number of hard photons along with the soft photon.   This derivation of the soft theorem from large gauge invariance clearly contains the same basic ingredients as the original derivation \cite{He:2014cra}, the difference being the use of a functional formalism rather than the  canonical formalism of  \cite{He:2014cra}.

\subsection{Subleading soft theorem}

The subleading soft theorem gives a relation between the $O(\omega^0)$ part of the amplitude to emit a photon with energy $\omega$ and the same amplitude without the photon.  This relation then implies a relation between the corresponding terms in the partition function.  This can be written as an invariance of the partition function under a transformation of the boundary data, which one can regard as a symmetry.  For the reasons noted above, this symmetry only holds as a statement regarding the transformation of the on-shell tree level action with respect to boundary data.  How to identify this symmetry directly, i.e. without reverse engineering it from the amplitudes, is somewhat mysterious,  as we discuss at the end of this section.  The analysis that follows draws heavily on \cite{Lysov:2014csa}. 

In position space, the $O(\omega^0)$ part of the photon emission amplitude is written as \newline $\int_{-\infty}^\infty \! du G^A_{n,1}(u_I,z_I;u,z) $.   Converting the $O(\omega^0)$  part of \rf{d19} into a relation between boundary correlators eventually gives
\eq{d28}{ \sqrt{\gamma(z)} \int\! du G_{n,1}^z(u_I,z_I;u,z) &  = {ie \over 2\pi}{1\over 1+z\zb}  \sum_{i=1}^n {Q_i\over \zb-\zb_i} \Bigg[ -(1+z\zb_i)  u_i G_n(u_I,z_I) \cr
&  +(z-z_i)(1+z_i\zb_i) \nabh_{z_i} \int^{u_i} du'_i G_n(u_1,z_1,\ldots ,u'_i,z_i, \ldots u_n,z_n) \Bigg]\cr}
and similarly for $G^{\zb}_{n,1}$. This is the subleading analog of \rf{d18}.   Using this relation we can now engineer a transformation of the asymptotic data under which the transformations of $Z_n[\phib]$ and $Z_{n,1}[\phib,\Ab]$  cancel.  For the $G_{n,1}^z$ part of $Z_{n,1}$ this works out to be, after somewhat lengthy manipulations,  
\eq{d29}{    \delta \Ab_{z,0}(u,z) &=    u\nabh_{z} \nabh_A\lambda^A(z,\zb) \cr
  \delta \Ab_{\zb,0}(u,z) &=  0 \cr
\delta \p_{u_i} \phib_{1i} (u_i,z_i)& = ieQ_i  \Big[ \p_{u_i} \Big(u_i \nabh_{A}\lambda^{A}(z_i.\zb_i) \phib_1(u_i,z_i) \Big) + \lambda^A(z_i,\zb_i) \nabh_{A} \phib_1(u_i,z_i)    \Big]      }
where $\lambda^A$ is a vector field on $S^2$ with $\lambda^{\zb}=0$.   For $G_{n,1}^{\zb}$  we have the analogous transformations, where now $\lambda^z=0$ and  $\Ab_{\zb,0}$ transforms via a nonzero $\lambda^{\zb}$.

The first line and the first term on the right hand side of the third line in \eqref{d29} look like a gauge transformation with parameter $\lambda = u \nabh_A\lambda^A$, but the other terms spoil this interpretation.  Most notably, the transformation of $\phib_{1i}$ is nonlocal since what appears on the right hand side of the third line is not a total $u_i$ derivative.    The symmetry corresponding to this transformation is certainly not manifest in the scalar QED action, but this can be accounted for by the nonlocal nature of the transformation together with the fact that it has only been defined to act on the on-shell classical action expressed in terms of asymptotic data, so the extension of this transformation into the bulk remains unclear\footnote{We note, however, that a non-local transformation of the boundary data does not preclude an extension to a local action on the bulk fields. In appendix \ref{Sec:Asymptotic special conformal transformations} we show as an example that special conformal transformations also have this property.}.  We should also note that we have not kept sufficiently careful track of possible boundary terms at the endpoints of $\Ic$; these are likely to be relevant if one tries to directly verify that \rf{d29} leaves the action invariant.   For all these reasons, the status and implications of the subleading symmetry within our framework is rather murky.  

Our analysis is based on looking for symmetries that preserve our asymptotic boundary conditions.  We should note that there is an alternative approach \cite{Campiglia:2016hvg,AtulBhatkar:2019vcb,Peraza:2023ivy,Choi:2024ygx,Nagy:2024dme} based on extending the phase space to allow for gauge transformations that grow at large $r$.  This approach does lead to a symmetry based derivation of subleading soft theorems, and so would be interesting to accommodate within our framework.

\section{Flat space limit of AdS}
\label{Flat space limit of AdS}

The formulation of the Minkowski S-matrix in terms of the AFS path integral is directly analogous to the GKP/W version of the AdS/CFT dictionary. Since S-matrix elements are obtained from the large AdS radius limit of the  AdS/CFT partition function, it is natural to ask for the relation between these generating functionals.

The flat space S-matrix is extracted from AdS boundary correlators by generating highly collimated wavepackets at the AdS boundary which fall freely into the bulk before colliding in a region much smaller than the AdS radius \cite{Polchinski:1999ry, Susskind:1998vk}. We work  in global coordinates $(\rho, \tau, \hat x)$ on global AdS$_{d+1}$ where the metric is
\eq{7a}{
    \dr s^2_{\text{AdS}} = \frac{R^2}{\cos^2\rho}(-\dr\tau^2 + \dr\rho^2 + \sin^2\rho \dr\Omega_{d-1} )
}
and $0\leq \rho \leq \frac{\pi}{2}$ is the radial coordinate with AdS boundary at $\rho = \frac{\pi}{2}$, $\tau$ the global time, and $R$ the AdS radius. The flat region can be placed at the center of AdS and identified by taking $R\arrow\infty$ while holding
\eq{7b}{
    r= \rho R,\ \ \ t = \tau R
}
fixed. The metric in this region indeed becomes flat,
\eq{metric flat limit}{
    \lim_{R\rightarrow\infty}\dr s^2_\text{AdS} = - \dr t^2 + \dr r^2 + r^2 \dr\Omega_{d-1}^2.
}
A highly collimated wavepacket having width on the order of $\delta\tau = \frac{1}{R}$ at the AdS boundary has high energy content with respect to global time. Hence, these excitations travel radially inwards and outwards concentrated along null geodesics and collide at the bulk point singularity \cite{Gary:2009ae}. The travel time for a null geodesic between the origin and the AdS boundary is $\Delta\tau = \frac{\pi}{2}$, so for collisions to occur within the flat region defined by \eqref{7b}, we center our thin packets about $\tau = \pm\frac{\pi}{2}$. We summarize the setup in figure \ref{Lorentzian pill}.   This construction \cite{Polchinski:1999ry, Susskind:1998vk}, or modifications thereof, is adopted in most work on this subject \cite{Gary:2009ae,Gary:2009mi,Giddings:1999jq,Fitzpatrick:2011jn,Hijano:2019qmi,Hijano:2020szl,Komatsu:2020sag,Duary:2022pyv,Bagchi:2023fbj,deGioia:2024yne,Alday:2024yyj,Marotta:2024sce}. See \cite{Li:2021snj} for a survey of approaches.

Since AdS acts as a box, one should worry that the wavepackets will collide with periodicity $2\pi$, muddying our ability to extract the S-matrix elements of a single collision. However, this can be avoided by truncating the cylinder with Euclidean ``caps'' at $\tau = \pm \pi$, as shown in figure \ref{Lorentzian pill} and emphasized in \cite{Hijano:2020szl}, projecting us onto the vacuum state at early and late times, but before a repeat collision can occur. In perturbation theory, this amounts to using the bulk-bulk and bulk-boundary propagator obeying Feynman boundary conditions.

In order to obtain the flat S-matrix, it's necessary to assume that evolution is free everywhere except within the flat region. This assumption should be thought of as the AdS analog of the basic assumption of LSZ reduction that all particles are asymptotically free. In sections \ref{Formal Argument: Witten idagrams} and \ref{Formal Argument: Path integral} we show more precisely how this assumption together with the setup described above allows one to extract flat S-matrix elements from AdS boundary correlators.

Though the present approach to taking the flat limit is capable of producing both massive and massless particles in the flat region, we focus on the massless case for simplicity. In the bulk AdS, this corresponds to fixing any $\Delta \sim O(1)$ while a massive flat particle would descend from taking $\Delta \sim O(R)$ since $m^2R^2 = \Delta(\Delta - d)$. See \cite{Hijano:2019qmi,Hijano:2020szl} for comments on the massive case, and \cite{Komatsu:2020sag} for some subtleties special to the massive limit.

Our main results for this paper are the data map under which the AdS and AFS partition functions are equal in the case of spin 0, in section \ref{Flat limit scalar fields}, and spin 1 fields, in section \ref{Flat limit gauge fields}. For the scalar field we find
\eq{7flat1}{
    Z_\text{AdS}[J^+(\tau, \hat x), J^-(\tau, \hat x)] = Z_\text{AFS}[\bar\phi^-_1(u, \hat x), \bar\phi_1^+(v, \hat x)]
}
where the data are related by
\eq{7flat2}{
    \bar\phi_1^-(u, \hat x) = -R^2\p_u J^-\Big(\frac{\pi}{2} + \frac{u}{R}, \hat x\Big),\ \ \ \bar\phi^+_1(v, \hat x) = R^2\p_v J^+\Big(-\frac{\pi}{2} + \frac{v}{R}, \hat x\Big).
}
The mapping for the gauge field is even simpler,
\eq{7flat3}{
    Z_\text{AdS}[a_A^+(\tau, \hat x), a_A^-(\tau, \hat x)] = Z_\text{AFS}[\hat{\overline A}^-_{A,0}(u, \hat x), \hat{\overline A}_{A,0}^+(v, \hat x)]
}
with the data map\footnote{It's worth noting that working with the Carroll data has the advantage of avoiding potential headaches to do with matching polarization vectors on both sides of the flat limit.}
\eq{7flat4}{
    \hat{\overline A}^-_{A,0}(u, \hat x) = a^-_A\Big(\frac{\pi}{2} + \frac{u}{R}, \hat x\Big),\ \ \ \hat{\overline A}^+_{A,0}(v, \hat x) = a^+_A\Big(-\frac{\pi}{2} + \frac{v}{R}, \hat x\Big).
}

\subsection{Witten diagrams}
\label{Formal Argument: Witten idagrams}

Though less in the spirit of the generating functionals of interest in this work, the structure of Witten diagrams offer the most straightforward path to the flat limit. A generic Witten diagram, smeared against boundary sources, takes the form
\eq{generic Witten diagram}{
    \mathcal{A} = \int_\text{AdS} \sqrt{-g_1}\dr^Dx_1\cdots \sqrt{-g_1}\dr^Dx_n G_n(x_1,\ldots,x_n)\psi_1(x_1) \cdots \psi_n(x_n)
}
where $G_n$ is the sum of bulk diagrams with external legs stripped and the wavefunctions are
\eq{7d}{
    \psi_k(x) = \int_{\p\text{AdS}}\sqrt{-h}\dr^dy K(x; y)J_k(y)
}
with $K$ the bulk-boundary propagator and $h_{ij}$ the CFT metric. We assume that the sources $J_k$ do not overlap so that the particles represented by the $\Psi_k$ do not collide at the boundary of AdS.

The key assumption to obtain the flat limit is that the dominant contribution to \eqref{generic Witten diagram} is given when all bulk points in the Witten diagram -- both those explicitly appearing in \eqref{generic Witten diagram} and those appearing inside $G_n$ -- are contained within the flat region. If this is the case, then the flat limit is immediate due to \eqref{metric flat limit}, which also implies that the wavefunctions $\psi_k$ appearing in the integration obey the flat space wave equation. In words, \eqref{generic Witten diagram} would now be the sum of Feynman diagrams with external legs stripped and replaced by external wavefunctions obeying the flat wave equation, which is nothing more than a flat S-matrix element obtained by LSZ reduction written in position space.

In order for the Witten diagrams to plausibly be dominated by bulk points contained in the flat region, we must tune the boundary data $J_k$ such that the wavefunctions $\psi_k$ are well-collimated outside, and hence only collide inside the flat region.

To see how this works, recall the Lorentzian bulk-boundary propagator corresponding to a scalar boundary operator of dimension $\Delta$, corresponding to a bulk scalar field of mass $m^2R^2 = \Delta(\Delta - d)$, is\footnote{The factor of $i$ is a result of working in Lorentzian signature; see e.g. \cite{Skenderis:2008dh}. Note also that the $i\epsilon$ prescription here is the one valid globally in AdS.}
\eq{7e}{
    K(x; x') = \frac{i\Gamma(\Delta)}{2^\Delta \pi^{\frac{d}{2}}\Gamma(\Delta - \frac{d}{2})}\left( \frac{\cos\rho}{\cos((1-i\epsilon)(\tau - \tau')) - \sin\rho \hat x\cdot\hat x'} \right)^\Delta
}
where $x^\mu = (\rho,\tau,\hat x)$ denotes a bulk point and $x'^\mu = (\tau', \hat x')$ is a boundary point. This has a  light cone singularity regulated by the $i\epsilon$, which is smoothed out when integrating against $J_k$ along the time direction. It follows that when the source's support in $\tau$ is thin, the resulting wavepacket $\psi_k$ will be similarly well-localized as it falls towards the flat region. Hence the dominant contribution from interaction vertices with external legs will come when the vertex is within the flat region. The remaining vertices find their dominant contributions from the flat region due to the decay of the bulk-bulk propagator for proper distances of order $O(R)$\footnote{One might worry about internal vertices which are light-like separated, but the Feynman $i\epsilon$ regulating the lightcone singularity forces the Green's function to decay in these directions as well.}.

The simplest way to build a source with the desired support is to construct a wavepacket in the frequency domain\footnote{If one has already constructed a boundary source with the required support on the strip, it is possible to avoid using momentum space for integer $\Delta\geq 2$ by comparison with the Carrollian bulk-boundary propagators \eqref{aa2}. We demonstrate how to perform this calculation in section \ref{position space flat limit} but otherwise focus on the momentum space computation since it makes the desired support automatic.} after rescaling away $R$. For example, in the strip about $\tau = -\frac{\pi}{2}$ we may define the rescaled variable $v = R(\tau + \frac{\pi}{2})$. If the wavepacket had width of order $1/R$ in $\tau$ about $-\frac{\pi}{2}$, then it is free to have any $O(R^0)$ width in $v$, and hence also have any $O(R^0)$ width in $\omega$:\footnote{We write a strictly positive frequency source here, but this unnecessary. It will be simple to see later that, in the strip about $\tau = -\frac{\pi}{2}$, the negative frequency component of $J$ will not contribute to the wavefunction in the flat limit.}
\eq{7f}{
    J(\tau, \hat x) = \int_{0}^\infty\frac{\dr\omega}{2\pi}\tilde J(\omega,\hat x) e^{-i\omega R(\tau + \frac{\pi}{2})}.
}

Since the source specification $\tilde J$ automatically produces the type of wavepacket which localizes us to the flat region, and $\psi_k$ becomes a free field in the free region, it only remains to compute the $\dr\tau$ integral in 
\eq{flat limit wavefunction}{
    \psi(x) = \int_0^\infty\frac{\dr\omega}{2\pi}\int_{-\pi}^0\sqrt{-h}\dr \tau' \dr^{d-1}\hat x' K(x; x')e^{-i\omega R(\tau' + \frac{\pi}{2})}\tilde J(\omega, \hat x')
}
when $x$ is in the flat region to determine the map between the $\tilde J$ and the corresponding flat space data. There is a completely analogous expression for the outgoing wavepackets. We perform this computation for spin 0 and spin 1 fields in sections \ref{Flat limit scalar fields} and \ref{Flat limit gauge fields}, respectively.

This argument establishes the connection between Witten and Feynman diagrams for some map between AdS data and flat external wavefunctions. Since the AdS and AFS path integrals are the generating functionals for these objects, it follows that the generating functionals themselves must also be equal, with the possible exception of contact terms where AdS sources overlap.

It's important to note that in the above we have neglected 2-point corrections on the external legs. Such corrections would not be suppressed by our arguments about localization onto the bulk point singularity. However, just like in LSZ, such corrections amount to a wavefunction renormalization $Z$. In any perturbative evaluation the prescription to avoid these factors entirely is identical to LSZ: do not include any diagrams with bubbles on the external legs.

\subsection{Path integral}
\label{Formal Argument: Path integral}

The argument of the previous section establishes the equivalence of the AdS and AFS path integrals indirectly, by instead relating the objects they generate. This relation can be established more directly using a combination of operator and path integral methods. To do this, first define the time evolution operator in the presence of boundary sources  
\eq{zz99}{
    \hat U(\Sigma_f, \Sigma_i) = T\exp\left( -i\int_{\Sigma_i}^{\Sigma_f}\hat H\dr \tau + i\int_{\Sigma_i}^{\Sigma_f}J\hat O\dr \tau \right) }
where $\Sigma_i$ and $\Sigma_f$ are a pair of initial and final time slices. The explicit details of this operator are largely unimportant for our present purposes. The only important properties of this object are that it obeys the usual composition property
\eq{flat2}{
    \hat U(\Sigma_f, \Sigma_i) = \hat U(\Sigma_f, \Sigma')\hat U(\Sigma', \Sigma_i) 
}
for any slice $\Sigma'$ between $\Sigma_i$ and $\Sigma_f$. The AdS partition function, with Euclidean caps as in figure \ref{Lorentzian pill}, may be expressed in terms of this operator by
\eq{flat3a1}{
    Z[J] = \langle0|\hat U(\Sigma_{\pi}, \Sigma_{-\pi})|0\rangle
}
where $\Sigma_{\pm\pi}$ are the constant $\tau = \pm \pi$ slices.

\begin{figure}[h]
    \centering
    \includegraphics[scale=0.4]{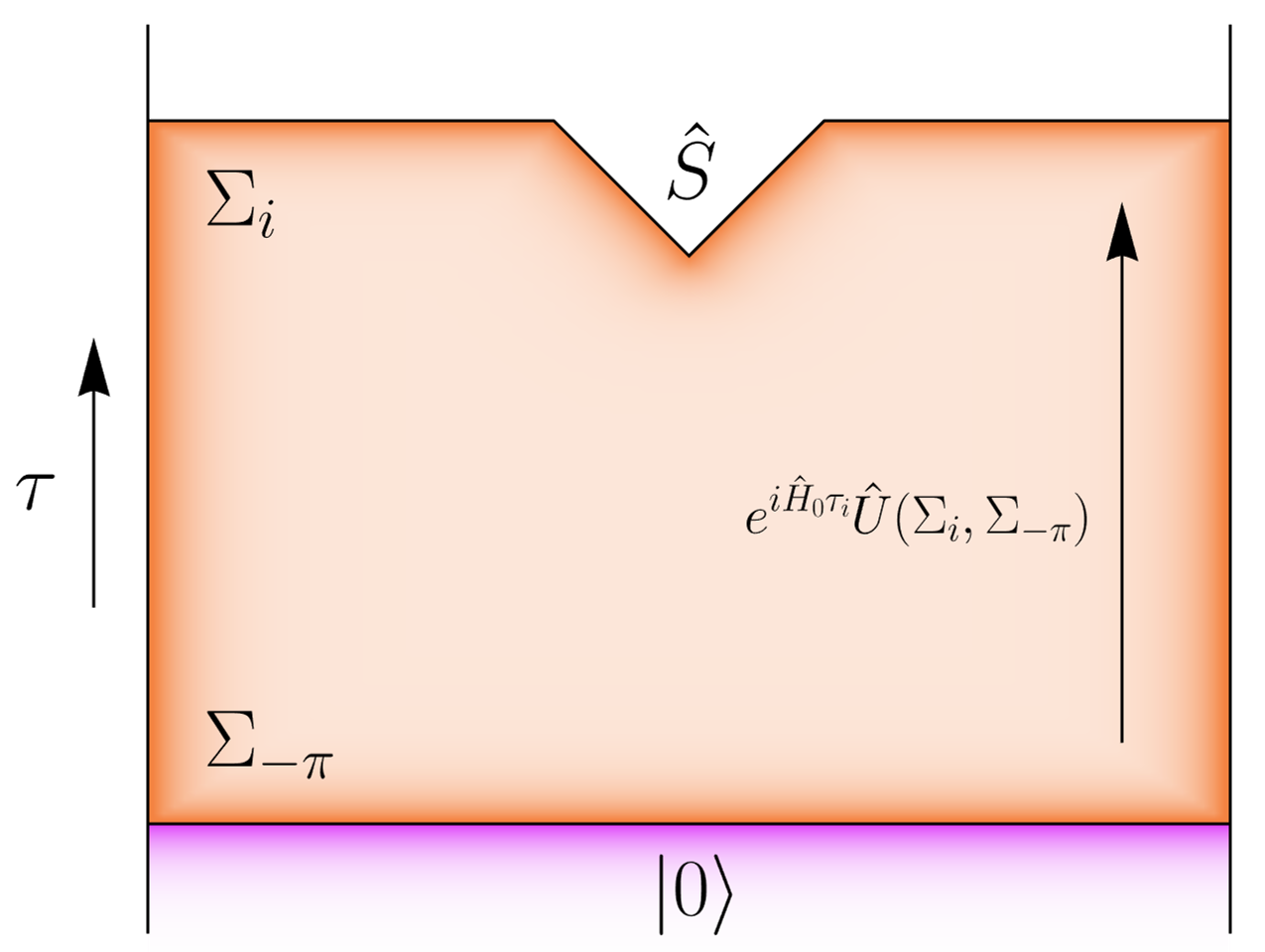}
    \caption{An Euclidean cap prepares the vacuum below $\Sigma_{-\pi}$, then the evolution operator $e^{i\hat H_0\tau_i}\hat U(\Sigma_i, \Sigma_{-\pi})$ prepares the in state that $\hat S = e^{i\hat H_0\tau_f}\hat U(\Sigma_f, \Sigma_i)e^{-i\hat H_0\tau_i}$ acts upon. A symmetric construction at future times prepares an out state in terms of a region whose lower boundary is  the surface $\Sigma_f$. If the AdS data is chosen so the in and out states are non-trivial only on the diamond, the result coincides with a matrix element of the flat S-matrix.}
    \label{Fig:free evol diagram}
\end{figure}

We now introduce the slices $\Sigma_i,\Sigma_f$ as in figure \ref{Fig:free evol diagram}, which bound the flat region, and write\footnote{We write $\tau_i,\tau_f$ as a shorthand in the free evolution operators for free evolution between $\tau = 0$ and the surfaces $\Sigma_i,\Sigma_f$. }
\eq{flat3}{
    Z[J] = \langle 0|\hat U(\Sigma_\pi, \Sigma_f) e^{-i\hat H_0 \tau_f}e^{i\hat H_0\tau_f}\hat U(\Sigma_f, \Sigma_i)e^{-i\hat H_0 \tau_i}e^{i\hat H_0\tau_i}\hat U(\Sigma_i, \Sigma_{-\pi})|0\rangle.
}
Inserting an overcomplete basis of AdS coherent states between the free evolution operators gives
\eq{flat4}{
    Z[J] = \int&[\D\phi_+^\alpha\D\phi_-^\alpha][\D\phi_+^\beta\D\phi_-^\beta]e^{-2i(\phi^\alpha_-,\phi_+^\alpha)_{\Sigma_f}}e^{-2i(\phi^\beta_-, \phi^\beta_+)_{\Sigma_i}}\\
    &\cdot \langle 0|\hat U(\Sigma_\pi, \Sigma_f)|\phi^\alpha_+(\tau_f)\rangle\langle \phi_-(\tau_f)|\hat U(\Sigma_f, \Sigma_i) |\phi^\beta_+(\tau_i)\rangle\langle\phi^\beta_-(\tau_i)|\hat U(\Sigma_i, \Sigma_{-\pi})|0\rangle.
}
This is a lengthy expression, but the measure is just the norm of the coherent states written covariantly as in \eqref{a3c}. The first and last $\hat U$ factors prepare the state as it falls from the AdS boundary towards the flat region, and the factor
\eq{flat5}{
    S[\phi_-^\alpha, \phi^\beta_+] = \langle\phi^\alpha_-(\tau_f)|\hat U(\Sigma_f, \Sigma_i)|\phi^\beta_+(\tau_i)\rangle
}
will become the AFS generating functional.

To this point, we have not made any assumptions about the boundary data that would allow us to conclude that $S$ should be identified with the AFS generating functional. Indeed, it is only equal to the AFS generating functional on the subspace of coherent data where $\phi_-^\alpha$ and $\phi^\beta_+$ have support only at the boundary of the flat region. To get a handle on how the kinematic choice of boundary source is transported to the coherent data on $\Sigma_i,\Sigma_f$, we use path integral expressions for the infalling and outfalling factors. Focusing on the infalling factor, we are instructed to compute the path integral
\eq{flat6}{
    \langle\phi^\beta(\tau_i)|\hat U(\Sigma_i, \Sigma_{-\pi})|0\rangle = \int\D\phi e^{iI[\phi] + iI_\p[\phi]}
}
over fields obeying
\eq{flat6a}{
    \phi_-(\tau_i) &= \phi^\beta_-(\tau_i)\cr
    \phi(\rho, \tau, \hat x) &\arrow \cos^{d-\Delta}\rho J(\tau, \hat x)\cr
    \phi_+(\Sigma_{-\pi}) &= 0
}
That is, the path integral over fields obeying the usual boundary conditions at the AdS boundary in addition to the boundary conditions implied by the vacuum at $\Sigma_{-\pi}$ and the coherent state at $\Sigma_i$. The boundary term is determined by demanding a good variational principle compatible with these boundary conditions,
\eq{flat7}{
    I_\p[\phi] = (\phi_-, \phi_+)_{\Sigma_i} - (\phi_-, \phi_+)_{\Sigma_{-\pi}} + I_{\p\text{AdS}}[\phi]
}
where we have assumed that the kinetic term is written in the form $\phi\nabla^2\phi$. The inner product at $\Sigma_{-\pi}$ vanishes since $\phi_+ = 0$ there.

As usual, it is useful to move the boundary data dependence into the integrand by shifting the integration variable to
\eq{flat8}{
    \phi = \bar\phi + \tilde\phi, \ \ \ \ \bar\phi(x) = \underbrace{\int_{-\pi}^0\sqrt{-h}\dr^d y K(x; y)J(y)}_{\psi_\text{in}(x)} + \phi^\beta_-(x).
}
Here $\tilde\phi$ is required to have zero negative frequency on $\Sigma_i$ and zero positive frequency on $\Sigma_{-\pi}$. Hence the boundary terms reduce to
\eq{flat9}{
    I_\p[\phi] = (\phi^\beta_-, \psi_\text{in} + \tilde \phi_+)_{\Sigma_i} + I_{\p\text{AdS}}[\bar\phi + \tilde \phi].
}

Using this shift in the bulk action,
\eq{flat10}{
    I[\phi] = \frac{1}{2}\int\dr^Dx \sqrt{-g}\,\bar\phi\nabla^2\tilde\phi + \int\dr^Dx\sqrt{-g}\left[ \frac{1}{2} \tilde\phi\nabla^2\tilde\phi - V(\bar\phi + \tilde\phi) \right].
}
The first term is bulk free action linearized around the free solution $\bar\phi$. As such, the demand of a good variational principle, or direct calculation, implies that this term exactly cancels\footnote{More generally, only the linear dependence on $\tilde\phi$ would be guaranteed to cancel, but $I_\p$ is a linear functional of $\tilde \phi$ here.} the $\tilde\phi$ dependence in $I_\p$. The total action is therefore
\eq{flat11}{
    I + I_\p = (\phi_-^\beta, \psi_\text{in})_{\Sigma_i} + I_{\p\text{AdS}}[\bar\phi] + \int\dr^Dx\sqrt{-g}\left[ \frac{1}{2}\tilde\phi\nabla^2\tilde\phi - V(\bar\phi + \tilde\phi) \right].
}
By a straightforward calculation, the details of which are included in appendix \ref{Sec:bndy terms}, we may evaluate the AdS boundary terms to find
\eq{exterior action}{
    I + I_\p = 2(\phi_-^\beta, \psi_\text{in})_{\Sigma_i} + \int\dr^Dx\sqrt{-g}\left[ \frac{1}{2}\tilde\phi\nabla^2\tilde\phi - V(\bar\phi + \tilde\phi) \right].
}

If the $\tilde\phi$ path integral were trivial, the boundary term above would imply\footnote{The result of the coherent state integrals are simplest to see in terms of the mode data where they take the schematic form $\int \frac{\dr z \dr \overline z}{2\pi i} e^{-\overline z(z - w)}$.}
\eq{flat13}{
    Z[J] &= \int[\D\phi_+^\alpha\D\phi_-^\alpha][\D\phi_+^\beta\D\phi_-^\beta]e^{-2i(\phi^\alpha_-,\phi_+^\alpha - \psi_\text{out})_{\Sigma_f}}e^{-2i(\phi^\beta_-, \phi^\beta_+ - \psi_\text{in})_{\Sigma_i}}S[\phi_-^\alpha,\phi^\beta_+]\\
    &= S[\psi_\text{out}, \psi_\text{in}].
}
As already discussed, this object is the AFS generating functional in the special case where the data $\psi_\text{in}$ and $\psi_\text{out}$ is supported only in the flat region. But since $\psi$ is defined by the bulk-boundary propagator, we know that this support can be engineered by placing the sources within a width $1/R$ window of $\tau = \pm\frac{\pi}{2}$, exactly as in \eqref{flat limit wavefunction}.

To argue that the interactions in \eqref{exterior action} can be neglected, we can first imagine the boundary source decomposing as $J = \sum_k J_k$ into sources with disjoint support on the boundary. Each source produces a well-collimated particle track falling towards the flat region. Since these tracks are well-collimated, we expect that the $\tilde\phi$ path integral factors into integrals over fluctuations about each particle track. This is the path integral analog of the diagrammatic observation that the well-separated and  well-localized  particles are not expected to interact with each other until they enter the flat region.  

This still leaves, however, the possibility of self-interactions, but this is identical to the discussion of diagrams with bubbles on external legs at the end of section \ref{Formal Argument: Witten idagrams}. Hence these interactions just amount to a wavefunction renormalization and can be ignored.

\section{Flat space limit for scalar fields}
\label{Flat limit scalar fields}

With the formal comments in the previous section in mind, we turn to computing the map between the AdS and AFS data by computing the flat limit of \eqref{flat limit wavefunction}. To do this, we first recall some explicit formulas for scalar fields in AdS.

\subsection{Basic formulas}

The bulk-boundary propagator corresponding to a scalar boundary operator of dimension $\Delta$, corresponding to a bulk scalar field of mass $m^2 R^2 = \Delta (\Delta - d)$, is
\eq{fs1}{
    K(x; x') = C_\Delta \left( \frac{\cos\rho}{\cos((1-i\epsilon)(\tau - \tau')) - \sin\rho\hat x\cdot \hat x'} \right)^\Delta
}
where $x^\mu = (\rho, \tau, \hat x)$ denotes a bulk point and $x'^\mu = (\tau', \hat x')$ is a boundary point. We have also defined\footnote{The factor of $i$ is a result of working in Lorentzian signature; see e.g. \cite{Skenderis:2008dh}.}
\eq{fs2}{
    C_\Delta = \frac{i\Gamma(\Delta)}{2^\Delta \pi^{\frac{d}{2}}\Gamma(\Delta - \frac{d}{2})}.
}
The bulk-boundary propagator obeys $(\nabla^2 - m^2R^2 + i\epsilon)K(x; x') = 0$,  after a redefinition of $\epsilon$, and has leading boundary behavior 
\eq{fs3}{
    K(x; x') \rightarrow \frac{1}{\sqrt{-h}}\delta^{(d)}(x - x') \cos^{d-\Delta}\rho \ \ \ \text{as}\ \ \ \rho\arrow\frac{\pi}{2}
}
where $\frac{1}{\sqrt{-h}}\delta^{(d)}(x - x')$ is the invariant delta function with respect to the boundary metric $\dr s_b^2 = -\dr\tau^2 + \dr\Omega_{d-1}^2$. As is standard, given a boundary source $J(x')$ we define a bulk free field solution as
\eq{fs4}{
    \phi(x) = \int\dr^dx' \sqrt{-h(x')}K(x; x')J(x')
}
which has leading boundary behavior $\phi(x) \rightarrow J(x)\cos^{d-\Delta}\rho$.

The flat limit of the bulk-boundary propagator is obtained by taking the bulk point to lie in the flat region, i.e. at fixed $(r, t, \hat x)$, defined as in \eqref{7b}, as $R\rightarrow\infty$,
\eq{fs5}{
    K^f(x, x') = C_\Delta R^\Delta \left( \frac{1}{R\cos\tau' + t\sin\tau' - r\hat x\cdot\hat x' + i\epsilon R|\tau - \tau'|\sin|\tau - \tau'|} \right)^\Delta.
}

As argued in the previous section, it's important that we integrate $K^f$ against only sources localized to the vicinity of $\tau' = \pm \frac{\pi}{2}$. To zoom in on these regions we write
\eq{fs61}{
    \tau' = \begin{cases}
        ~~\frac{\pi}{2} + \frac{u}{R} & (\text{future source})\\
        -\frac{\pi}{2} + \frac{v}{R} & (\text{past source})
    \end{cases}
}
It's simple to check that $\sin|\tau - \tau'|>0$ so long as $t-v,t-u \sim O(R^0)$. Hence we can redefine the $\epsilon$ to simplify the flat limit:
\eq{fs6}{
    K^f(x, x') = C_\Delta R^\Delta \left( \frac{1}{R\cos\tau' + t\sin\tau' - r\hat x\cdot\hat x' + i\epsilon} \right)^\Delta.
}

\subsection{Past sources}
\label{Scalar past source}

We first focus on past sources localized near $\tau' = -\frac{\pi}{2}$. Writing $\tau' = -\frac{\pi}{2} + \frac{v}{R}$, the flat limit of the bulk-boundary propagator becomes
\eq{fs7}{
    K^f(x; x') = C_\Delta (\omega R)^\Delta \left( \frac{1}{\omega v + p\cdot x + i\epsilon} \right)^\Delta
}
where we have defined $p^\mu = \omega(1, -\hat x')$ for some arbitrary frequency $\omega$ that we will identify with the integrated frequency in \eqref{flat limit wavefunction}.

Indeed, the flat limit of the wavefunction \eqref{flat limit wavefunction} is given by
\eq{fs8}{
    \psi(x) = C_\Delta R^{\Delta -1}\int_0^\infty\frac{\dr\omega}{2\pi}d^{d-1}\hat{p} \omega^{\Delta - 1}\tilde J^+(\omega, -\hat p)e^{ip\cdot x}\int_{-\frac{\pi}{2}R}^{\frac{\pi}{2}R}\dr v \frac{e^{-iv}}{(v + i\epsilon)^\Delta}~.
}
Here we have only written the contribution due to the positive frequency component of $\tilde J$, but the $i\epsilon$ in the $\dr v$ integration will annihilate any negative frequency component of $\tilde J$ at leading order in $R$. The positive frequency integral above can be computed using a simple contour analysis to find
\eq{fs9}{
    \int_{-\infty}^\infty \dr v \frac{e^{-iv}}{(v + i\epsilon)^\Delta} = \frac{2\pi}{\Gamma(\Delta)}e^{-i\frac{\pi}{2}\Delta}~,
}
so the wavefunction becomes
\eq{fs10}{
    \psi(x) = C_\Delta R^{\Delta - 1}\frac{2\pi}{\Gamma(\Delta)}e^{-i\frac{\pi}{2}\Delta}\int_0^\infty \frac{\dr\omega}{2\pi}d^{d-1}\xh \omega^{\Delta - 1}\tilde J^+(\omega, -\hat p)e^{ip\cdot x}~.
}

We can put the wavefunction into the standard form of a positive frequency flat space free solution by noting $\dr^dp = \omega^{d-1}\dr\omega d^{d-1}\hat{p}$ to find
\eq{fs11}{
    \psi(x) &= C_\Delta R^{\Delta - 1}\frac{2(2\pi)^d}{\Gamma(\Delta)}e^{-i\frac{\pi}{2}\Delta}\int\frac{\dr^dp}{(2\pi)^d}\frac{1}{2\omega} \omega^{\Delta - d + 1}\tilde J^+(\omega, -\hat p)e^{ip\cdot x}\\
    &\equiv \int \frac{\dr^dp}{(2\pi)^d}\frac{1}{2\omega}b(\vec p)e^{ip\cdot x}~.
}
This immediately gives the data map in terms of mode data,
\eq{fs12}{
    b(\vec p) &= C_\Delta R^{\Delta - 1}\frac{2(2\pi)^d}{\Gamma(\Delta)}e^{-i\frac{\pi}{2}\Delta} \omega^{\Delta - d + 1}\tilde J^+(\omega, -\hat p)\\
    &= \frac{-i\omega}{2\pi} R^2 \tilde J^+(\omega, -\hat p),\ \ \ \Delta = d = 3.
}
We can now express this map in terms of the Carroll data \eqref{a6} in the special case $d = \Delta = 3$ to find
\eq{fs13}{
    \bar \phi_1^+(v, \hat x) = R^2 \p_v J^+\Big(-\frac{\pi}{2} + \frac{v}{R}, \hat x\Big).
}
It is perhaps worth mentioning that the choice $\Delta = 3$ is not crucial here, and for example $\Delta = 2$ would have produced the map $\bar\phi_1^+(v, \hat x) = RJ^+(-\frac{\pi}{2} + \frac{v}{R}, \hat x)$.

\subsection{Future sources}
\label{Scalar future sources}

We now write $\tau' = \frac{\pi}{2} + \frac{u}{R}$ to consider a negative frequency source
\eq{fs14}{
    J^-(u, \hat x') = \int_0^\infty \frac{\dr\omega}{2\pi} \tilde J^-(\omega, \hat x')e^{iR\omega(\tau' - \frac{\pi}{2})}
}
in the vicinity of $\tau' = +\frac{\pi}{2}$. As with the sources near $\tau' = -\frac{\pi}{2}$, the bulk-boundary propagator simplifies to
\eq{fs15}{
    K(x, x') \arrow C_\Delta (\omega R)^\Delta \frac{1}{(-\omega u - p\cdot x + i\epsilon)^\Delta}
}
with the notable distinction that now
\eq{fs16}{
    p^\mu = \omega(1, \hat x')
}
includes no antipode in its definition.

The wavefunction in the flat region therefore reduces to
\eq{fs17}{
    \psi(x) &= C_\Delta \int_0^\pi\dr\tau'd^{d-1}\hat{p}'\int_0^\infty\frac{\dr\omega}{2\pi}\frac{\omega^\Delta e^{i\omega R(\tau' - \frac{\pi}{2})}}{(-\omega u - p\cdot x + i\epsilon)^\Delta}\tilde J^-(\omega, \hat p)\\
    &= C_\Delta R^{\Delta - 1}\int_0^\infty \frac{\dr\omega}{2\pi}d^{d-1}\hat{p}\omega^{\Delta - 1}\tilde J^-(\omega, \hat p)e^{-ip\cdot x}\int_{-\frac{\pi}{2}R}^{\frac{\pi}{2}R}\dr u\frac{e^{iu}}{(-u + i\epsilon)^\Delta}.
}
Once again this takes the form of a flat space free solution, but now of negative frequency content. The remaining $\dr u$ integral is identical to the \eqref{fs9} under $u \arrow -u$, and so we find
\eq{fs18}{
    \psi(x) = C_\Delta R^{\Delta - 1} \frac{2(2\pi)^d}{\Gamma(\Delta)}e^{-i\frac{\pi}{2}\Delta}\int \frac{\dr^d p}{(2\pi)^d}\frac{1}{2\omega} \omega^{\Delta - d + 1}\tilde J^-(\omega, \hat p) e^{-ip\cdot x}
}
from which we read off the mode data map
\eq{fs19}{
    b^\dag(\vec p) &= C_\Delta R^{\Delta - 1}\frac{2(2\pi)^d}{\Gamma(\Delta)} e^{-i\frac{\pi}{2}\Delta} \omega^{\Delta - d + 1}\tilde J^-(\omega, \hat p)\\
    &= -\frac{i\omega}{2\pi}\tilde J^-(\omega, \hat p),\ \ \ \Delta = d = 3.
}

Comparing with the Carroll data \eqref{a6} and specializing to $d = \Delta = 3$,
\eq{fs20}{
    \bar\phi^-_1(u, \hat x) = - R^2\p_u J^-\Big( \frac{\pi}{2} + \frac{u}{R}, \hat x\Big).
}
As with the past sources, had we chosen, for example, $\Delta = 2$ we would have found the map $\bar\phi_1^-(u, \hat x) = RJ^-(\frac{\pi}{2} + \frac{u}{R}, \hat x)$.

\subsection{Approach in position space}
\label{position space flat limit}

The momentum space representation of the boundary source was introduced as a convenient tool for constructing boundary sources with the required support on the thin strips about $\tau = \pm\frac{\pi}{2}$. This isn't strictly necessary, and for the special case of integer scaling dimension with $\Delta \geq 2$ one can slightly simplify the calculations in the previous two subsections by comparing the flat limit of $K$ with the Carroll bulk-boundary propagators \eqref{aa2}.

In the special case $\Delta = 2$ the bulk-boundary propagator for a source at $\tau' = -\frac{\pi}{2} + \frac{v'}{R}$, renaming $v\arrow v'$ compared with \eqref{fs7}, may be written
\eq{ps1}{
    K^f(x; x') = \frac{i}{(4\pi)^2}R^2\frac{1}{(v' + q(-\hat x')\cdot x + i\epsilon)^2}
}
where we have defined $q^\mu(\hat x') = (1, \hat x')$ as below \eqref{aa2}. Then in the flat region the wavefunction \eqref{7d} becomes
\eq{ps2}{
    \psi(x) = \int\dr v'\dr^2\hat x' \left( \frac{i}{(2\pi)^2}\frac{1}{(v' + q(-\hat x')\cdot x + i\epsilon)^2} \right) RJ^+\Big(-\frac{\pi}{2} + \frac{v'}{R}, \hat x'\Big)~.
}
Comparing with the Carroll bulk-boundary propagator \eqref{aa2}, we immediately find $\bar\phi_1^+(v, \hat x) = RJ^+(-\frac{\pi}{2} + \frac{v}{R}, \hat x)$. This is the same map found for $\Delta = 2$ via the momentum space computation in section \ref{Scalar past source}.

For integer $\Delta \geq 2$ the same approach can be used by first writing $K^f$ as some number of derivatives acting on the $\Delta = 2$ case. For non-integral $\Delta$ Fourier space seems unavoidable since to apply the argument in this subsection one would need to introduce Fourier space to define a fractional derivative of the $\Delta = 2$ case. The calculation for future sources is essentially identical and leads to the $\Delta = 2$ map derived in section \ref{Scalar future sources}.

\section{Flat space limit for gauge fields}
\label{Flat limit gauge fields}

Now that we have oriented ourselves with scalar fields, we turn to the flat limit in the case of gauge fields. Here it will be useful to adopt stereographic coordinates on the sphere so that
\eq{fs21}{
    \dr s^2 = \frac{R^2}{\cos^2\rho}\Big( -\dr\tau^2 + \dr\rho^2 + 2\sin^2\rho \gamma_{z\overline z}\dr z\dr\overline z \Big)
}
where $\gamma_{z\overline z} = \frac{2}{(1 + z\overline z)^2}$. These coordinates relate to the unit vector representation of the sphere by
\eq{fs22}{
    \hat x = \frac{1}{1 + z\overline z}(z + \overline z, -i(z - \overline z), 1 - z\overline z).
}

We are interested in gauge fields $A_\mu$ that are  solutions of the source free Maxwell equations with specified AdS boundary conditions.  Such solutions may be  expressed in terms of the 1-form valued bulk-boundary propagators $K^i$ as
\eq{fs23}{
    A_\mu(x)\dr x^\mu = \int\dr^3 x' \sqrt{-h} \left( K^\tau(x; x')a_\tau(x') + K^z(x; x')a_z(x') + K^{\overline z}(x; x')a_{\overline z}(x') \right)
}
which obey the Maxwell equations $\nabla_\mu F^{\mu\nu} = 0$, Lorenz gauge $\nabla^\mu A_\mu = 0$, and which behave as
\eq{fs24}{
    A_\mu(x) \dr x^\mu \arrow a_\tau(x)\dr\tau + a_z(x)\dr z + a_{\overline z}\dr \overline z
}
near the AdS boundary.

We review how to obtain the bulk-boundary propagators satisfying these requirements from the more familiar Euclidean Poincar\'e coordinate expression in appendix \ref{Sec:spin-1 bulk-boundary propagator}. The expressions are
\eq{fs25}{
    K^\tau(x; x') &= \frac{i}{2\pi^2} \frac{\cos\rho}{\cos(\tau - \tau') - \sin\rho \hat x\cdot\hat x' + i\epsilon} \dr\left(\frac{\sin(\tau - \tau')}{\cos(\tau -\tau') - \sin\rho \hat x\cdot \hat x' + i\epsilon} \right),\\
    K^z(x; x') &= \frac{i}{4\pi^2}\frac{\cos\rho}{\cos(\tau - \tau') - \sin\rho \hat x\cdot \hat x' + i\epsilon} \dr\left( \frac{\sin\rho \gamma^{z'\overline z'}\p_{\overline z'}(\hat x\cdot \hat x')}{\cos(\tau - \tau') - \sin\rho \hat x\cdot \hat x' + i\epsilon} \right),\\
    K^{\overline z}(x; x') &= \frac{i}{4\pi^2}\frac{\cos\rho}{\cos(\tau - \tau') - \sin\rho \hat x\cdot \hat x' + i\epsilon} \dr\left( \frac{\sin\rho \gamma^{z'\overline z'}\p_{z'}(\hat x\cdot \hat x')}{\cos(\tau - \tau') - \sin\rho \hat x\cdot \hat x' + i\epsilon} \right).
}
The $i\epsilon$ prescription here is valid only for $\sin|\tau - \tau'| > 0$, but this is all we will require, as was the case for the scalar field. We also note
\eq{fs26}{
    \gamma^{z'\overline z'}\p_{\overline z'}(\hat x\cdot \hat x') &= \frac{1 + z'\overline z}{1 + z\overline z}(z - z'),\\
    \gamma^{z'\overline z'}\p_{z'}(\hat x\cdot \hat x') &= \frac{1 + z\overline z'}{1 + z\overline z}(\overline z - \overline z').
}

Taking the bulk point in \eqref{fs25} to lie in the flat region and taking $R \arrow \infty$, we find
\eq{fs27}{
    K^\tau & \arrow \frac{i}{2\pi^2}\frac{R}{R \cos\tau' + (t\sin\tau' - r\hat x\cdot\hat x') + i\epsilon} \dr \left( \frac{-R\sin\tau'}{R\cos\tau' + (t \sin\tau' - r\hat x\cdot \hat x') + i\epsilon} \right)\\
    &= -\frac{iR^2}{4\pi^2}\sin\tau' \dr\left[ \left( \frac{1}{R\cos\tau' + (t \sin\tau' - r\hat x\cdot \hat x') + i\epsilon} \right)^2 \right],\\
    K^z &\arrow \frac{i}{4\pi^2}\frac{R}{R\cos\tau' + (t\sin\tau' - r\hat x\cdot \hat x') + i\epsilon}\dr\left( \frac{r\gamma^{z'\overline z'}\p_{\overline z'}(\hat x\cdot \hat x')}{R\cos\tau' + (t\sin\tau' - r\hat x\cdot \hat x') + i\epsilon} \right),\\
    K^{\overline z} &\arrow \frac{i}{4\pi^2}\frac{R}{R\cos\tau' + (t\sin\tau' - r\hat x\cdot \hat x') + i\epsilon}\dr\left( \frac{r\gamma^{z'\overline z'}\p_{ z'}(\hat x\cdot \hat x')}{R\cos\tau' + (t\sin\tau' - r\hat x\cdot \hat x') + i\epsilon} \right).
}
We see that in the flat region the $\tau$ component of the bulk-boundary propagator is an exact form. It's straightforward to see that for either $\tau' = -\frac{\pi}{2} + \frac{v}{R}$ or $\tau' = +\frac{\pi}{2} + \frac{u}{R}$ this total derivative takes the form $\dr e^{\pm ip\cdot x}$, and hence is pure gauge, removable by a small gauge transformation. The physically interesting solutions thus correspond to $K^z$ and $K^{\overline z}$, so we restrict attention to these.

\subsection{Past sources}

As with the scalar field we write $\tau' = -\frac{\pi}{2} + \frac{v}{R}$ for sources in the past. As already discussed, we may ignore $K^\tau$, and $K^{\overline z}$ is the same as $K^z$ under swapping $z\leftrightarrow\overline z$ and $z' \leftrightarrow \overline z'$. Focusing on the $K^z$ contribution, in the flat limit we have
\eq{fsp1a1}{
    K^z \arrow \frac{iR\omega^2}{4\pi^2}\frac{1}{\omega v + p\cdot x + i\epsilon}\dr\left( \frac{r \gamma^{z'\overline z'}\p_{\overline z'}(\hat x\cdot \hat x')}{\omega v + p\cdot x + i\epsilon} \right)
}
where $p^\mu = \omega(1, -\hat x')$. As with the scalar field, we write the boundary source as
\eq{fsp1}{
    a_z^+(\tau', \hat x') = \int\frac{\dr\omega}{2\pi}\tilde a^+_z(\omega, \hat x')e^{-i\omega R(\tau' + \frac{\pi}{2})}
}
so the bulk wavefunction in the flat region is given by
\eq{fsp2}{
    \psi_z(x) = \frac{i}{4\pi^2}\int\frac{\dr\omega}{2\pi} \dr^2\hat p\omega^2 \tilde a_z^+(\omega, -\hat p)\int_{-\infty}^\infty \dr v \frac{e^{-i\omega v}}{\omega v + p\cdot x + i\epsilon}\dr\left( \frac{r \gamma^{z'\overline z'}\p_{\overline z'}(\hat x\cdot \hat x')}{\omega v + p\cdot x + i\epsilon} \right).
}

Three type of  terms will arise from the differential: $\dr(p\cdot x)$, $\dr r$, and $\dr(\hat x\cdot \hat x')$. Our goal here is find the asymptotic Carroll data specified by this field configuration. As discussed in section \ref{ScalarQED}, in Lorenz gauge this data is entirely contained in the transverse angular components. Hence for the sake of obtaining the data map we can ignore the longitudinal $\dr(p\cdot x)$ and radial $\dr r$ terms, since they are fixed in terms of the transverse angular components. The remaining angular terms are
\eq{fsp2a1}{
    \psi^{\text{ang}}_z &= \frac{ir}{4\pi}\int\frac{\dr\omega}{2\pi}\dr^2\hat p \omega \tilde a_z^+(\omega, -\hat p)e^{ip\cdot x}\dr(\gamma^{z'\overline z'}\p_{\overline z'}(\hat x\cdot \hat x'))\int\dr v \frac{e^{-iv}}{(v + i\epsilon)^2}\\
    &= -\frac{ir}{2\pi}\int\frac{\dr\omega}{2\pi}\dr^2\hat p \omega \tilde a^+_z(\omega, -\hat p)\dr(\gamma^{z'\overline z'}\p_{\overline z'}(\hat x\cdot \hat x')).
}
We note that
\eq{fsp3}{
    \dr(\gamma^{z'\overline z'}\p_{\overline z'}(\hat x\cdot \hat x')) = \left( \frac{ 1 + z'\overline z}{1 + z\overline z}\right)^2 \dr z - \left( \frac{z - z'}{1 + z\overline z} \right)^2 \dr\overline z
}
so together with the saddle point approximation \eqref{a4a}, the angular part of the wavefunction on $\scrI^-$ is given by
\eq{fsp3a1}{
    \psi^\text{ang}_z \arrow a^+_z\Big( -\frac{\pi}{2} + \frac{v}{R}, \hat x\Big)\dr z.
}
The complete map to the flat Carroll data is therefore given by
\eq{fsp4}{
    \hat{\overline A}_0^+(v, \hat x) = a_z^+\Big(- \frac{\pi}{2} + \frac{v}{R}, \hat x\Big)\dr z + a_{\overline z}^+\Big( - \frac{\pi}{2} + \frac{v}{R}, \hat x\Big).
}
We see that the flat Carroll data is precisely the background gauge field of the CFT, suitably localized on the AdS boundary.  This is the simplest possible relation between the AdS and Carroll data. Indeed, working with the Carroll data has allowed us to avoid a potential headache in trying to match the polarization induced by the flat limit to a polarization choice in flat space.

\subsection{Future sources}

The flat limit of $K^z$ with a source spread over $\tau' = + \frac{\pi}{2} + \frac{u}{R}$ is given by
\eq{fsf1}{
    K^z \arrow \frac{iR\omega^2}{4\pi^2}\frac{1}{-\omega u - p\cdot x + i\epsilon}\dr\left( \frac{r\gamma^{z'\overline z'}\p_{\overline z'}(\hat x \cdot \hat x')}{-\omega u - p\cdot x + i\epsilon} \right)
}
where now $p^\mu = \omega(1, \hat x')$, as was the case for the scalar field. We write the boundary source
\eq{fsf2}{
    a_z^-(\tau', \hat x') = \int\frac{\dr\omega}{2\pi}\tilde a^-_z(\omega, \hat x') e^{i\omega R(\tau' - \frac{\pi}{2})}.
}
The remainder of the calculation is essentially identical to the case of past sources, and the result for the Carroll data map is
\eq{fsf3}{
    \hat{\overline A}_{0}^- &= a^-_z\Big(\frac{\pi}{2} + \frac{u}{R}, \hat x\Big)\dr z + a^-_{\overline z}\Big(\frac{\pi}{2} + \frac{u}{R}, \hat x\Big)\dr \overline z.
}

\section{Flat space vs AdS symmetries}
\label{Flat space vs AdS symmetries}

\subsection{Flat space Lorentz transformations at the AdS boundary}

It's useful to verify how Lorentz transformations acting on the flat space boundary data are realized in the AdS boundary description.  Recall that the former are, on $\Ic^+$,
\eq{qq1}{ & r~\rt ~ r_\Lambda=   {r \over  f_\Lambda(z,\zb) } \cr
& u ~\rt ~u_\Lambda=  f_\Lambda(z,\zb)  u  \cr
& z~\rt ~ z_\Lambda= {az+b\over cz +d}  \cr
& \zb~\rt ~ \zb_\Lambda= {\ab\zb+\bb\over \cb\zb +\db}~.}
We now wish to write down  AdS isometries which reduces to \rf{qq1} in the flat region and then analyze their behavior at the AdS boundary.  Recall that the AdS$_4$ hyperboloid condition
\eq{qq2}{ (X^{-1})^2 +(X^0)^2 -(X^1)^2 - (X^2)^2- (X^3)^2 =R^2}
is solved using global coordinates as
 \eq{qq3}{ X^{-1}& = R {\cos \tau \over \cos \rho}\cr
 X^0 &= R{\sin \tau \over \cos \rho} \cr
 X^1& = R \tan\rho  {(z+\zb) \over 1+z\zb} \cr
 X^2& =  R \tan\rho  {-i(z-\zb) \over 1+z\zb} \cr
X^3& =   R \tan\rho  {1-z\zb\over 1+z\zb} }
Since the flat region is obtained by writing $\tau = {t\over R}$ and $\rho = {r\over R}$ and then taking $R\rt \infty$, we see that $X^{-1} \rt 1$ in this limit.   Lorentz transformation thus leave $X^{-1}$ invariant.  This suggests that we combine the remaining embedding coordinates into a matrix as 
\eq{qq4}{ X =
\left(
\begin{array}{cc}
  X^0-X^3  &  X^1+iX^2    \\
  X^1- iX^2 &    X^0 +X^3
\end{array}
\right)
 }
and consider the AdS isometries 
\eq{qq4a}{X\rt X_\Lambda =SX S^\dagger}
with  $S\in SL(2,C)$ as in \rf{b15}. Given that $X^\mu \rt x^\mu= (t,r\xh^i)$ in the large $R$ limit, it's immediately clear that we recover \rf{qq1} in this limit. 

Next we ask how such AdS isometries act on the relevant parts of the AdS boundary.  Let's focus on the thin strip near $\tau={\pi\over 2}$ and choose suitable notation to make clear the relation to the Lorentz transformations in the flat region.   To zoom in on this boundary strip we write 
\eq{qq5}{ U=X^0 -{R\over \cos \rho}}
and take $\cos \rho \rt 0$ at fixed $(U,z,\zb)$.  This forces $\tau \rt {\pi\over 2} $ as desired. We can also trade $\rho$ for $r$ by defining $r={R\over \cos \rho}$.   The transformation \rf{qq4a} in general acts   in a messy way on the AdS coordinates but simplifies near  our  boundary region to give
\eq{qq6}{ & r~\rt ~r_\Lambda   =  {r\over f_\Lambda(z,\zb) }  \cr
& U ~\rt ~U_\Lambda=  f_\Lambda(z,\zb)  U  \cr
& z~\rt ~ z_\Lambda= {az+b\over cz +d}  \cr
& \zb~\rt ~ \zb_\Lambda= {\ab\zb+\bb\over \cb\zb +\db}  }
Finally, we  write $\tau = {\pi\over 2} +{u\over R}$ in which case the $U$ transformation at large $R$ can be worked out to give
\eq{qq7}{u~\rt ~ u_\Lambda = f_\Lambda(z,\zb) u}
and we recover the flat space formulas \rf{qq1}.  The same analysis goes through relating the transformations on $\Ic^-$ to those in the strip at $\tau=-{\pi\over 2}.$  These AdS isometries acts as  conformal transformations on the AdS boundary on account of the rescaling of $r$.    The conformal invariance of the boundary CFT thus implies, under the data map relating the flat space and AdS generating functionals, the Lorentz invariance of the flat space S-matrix.  In more detail, a CFT source $J$ is identified by the non-normalizable asymptotics of the bulk scalar field via 
\eq{qq8}{ \phi(\rho,u,z,\zb)  \sim J(u,z,\zb) (\cos \rho)^{d-\Delta} + \ldots }
where $d=3$ for AdS$_4$.   Since the scalar field obeys $\phi_\Lambda(\rho,u,z,\zb) = \phi(\rho_\Lambda,u_\Lambda,z_\Lambda,\zb_\Lambda) $ the boundary source transforms as 
\eq{qq9}{ J_\Lambda(u,z,\zb) =  [f_\Lambda(z,\zb)]^{\Delta-d}  J(u_\Lambda,z_\Lambda,\zb_\Lambda)}
The invariance of $Z_{\rm AdS}[J]$ under this transformation translates into the invariance of $Z_{\rm AFS}$ under Lorentz transformations.

\subsection{Large gauge transformations}

We wish to relate large gauge transformations in Minkowski space and AdS.  We restrict  to $4$ spacetime dimensions and draw on formulas in \cite{Hijano:2020szl}.

\subsubsection{Large gauge transformations in Minkowski space}

A large gauge transformation in Minkowski space is a solution of $\nabla^2 \lambda=0$  with boundary conditions on $\Ic$ given by \rf{d13}.   Taking $\lambda_0(\xh) = Y_\ell^m(\xh)$ and writing the ansatz
\eq{g1}{ \lambda(x) = f_\ell\left( {t\over r}\right) Y_\ell^m(\xh) }
we find, with $y={t\over r}$,
\eq{g2}{ \nabla^2 \lambda = {1\over r^2} \left[ (y^2-1)f_\ell''(y)-\ell(\ell+1) f_\ell(y)   \right]Y_\ell^m =0 }
The solution that is smooth at $r=0$ is
\eq{g3}{  \lambda(y) =  (1-y^2) y^{-2-\ell} F\left( {2+\ell\over 2},{3+\ell\over 2},{3+2\ell\over 2}, {1\over y^2}\right) Y_\ell^m(\xh)}
where $F={_2}F_1$ is the usual  hypergeometric function.  The asymptotics on $\Ic^+$ are given by taking $y= {t\over r} = {u+r\over r} \rt 1$.   To read off these asymptotics its convenient to use a hypergeometric transformation to obtain 
\eq{g4}{ \lambda(y) = -(1+y)^{-\ell} F\left( \ell,\ell+1,2\ell+2,{2\over 1+y} \right) Y_\ell^m(\xh) }
which gives
\eq{g5}{\lambda_0(\xh) =  \lambda\big|_{y=1} =  -{2^{-\ell} \Gamma(2\ell+2) \over \Gamma(\ell+1)\Gamma(\ell+2) } Y_\ell^m(\xh)  }
For the asymptotics on $\Ic^-$ we use another transformation to write 
\eq{g6}{ \lambda(y) = (-1)^{\ell+1} ( 1-y)^{-\ell} F\left(\ell,\ell+1,2\ell+2, {2\over 1-y}\right) Y_\ell^m(\xh)  }
so that the limit  $y= {v-r\over r} \rt -1  $ gives
\eq{g7}{ \lambda\big|_{y=-1} =  (-1)^{\ell} \lambda_0(\xh)= \lambda_0(-\xh)  }
in terms of \eqref{g5}, where $\xh\rt -\xh$ is the antipodal map.  This is in accord with \rf{d13}. 

\subsubsection{Large gauge transformations in AdS$_4$}

We now wish to write down a solution of $\nabla^2 \lambda =0 $ in AdS global coordinates that reduces to \rf{g3} in the flat space limit.  This turns out to be surprisingly easy to achieve.   In particular if we consider the ansatz 
\eq{g8}{ \lambda(x) = f_\ell\left( {\sin \tau\over \sin \rho}\right) Y_\ell^m(\xh) }
then 
\eq{g9}{R^2 \nabla^2 \lambda=  \cot^2 \rho \big[ (y^2-1)f''(y) - \ell(\ell+1)f(y)\big]Y_\ell^m(\xh) }
so the solution is the same as in \rf{g4} but now with $y={\sin \tau \over \sin \rho}$, which manifestly has the desired flat space limit, since $y\rt {t\over r} $ in this limit. Furthermore, taking $\rho \rt {\pi\over 2} $  with $\tau = \pm {\pi\over 2} $ takes $y\rt \pm 1$.  So the large gauge transformation takes the same form on $\Ic^+$ as it does at the   $\tau={\pi\over 2}$ thin strip boundary region, and likewise for $\Ic^-$ and the $\tau=-{\pi\over 2}$ thin strip boundary region.

This result could have been foreseen given that the AdS $\rt $ Carroll map essentially equates the gauge fields on the respective boundaries.  The implication is that large gauge invariances of the AdS partition function immediately carry over to invariances of the Carroll partition function, and vice versa. However, we should note that the gauge transformation \rf{g8} is nontrivial on the entire AdS boundary, not just in the strips $\Ic^+$ and $\Ic^-$. But since the boundary partition function is separately invariant under gauge transformations that act outside the strips, we  can effectively ignore  contributions outside the strips.  It follows that the leading soft theorem in flat space, when embedded in AdS, converts a standard statement about the gauge invariance of the AdS partition function, which in turn implies Ward identities.  This was the conclusion reached in \cite{Hijano:2020szl} by a slightly different chain of argument.

\subsubsection{Boundary gauge modes}

Since the Minkowski space null boundaries $\Ic^+$ and $\Ic^-$  are Cauchy surfaces the large gauge transformations that act there act on phase space.  Under the embedding into AdS these surfaces map to portions of the timelike AdS boundary, so the above extension of the large gauge transformations act as changes in boundary conditions (i.e. change the theory) rather than in phase space, so we do not obtain ``boundary photons" on the AdS boundary via this construction.   This is to be contrasted with the behavior found in \cite{DHoker:2010xwl,Kim:2023vbj}  where boundary photon/gravitons of an ``IR" AdS space map, via an extension of the large gauge transformations/diffeomorphims, to boundary gravitons of the full UV AdS.  The difference is that in the latter case the extension lead to normalizable behavior at the UV boundary rather than a change of boundary conditions.  However, it is possible that there exists an analogous extension in the present setup.

\section*{Acknowledgments}
We thank  Kristan Jensen, Ruben Monten, Andrea Puhm, Balt van Rees, and Romain Ruzziconi for discussions. P.K. is supported in part by the National Science Foundation grant PHY-2209700.  P.K. acknowledge Institut Pascal at Université Paris-Saclay for support during the completion of this work through  the program “Investissements d’avenir” ANR-11-IDEX-0003-01. R.M.M. thanks the CERN theory department for hospitality during a portion of this work.

\appendix

\section{Asymptotic special conformal transformations}
\label{Sec:Asymptotic special conformal transformations}

Here we work out the action of special conformal transformations\footnote{The closely related example of inversions have been studied in the context of the celestial basis in \cite{Jorstad:2023ajr} where inversions were related to the shadow transform on the celestial sphere.} (SCTs) on the asymptotic data of a real scalar field, focusing on the transformation at $\scrI^+$. We begin by recalling that a finite SCT with parameters $b_\mu$ acts as
\eq{sct1}{
    \phi(x)\arrow\phi'(x) = \Omega^{-1}(x) \phi(x')
}
where
\eq{sct2}{
    \Omega(x) &= 1 - 2b\cdot x + b^2 x^2\cr
    x'^\mu &= \frac{x^\mu - x^2 b^\mu}{\Omega(x)}.
}
We may express this in terms of $u = t - r$ as
\eq{sct3}{
    u' &= \frac{u + r - x^2 b^t - \sqrt{r^2 + 2x^2 r\vec b\cdot \hat x + x^4 \vec b^2}}{\Omega(x)}\cr
    r'^2 &= \frac{r^2 - 2x^2 r \vec b\cdot x + x^4 \vec b^2}{\Omega^2(x)}\cr
    \hat x' &= \frac{r \hat x^i - x^2 b^i}{\sqrt{r^2 - 2x^2 r\vec b\cdot \hat x + x^4 \vec b^2}}.
}
At leading large $r$, these yield
\eq{sct4}{
    u' &\approx \frac{1 + 2 b^t u - \sqrt{1 + 4u\vec b\cdot \hat x + 4u^2 \vec b^2}}{-2(-b^t + \vec b\cdot \hat x) - 2b^2 u},\cr
    r' &\approx \frac{\sqrt{1 + 4u\vec b\cdot \hat x + 4u^2 \vec b^2}}{-2(-b^t + \vec b \cdot \hat x) - 2b^2 u},\cr
    \hat x'^i &\approx \frac{\hat x^i + 2ub^i}{1 + 4u\vec b\cdot \hat x + 4u^2 \vec b^2}.
}
Note that for finite $b_\mu$, $r'$ does not diverge as $r\arrow\infty$. This is because finite SCTs move the location of null infinity to a finite coordinate location. Expanding in small $b_\mu$
\eq{sct5}{
    u' &\approx u - \frac{b^2 - \vec b^2 + (\vec b\cdot \vec x)^2}{-b^t + \vec b\cdot \hat x}u^2,\cr
    r' &\approx - \frac{1}{2}\frac{1}{-b^t + \vec b\cdot \hat x} + \frac{1}{2}\frac{b^2 - 2(b^t + \vec b\cdot \hat x)\vec b\cdot \hat x}{-b^t + \vec b\cdot \hat x}u,\cr
    \hat x'^i &\approx \hat x^i + 2u(b^i - \vec b \cdot \hat x \hat x^i)
}
we see a pole develop in $r'$. Hence the boundary of flat space remains at null infinity only in the limit where the components of $b^\mu$ go to zero.

Our interest is in the transformation of the leading component of $\phi$ at large $r$, namely $\phi_1(u, \hat x)$ where $\phi(r, u, \hat x)\arrow \frac{1}{r}\phi_1(u, \hat x) + \frac{1}{r^2}\phi_2(u, \hat x) + \cdots$. Hence we seek to compute
\eq{sct6}{
    r\phi'(x) = \phi'_1(x) = \frac{r}{\Omega(x)}\phi(r', u', \hat x')
}
by first taking $r$ large, and then $b_\mu$ small. With the expansions above it's straightforward to compute
\eq{sct7}{
    \phi_1'(x) = & \left[ - \frac{1}{2(-b^t + \vec  b \cdot \hat x)} + \frac{b^2 u}{2(-b^t + \vec b \cdot \hat x)^2} \right]\phi\left( - \frac{1}{2}\frac{1}{-b^t + \vec b\cdot \hat x} + \frac{1}{2}\frac{b^2 - 2(b^t + \vec b\cdot \hat x)\vec b\cdot \hat x}{-b^t + \vec b\cdot \hat x}u, u', \hat x' \right).
}
The small $b_\mu$ expansion, using the $1/r$ expansion of $\phi$, then produces at leading order
\eq{sct8}{
    \phi'_1(x) = (1 - 2\vec b\cdot \hat x u)\phi_1(u', \hat x') - 2(-b^t + \vec b\cdot \hat x)\phi_2(u', \hat x').
}

The appearance of $\phi_2$ in \eqref{sct8} makes this transformation appear non-local on $\scrI^+$. However, using the asymptotic equation of motion $\nabla^2\phi = 0$ we note
\eq{sct9}{
    \p_u \phi_2 = - \frac{1}{2}\hat\nabla^2 \phi_1
}
where $\hat\nabla^2$ is the Laplacian on the sphere. Taking a $u$-derivative of \eqref{sct8} we may use the equation of motion to write the local form
\eq{sct10}{
    \p_u \phi'_1 = \Big[ (1 - 2\vec b \cdot \hat x u)\p_u + (-b^t + \vec b \cdot \hat x)\hat \nabla^2 - 2\vec b\cdot \hat x \Big] \phi_1(u', \hat x')
}
where $u'$ and $\hat x'$ are as in \eqref{sct5}, so in particular the $\p_u$ and $\hat\nabla^2$ will generally act on both arguments of $\phi_1$. We note that a similar relation between apparent non-locality in asymptotic transformations and mixing with subleading terms in the $1/r$ expansion has been observed in relation to the subleading soft theorem \cite{Himwich:2019dug}.

\section{Equality of flat and AdS boundary terms}\label{Sec:bndy terms}

For concreteness we consider a massless real scalar field.  As usual we should distinguish between the normalizable and non-normalizable behavior of the field near the AdS boundary at $\rho=\pi/2$,
\eq{e7a}{ {\rm non-normalizable:}\quad &\phi \sim J(\tau,\xh) + \ldots \cr
{\rm normalizable:}\quad &\phi \sim f(\tau,\xh) \cos^3 \rho + \ldots} 
We will refer to $J(\tau,\xh)$ as a ``source" due to its role as such on the CFT side of the AdS/CFT dictionary where it corresponds to adding to the CFT action a term $\int J{\cal O}$ where ${\cal O}$ is the CFT operator dual to $\phi$.   Our AdS boundary conditions will consist of fixing the positive frequency part of the source in the boundary region near $\tau=-\pi/2$, and the negative frequency part of the source in the boundary region near $\tau=+\pi/2$.  As should be clear, this also corresponds to the type of boundary conditions we impose at $\Ic^\pm$ in flat spacetime. 

In the main body of the text we  worked out the relation between the AdS and flat space boundary conditions. In order to establish a relation between the AdS and flat space path integrals we need to understand how the respective actions, including boundary terms, are related to each other.  On the flat space side, as reviewed in section \ref{AFSreview} the appropriate action is 
\eq{e7b}{ I_{\rm flat}[\phi,\phib] & = \int_{\rm flat}\! d^4x \left( \frac{1}{2} \phi \nabla^2 \phi -V(\phi) \right) +  (\phib^-,\phi)_{\Ic^+} - (\phib^+,\phi)_{\Ic^-}~,}
with
\eq{e7c}{  (\phib^-,\phi)_{\Ic^+} & = {1\over 2} \int_{\Ic^+} du d^2\xh  ( \phib_1^-\p_u \phi_1 - \p_u \phib_1^- \phi_1  )\cr
 (\phib^+,\phi)_{\Ic^-} & = {1\over 2} \int_{\Ic^-} dv d^2\xh  ( \phib_1^+\p_v \phi_1 - \p_v \phib_1^+ \phi_1  )   }
On the other hand, the AdS action takes the form 
\eq{e7d}{ I_{\rm AdS}= \int_{\rm AdS}\! d^4x \sqrt{g} \Big( -{1\over 2} (\nabla \phi)^2 -V(\phi) \Big) + I^{\rm loc}_{\rm bndy} [ J]   }
We have indicated the AdS boundary terms, which are a local functional of the source $J$.  These are needed to render the AdS action finite for general sources,  but they do not contribute to boundary correlation functions of operators at non-coincident points, and for the same reason they will not be relevant to our considerations and will be dropped henceforth. 
We also note that the kinetic terms in $I_{\rm flat}$ and $I_{\rm AdS}$ differ by an integration by parts; this will be relevant in the comparison.

We are interested in comparing the actions for field configurations in which the field is effectively in the outer region between the AdS boundary and the null boundary $\Ic$ of the embedded flat space region.  In this region we can therefore neglect the potential $V(\phi)$.  

For scattering we work in a basis of solutions which look like localized wavepackets traveling between $\Ic$ and the timelike boundary of AdS near $\tau  = \pm {\pi\over 2}$.   We assume negligible overlap of the wavepackets in the outer region.  We focus on a wavepacket in the future region, travelling between $\Ic^+$ and the $\tau={\pi \over 2} $ region of the AdS boundary which we denote as $\p {\rm AdS}^+$.  We  first rewrite the flat boundary term $ (\phib^-,\phi)_{\Ic^+} $ as an AdS boundary term; assuming a free field we have\footnote{The minus sign stems from the fact that the flat boundary is null, and best viewed as a limit of spacelike surface, while the AdS boundary is timelike. } 
\eq{e7e}{ (\phib^-,\phi)_{\Ic^+} = -{1\over 2} \int_{\p {\rm AdS}^+} d\tau d^2\xh {1\over \cos^2 \rho} ( \phib_- \p_\rho \phi_+ - \p_\rho \phib_- \phi_+ )  }
$\phib_i$ has non-normalizable behavior corresponding to a negative frequency source near $\tau=\pi/2$, while $\phi_+\sim \cos^3\rho$ has normalizable falloff.  Therefore only the first term on the right hand side survives as $\rho \rt \pi/2$.

On the other hand, integrating by parts the AdS kinetic term gives 
\eq{e7f}{ -{1\over 2} \int_{{\rm AdS}} \! d^4x \sqrt{g} (\nabla \phi)^2={1\over 2} \int_{\rm AdS} \! d^4x \sqrt{g}   \phi\nabla^2 \phi -{1\over 2}  \int_{\p {\rm AdS}^+} d\tau d^2\xh  {1\over \cos^2 \rho}   \phib_- \p_\rho \phi_+  }
where we again only retained the boundary terms that contributes as $\rho \rt \pi/2.$ We can do the same manipulations in the past region.

Under our assumptions it's now clear that the two actions $I_{\rm flat}$ and $I_{\rm AdS}$ agree, since they only differ by terms involving $V(\phi)$ and $\nabla^2 \phi$ in the outer region, both of which are assumed to be negligible.

\section{Lorentzian spin-1 bulk-boundary propagator}
\label{Sec:spin-1 bulk-boundary propagator}

Here we obtain the Lorentzian bulk-boundary propagator for spin-1 fields from the better known expression in Euclidean Poincar\'e AdS. Throughout we consider only the special case $d = 3$. The strategy will be to write the expression valid in Poincar\'e coordinates in terms of embedding space quantities, which can then be expressed in terms of global coordinates.

Euclidean AdS$_4$ can be defined as the hyperboloid
\eq{spin1}{
    -R^2 = \eta_{MN} X^M X^N = -(X^{-1})^2 + (X^0)^2 + (X^1)^2 + (X^2)^2 + (X^3)^2
}
embedded in $\R^{1,3}$. It's useful to define $X^\pm = X^{-1} \pm X^0$ in terms of which
\eq{spin2}{
    -R^2 &= \eta_{MN}X^M X^N \phantom{\dr\dr}= -X^+ X^- + (X^1)^2 + (X^2)^2 + (X^3)^2\\
    \dr s^2 &= \eta_{MN}\dr X^M \dr X^N = -\dr X^+ \dr X^- + (\dr X^1)^2 + (\dr X^2)^2 + (\dr X^3)^2.
}
Euclidean Poincar\'e coordinates are defined by solving \rf{spin1} as
\eq{spin3}{
    X^+ = R \frac{x_0^2 + |\vec x|^2}{x_0},\ \ \ X^- = \frac{R}{x_0},\ \ \ X^i = \frac{R}{x_0}x^i,\ \ \ i = 1,2,3.
}

Points in the boundary of AdS are labeled by null vectors in embedding space. In practice this means taking the leading coefficient of $X^M$ as $x_0 \arrow 0$,
\eq{spin4}{
    P^+ = R |\vec x'|^2,\ \ \ P^- = R,\ \ \ P^i = Rx'^i.
}
The freedom to rescale the null vectors corresponds to a choice of conformal frame on the boundary.  A  rescaling $P \arrow e^{\omega(x')}P$ acts on the boundary metric by $h_{ij} \arrow e^{2\omega(x')}h_{ij}$.  The choice here corresponds to selecting boundary metric
\eq{spin5}{
    h_{ij}\dr x^i\dr x^j = \dr \vec x\cdot \dr \vec x.
}

In this frame, the 3 1-form bulk-boundary propagators for a spin-1 field were written down in \cite{Witten:1998qj}
\eq{spin6}{
    K^i(x, x') = \frac{2}{\pi^2}\left( \frac{x_0}{x_0^2 + |\vec x - \vec x'|^2} \right) \dr\left( \frac{x^i - x'^i}{x_0^2 + |\vec x - \vec x'|^2} \right).
}
These obey Maxwell's equations, satisfy Lorenz gauge, and have boundary behavior
\eq{spin7}{
    K^i(x, x') \arrow \delta^3(\vec x - \vec x')\dr x^i.
}

It is simple to check that
\eq{spin8}{
    P\cdot X &= -\frac{R^2}{2x_0}(x_0^2 + |\vec x - \vec x'|^2)\\
    -\frac{1}{2}h^{ij}\frac{\p}{\p x'^j}\ln(-P\cdot X) &= \frac{x^i - x'^i}{x_0^2 + |\vec x - \vec x'|^2}
}
from which it follows
\eq{spin9}{
    K^i(x, x') = \frac{R^2}{2\pi^2}\frac{1}{P\cdot X}h^{ij}\frac{\p}{\p x'^j}\dr \ln(-P\cdot X).
}
Importantly, we note that $\sqrt{-h}K^i$ does not depend on our choice of conformal frame.  This is important as one can check that our choice of conformal frame in global coordinates differs from the choice made in Poincar\'e coordinates.

Global coordinates on AdS$_4$ are defined by
\eq{spin10}{
    X^+ = R\frac{e^{\tau_E}}{\cos\rho},\ \ \ X^- = R\frac{e^{-\tau_E}}{\cos\rho},\ \ \ X^i = R\tan\rho \hat x^i
}
and the boundary given by
\eq{spin11}{
    P^+ = R e^{\tau'_E},\ \ \ P^- = R e^{-\tau'_E},\ \ \ P^i = R\hat x'^i.
}
With this we have
\eq{spin12}{
    h_{ij}\dr x^i\dr x^j &= \dr\tau_E^2 + \dr \hat x\cdot \dr\hat x,\\
    P\cdot X &= - R^2 \frac{\cosh(\tau_E - \tau'_E) - \sin\rho \hat x\cdot \hat x'}{\cos\rho}.
}
which implies
\eq{spin13}{
    K^{\tau_E}(x, x') &= \frac{1}{2\pi^2}\frac{\cos\rho}{\cosh(\tau_E - \tau'_E) - \sin\rho \hat x\cdot \hat x'}\dr \left( \frac{\sinh(\tau_E - \tau'_E)}{\cosh(\tau_E - \tau'_E) - \sin\rho \hat x\cdot \hat x'} \right),\\
    K^z(x, x') &= \frac{1}{4\pi^2}\frac{\cos\rho}{\cosh(\tau_E - \tau'_E) - \sin\rho \hat x\cdot \hat x'}\dr\left( \frac{\sin\rho \gamma^{z'\overline z'}\p_{\overline z'}(\hat x\cdot \hat x')}{\cosh(\tau_E - \tau'_E) - \sin\rho \hat x\cdot \hat x'} \right),\\
    K^{\overline z}(x, x') &= \frac{1}{4\pi^2}\frac{\cos\rho}{\cosh(\tau_E - \tau'_E) - \sin\rho \hat x\cdot \hat x'}\dr\left( \frac{\sin\rho \gamma^{z'\overline z'}\p_{z'}(\hat x\cdot \hat x')}{\cosh(\tau_E - \tau'_E) - \sin\rho \hat x\cdot \hat x'} \right).
}
We note that in the parametrization
\eq{spin14}{
    \hat x = \frac{1}{1 + z\overline z}(z + \overline z, -i(z - \overline z), 1 - z\overline z)
}
we have
\eq{spin15}{
    \gamma^{z'\overline z'}\p_{\overline z'}(\hat x\cdot \hat x') &= \frac{1 + z'\overline z}{1 + z\overline z}(z - z'),\\
    \gamma^{z'\overline z'}\p_{z'}(\hat x\cdot \hat x') &= \frac{1 + z\overline z'}{1 + z\overline z}(\overline z - \overline z').
}

The Lorentzian signature bulk-boundary propagators are determined by demanding that they produce the Lorentzian Dirac distribution on the boundary\footnote{Note that $1 = \int\dr\tau_E \delta(\tau_E) = \int\dr\tau i\delta(\tau_E)$ implies $\delta(\tau) = i\delta(\tau_E)$.},
\eq{spin16}{
    K_L^\tau &= K_E^{\tau_E} \arrow \delta(\tau_E - \tau'_E)\delta^{(2)}(\hat x - \hat x')\dr\tau_E = \delta(\tau - \tau') \delta^{(2)}(\hat x - \hat x')\dr\tau,\\
    K_L^z &= iK^z_E \arrow i\delta(\tau_E - \tau'_E)\delta^{(2)}(\hat x - \hat x')\dr z = \delta(\tau - \tau') \delta^{(2)}(\hat x - \hat x')\dr z.
}
Hence the Lorentzian signature bulk-boundary propagators are given by
\eq{spin17}{
    K_L^\tau(x, x') &= \frac{i}{2\pi^2} \frac{\cos\rho}{\cos(\tau - \tau') - \sin\rho \hat x\cdot\hat x' + i\epsilon} \dr\left(\frac{\sin(\tau - \tau')}{\cos(\tau -\tau') - \sin\rho \hat x\cdot \hat x' + i\epsilon} \right),\\
    K_L^z(x, x') &= \frac{i}{4\pi^2}\frac{\cos\rho}{\cos(\tau - \tau') - \sin\rho \hat x\cdot \hat x' + i\epsilon} \dr\left( \frac{\sin\rho \gamma^{z'\overline z'}\p_{\overline z'}(\hat x\cdot \hat x')}{\cos(\tau - \tau') - \sin\rho \hat x\cdot \hat x' + i\epsilon} \right),\\
    K_L^{\overline z}(x, x') &= \frac{i}{4\pi^2}\frac{\cos\rho}{\cos(\tau - \tau') - \sin\rho \hat x\cdot \hat x' + i\epsilon} \dr\left( \frac{\sin\rho \gamma^{z'\overline z'}\p_{z'}(\hat x\cdot \hat x')}{\cos(\tau - \tau') - \sin\rho \hat x\cdot \hat x' + i\epsilon} \right).
}
Here we have given the $i\epsilon$ prescription valid for $\sin|\tau - \tau'| > 0$ since that is all we will need in the main text. The globally valid prescription is to instead replace $\tau - \tau' \arrow (1 - i\epsilon)(\tau - \tau')$. In the main text we never need the Euclidean bulk-boundary propagator, so we drop the subscript $L$.

\bibliographystyle{bibstyle2017}
\bibliography{collection}

\providecommand{\href}[2]{#2}\begingroup\begin{thebibliography}{10}

\bibitem{Strominger:2017zoo}
A.~Strominger, {\it {Lectures on the Infrared Structure of Gravity and Gauge Theory}},  \href{http://arxiv.org/abs/1703.05448}{{\ttfamily arXiv:1703.05448 [hep-th]}}.

\bibitem{Pasterski:2021rjz}
S.~Pasterski, {\it {Lectures on celestial amplitudes}},  \href{http://dx.doi.org/10.1140/epjc/s10052-021-09846-7}{{\sf Eur. Phys. J. C} {\sf {81} }{\sf no.~12, }{\sf (2021) }{\sf 1062}}, \href{http://arxiv.org/abs/2108.04801}{{\ttfamily arXiv:2108.04801 [hep-th]}}.

\bibitem{Raclariu:2021zjz}
A.-M. Raclariu, {\it {Lectures on Celestial Holography}},  \href{http://arxiv.org/abs/2107.02075}{{\ttfamily arXiv:2107.02075 [hep-th]}}.

\bibitem{Pasterski:2021raf}
S.~Pasterski, M.~Pate, and A.-M. Raclariu, {\it {Celestial Holography}},  in {\sf {Snowmass 2021}}.
\newblock 11, 2021.
\newblock \href{http://arxiv.org/abs/2111.11392}{{\ttfamily arXiv:2111.11392 [hep-th]}}.

\bibitem{McLoughlin:2022ljp}
T.~McLoughlin, A.~Puhm, and A.-M. Raclariu, {\it {The SAGEX review on scattering amplitudes chapter 11: soft theorems and celestial amplitudes}},  \href{http://dx.doi.org/10.1088/1751-8121/ac9a40}{{\sf J. Phys. A} {\sf {55} }{\sf no.~44, }{\sf (2022) }{\sf 443012}}, \href{http://arxiv.org/abs/2203.13022}{{\ttfamily arXiv:2203.13022 [hep-th]}}.

\bibitem{Donnay:2023mrd}
L.~Donnay, {\it {Celestial holography: An asymptotic symmetry perspective}},  \href{http://dx.doi.org/10.1016/j.physrep.2024.04.003}{{\sf Phys. Rept.} {\sf {1073} }{\sf (2024) }{\sf 1--41}}, \href{http://arxiv.org/abs/2310.12922}{{\ttfamily arXiv:2310.12922 [hep-th]}}.

\bibitem{Duval:2014uoa}
C.~Duval, G.~W. Gibbons, P.~A. Horvathy, and P.~M. Zhang, {\it {Carroll versus Newton and Galilei: two dual non-Einsteinian concepts of time}},  \href{http://dx.doi.org/10.1088/0264-9381/31/8/085016}{{\sf Class. Quant. Grav.} {\sf {31} }{\sf (2014) }{\sf 085016}}, \href{http://arxiv.org/abs/1402.0657}{{\ttfamily arXiv:1402.0657 [gr-qc]}}.

\bibitem{Duval:2014uva}
C.~Duval, G.~W. Gibbons, and P.~A. Horvathy, {\it {Conformal Carroll groups and BMS symmetry}},  \href{http://dx.doi.org/10.1088/0264-9381/31/9/092001}{{\sf Class. Quant. Grav.} {\sf {31} }{\sf (2014) }{\sf 092001}}, \href{http://arxiv.org/abs/1402.5894}{{\ttfamily arXiv:1402.5894 [gr-qc]}}.

\bibitem{Hartong:2015xda}
J.~Hartong, {\it {Gauging the Carroll Algebra and Ultra-Relativistic Gravity}},  \href{http://dx.doi.org/10.1007/JHEP08(2015)069}{{\sf JHEP} {\sf {08} }{\sf (2015) }{\sf 069}}, \href{http://arxiv.org/abs/1505.05011}{{\ttfamily arXiv:1505.05011 [hep-th]}}.

\bibitem{Hartong:2015usd}
J.~Hartong, {\it {Holographic Reconstruction of 3D Flat Space-Time}},  \href{http://dx.doi.org/10.1007/JHEP10(2016)104}{{\sf JHEP} {\sf {10} }{\sf (2016) }{\sf 104}}, \href{http://arxiv.org/abs/1511.01387}{{\ttfamily arXiv:1511.01387 [hep-th]}}.

\bibitem{Bagchi:2016bcd}
A.~Bagchi, R.~Basu, A.~Kakkar, and A.~Mehra, {\it {Flat Holography: Aspects of the dual field theory}},  \href{http://dx.doi.org/10.1007/JHEP12(2016)147}{{\sf JHEP} {\sf {12} }{\sf (2016) }{\sf 147}}, \href{http://arxiv.org/abs/1609.06203}{{\ttfamily arXiv:1609.06203 [hep-th]}}.

\bibitem{Ciambelli:2018wre}
L.~Ciambelli, C.~Marteau, A.~C. Petkou, P.~M. Petropoulos, and K.~Siampos, {\it {Flat holography and Carrollian fluids}},  \href{http://dx.doi.org/10.1007/JHEP07(2018)165}{{\sf JHEP} {\sf {07} }{\sf (2018) }{\sf 165}}, \href{http://arxiv.org/abs/1802.06809}{{\ttfamily arXiv:1802.06809 [hep-th]}}.

\bibitem{Ciambelli:2019lap}
L.~Ciambelli, R.~G. Leigh, C.~Marteau, and P.~M. Petropoulos, {\it {Carroll Structures, Null Geometry and Conformal Isometries}},  \href{http://dx.doi.org/10.1103/PhysRevD.100.046010}{{\sf Phys. Rev. D} {\sf {100} }{\sf no.~4, }{\sf (2019) }{\sf 046010}}, \href{http://arxiv.org/abs/1905.02221}{{\ttfamily arXiv:1905.02221 [hep-th]}}.

\bibitem{Salzer:2023jqv}
J.~Salzer, {\it {An embedding space approach to Carrollian CFT correlators for flat space holography}},  \href{http://dx.doi.org/10.1007/JHEP10(2023)084}{{\sf JHEP} {\sf {10} }{\sf (2023) }{\sf 084}}, \href{http://arxiv.org/abs/2304.08292}{{\ttfamily arXiv:2304.08292 [hep-th]}}.

\bibitem{Bagchi:2019clu}
A.~Bagchi, R.~Basu, A.~Mehra, and P.~Nandi, {\it {Field Theories on Null Manifolds}},  \href{http://dx.doi.org/10.1007/JHEP02(2020)141}{{\sf JHEP} {\sf {02} }{\sf (2020) }{\sf 141}}, \href{http://arxiv.org/abs/1912.09388}{{\ttfamily arXiv:1912.09388 [hep-th]}}.

\bibitem{Hansen:2021fxi}
D.~Hansen, N.~A. Obers, G.~Oling, and B.~T. S\o{}gaard, {\it {Carroll Expansion of General Relativity}},  \href{http://dx.doi.org/10.21468/SciPostPhys.13.3.055}{{\sf SciPost Phys.} {\sf {13} }{\sf no.~3, }{\sf (2022) }{\sf 055}}, \href{http://arxiv.org/abs/2112.12684}{{\ttfamily arXiv:2112.12684 [hep-th]}}.

\bibitem{Donnay:2022aba}
L.~Donnay, A.~Fiorucci, Y.~Herfray, and R.~Ruzziconi, {\it {Carrollian Perspective on Celestial Holography}},  \href{http://dx.doi.org/10.1103/PhysRevLett.129.071602}{{\sf Phys. Rev. Lett.} {\sf {129} }{\sf no.~7, }{\sf (2022) }{\sf 071602}}, \href{http://arxiv.org/abs/2202.04702}{{\ttfamily arXiv:2202.04702 [hep-th]}}.

\bibitem{Bagchi:2022emh}
A.~Bagchi, S.~Banerjee, R.~Basu, and S.~Dutta, {\it {Scattering Amplitudes: Celestial and Carrollian}},  \href{http://dx.doi.org/10.1103/PhysRevLett.128.241601}{{\sf Phys. Rev. Lett.} {\sf {128} }{\sf no.~24, }{\sf (2022) }{\sf 241601}}, \href{http://arxiv.org/abs/2202.08438}{{\ttfamily arXiv:2202.08438 [hep-th]}}.

\bibitem{Donnay:2022wvx}
L.~Donnay, A.~Fiorucci, Y.~Herfray, and R.~Ruzziconi, {\it {Bridging Carrollian and celestial holography}},  \href{http://dx.doi.org/10.1103/PhysRevD.107.126027}{{\sf Phys. Rev. D} {\sf {107} }{\sf no.~12, }{\sf (2023) }{\sf 126027}}, \href{http://arxiv.org/abs/2212.12553}{{\ttfamily arXiv:2212.12553 [hep-th]}}.

\bibitem{Bagchi:2023fbj}
A.~Bagchi, P.~Dhivakar, and S.~Dutta, {\it {AdS Witten diagrams to Carrollian correlators}},  \href{http://dx.doi.org/10.1007/JHEP04(2023)135}{{\sf JHEP} {\sf {04} }{\sf (2023) }{\sf 135}}, \href{http://arxiv.org/abs/2303.07388}{{\ttfamily arXiv:2303.07388 [hep-th]}}.

\bibitem{deBoer:2023fnj}
J.~de~Boer, J.~Hartong, N.~A. Obers, W.~Sybesma, and S.~Vandoren, {\it {Carroll stories}},  \href{http://dx.doi.org/10.1007/JHEP09(2023)148}{{\sf JHEP} {\sf {09} }{\sf (2023) }{\sf 148}}, \href{http://arxiv.org/abs/2307.06827}{{\ttfamily arXiv:2307.06827 [hep-th]}}.

\bibitem{Saha:2023hsl}
A.~Saha, {\it {Carrollian approach to 1 + 3D flat holography}},  \href{http://dx.doi.org/10.1007/JHEP06(2023)051}{{\sf JHEP} {\sf {06} }{\sf (2023) }{\sf 051}}, \href{http://arxiv.org/abs/2304.02696}{{\ttfamily arXiv:2304.02696 [hep-th]}}.

\bibitem{Nguyen:2023vfz}
K.~Nguyen and P.~West, {\it {Carrollian Conformal Fields and Flat Holography}},  \href{http://dx.doi.org/10.3390/universe9090385}{{\sf Universe} {\sf {9} }{\sf no.~9, }{\sf (2023) }{\sf 385}}, \href{http://arxiv.org/abs/2305.02884}{{\ttfamily arXiv:2305.02884 [hep-th]}}.

\bibitem{Nguyen:2023miw}
K.~Nguyen, {\it {Carrollian conformal correlators and massless scattering amplitudes}},  \href{http://dx.doi.org/10.1007/JHEP01(2024)076}{{\sf JHEP} {\sf {01} }{\sf (2024) }{\sf 076}}, \href{http://arxiv.org/abs/2311.09869}{{\ttfamily arXiv:2311.09869 [hep-th]}}.

\bibitem{Bagchi:2023cen}
A.~Bagchi, P.~Dhivakar, and S.~Dutta, {\it {Holography in Flat Spacetimes: the case for Carroll}},  \href{http://arxiv.org/abs/2311.11246}{{\ttfamily arXiv:2311.11246 [hep-th]}}.

\bibitem{Mason:2023mti}
L.~Mason, R.~Ruzziconi, and A.~Yelleshpur~Srikant, {\it {Carrollian amplitudes and celestial symmetries}},  \href{http://dx.doi.org/10.1007/JHEP05(2024)012}{{\sf JHEP} {\sf {05} }{\sf (2024) }{\sf 012}}, \href{http://arxiv.org/abs/2312.10138}{{\ttfamily arXiv:2312.10138 [hep-th]}}.

\bibitem{Cotler:2024xhb}
J.~Cotler, K.~Jensen, S.~Prohazka, A.~Raz, M.~Riegler, and J.~Salzer, {\it {Quantizing Carrollian field theories}},  \href{http://arxiv.org/abs/2407.11971}{{\ttfamily arXiv:2407.11971 [hep-th]}}.

\bibitem{Bondi:1962px}
H.~Bondi, M.~G.~J. van~der Burg, and A.~W.~K. Metzner, {\it {Gravitational waves in general relativity. 7. Waves from axisymmetric isolated systems}},  \href{http://dx.doi.org/10.1098/rspa.1962.0161}{{\sf Proc. Roy. Soc. Lond. A} {\sf {269} }{\sf (1962) }{\sf 21--52}}.

\bibitem{Sachs:1962zza}
R.~Sachs, {\it {Asymptotic symmetries in gravitational theory}},  \href{http://dx.doi.org/10.1103/PhysRev.128.2851}{{\sf Phys. Rev.} {\sf {128} }{\sf (1962) }{\sf 2851--2864}}.

\bibitem{Barnich:2009se}
G.~Barnich and C.~Troessaert, {\it {Symmetries of asymptotically flat 4 dimensional spacetimes at null infinity revisited}},  \href{http://dx.doi.org/10.1103/PhysRevLett.105.111103}{{\sf Phys. Rev. Lett.} {\sf {105} }{\sf (2010) }{\sf 111103}}, \href{http://arxiv.org/abs/0909.2617}{{\ttfamily arXiv:0909.2617 [gr-qc]}}.

\bibitem{Barnich:2010eb}
G.~Barnich and C.~Troessaert, {\it {Aspects of the BMS/CFT correspondence}},  \href{http://dx.doi.org/10.1007/JHEP05(2010)062}{{\sf JHEP} {\sf {05} }{\sf (2010) }{\sf 062}}, \href{http://arxiv.org/abs/1001.1541}{{\ttfamily arXiv:1001.1541 [hep-th]}}.

\bibitem{Barnich:2010ojg}
G.~Barnich and C.~Troessaert, {\it {Supertranslations call for superrotations}},  \href{http://dx.doi.org/10.22323/1.127.0010}{{\sf PoS} {\sf {CNCFG2010} }{\sf (2010) }{\sf 010}}, \href{http://arxiv.org/abs/1102.4632}{{\ttfamily arXiv:1102.4632 [gr-qc]}}.

\bibitem{Campiglia:2014yka}
M.~Campiglia and A.~Laddha, {\it {Asymptotic symmetries and subleading soft graviton theorem}},  \href{http://dx.doi.org/10.1103/PhysRevD.90.124028}{{\sf Phys. Rev. D} {\sf {90} }{\sf no.~12, }{\sf (2014) }{\sf 124028}}, \href{http://arxiv.org/abs/1408.2228}{{\ttfamily arXiv:1408.2228 [hep-th]}}.

\bibitem{Campiglia:2015yka}
M.~Campiglia and A.~Laddha, {\it {New symmetries for the Gravitational S-matrix}},  \href{http://dx.doi.org/10.1007/JHEP04(2015)076}{{\sf JHEP} {\sf {04} }{\sf (2015) }{\sf 076}}, \href{http://arxiv.org/abs/1502.02318}{{\ttfamily arXiv:1502.02318 [hep-th]}}.

\bibitem{Flanagan:2019vbl}
E.~E. Flanagan, K.~Prabhu, and I.~Shehzad, {\it {Extensions of the asymptotic symmetry algebra of general relativity}},  \href{http://dx.doi.org/10.1007/JHEP01(2020)002}{{\sf JHEP} {\sf {01} }{\sf (2020) }{\sf 002}}, \href{http://arxiv.org/abs/1910.04557}{{\ttfamily arXiv:1910.04557 [gr-qc]}}.

\bibitem{Freidel:2021fxf}
L.~Freidel, R.~Oliveri, D.~Pranzetti, and S.~Speziale, {\it {The Weyl BMS group and Einstein\textquoteright{}s equations}},  \href{http://dx.doi.org/10.1007/JHEP07(2021)170}{{\sf JHEP} {\sf {07} }{\sf (2021) }{\sf 170}}, \href{http://arxiv.org/abs/2104.05793}{{\ttfamily arXiv:2104.05793 [hep-th]}}.

\bibitem{Low:1958sn}
F.~E. Low, {\it {Bremsstrahlung of very low-energy quanta in elementary particle collisions}},  \href{http://dx.doi.org/10.1103/PhysRev.110.974}{{\sf Phys. Rev.} {\sf {110} }{\sf (1958) }{\sf 974--977}}.

\bibitem{Weinberg:1965nx}
S.~Weinberg, {\it {Infrared photons and gravitons}},  \href{http://dx.doi.org/10.1103/PhysRev.140.B516}{{\sf Phys. Rev.} {\sf {140} }{\sf (1965) }{\sf B516--B524}}.

\bibitem{Strominger:2021mtt}
A.~Strominger, {\it {$w_{1+\infty}$ Algebra and the Celestial Sphere: Infinite Towers of Soft Graviton, Photon, and Gluon Symmetries}},  \href{http://dx.doi.org/10.1103/PhysRevLett.127.221601}{{\sf Phys. Rev. Lett.} {\sf {127} }{\sf no.~22, }{\sf (2021) }{\sf 221601}}, \href{http://arxiv.org/abs/2105.14346}{{\ttfamily arXiv:2105.14346 [hep-th]}}.

\bibitem{Kulish:1970ut}
P.~P. Kulish and L.~D. Faddeev, {\it {Asymptotic conditions and infrared divergences in quantum electrodynamics}},  \href{http://dx.doi.org/10.1007/BF01066485}{{\sf Theor. Math. Phys.} {\sf {4} }{\sf (1970) }{\sf 745}}.

\bibitem{Prabhu:2022zcr}
K.~Prabhu, G.~Satishchandran, and R.~M. Wald, {\it {Infrared finite scattering theory in quantum field theory and quantum gravity}},  \href{http://dx.doi.org/10.1103/PhysRevD.106.066005}{{\sf Phys. Rev. D} {\sf {106} }{\sf no.~6, }{\sf (2022) }{\sf 066005}}, \href{http://arxiv.org/abs/2203.14334}{{\ttfamily arXiv:2203.14334 [hep-th]}}.

\bibitem{Kim:2023qbl}
S.~Kim, P.~Kraus, R.~Monten, and R.~M. Myers, {\it {S-matrix path integral approach to symmetries and soft theorems}},  \href{http://dx.doi.org/10.1007/JHEP10(2023)036}{{\sf JHEP} {\sf {10} }{\sf (2023) }{\sf 036}}, \href{http://arxiv.org/abs/2307.12368}{{\ttfamily arXiv:2307.12368 [hep-th]}}.

\bibitem{Arefeva:1974jv}
I.~Y. Arefeva, L.~D. Faddeev, and A.~A. Slavnov, {\it {Generating Functional for the s Matrix in Gauge Theories}},  \href{http://dx.doi.org/10.1007/BF01038094}{{\sf Teor. Mat. Fiz.} {\sf {21} }{\sf (1974) }{\sf 311--321}}.

\bibitem{Gubser:1998bc}
S.~S. Gubser, I.~R. Klebanov, and A.~M. Polyakov, {\it {Gauge theory correlators from noncritical string theory}},  \href{http://dx.doi.org/10.1016/S0370-2693(98)00377-3}{{\sf Phys. Lett. B} {\sf {428} }{\sf (1998) }{\sf 105--114}}, \href{http://arxiv.org/abs/hep-th/9802109}{{\ttfamily arXiv:hep-th/9802109}}.

\bibitem{Witten:1998qj}
E.~Witten, {\it {Anti-de Sitter space and holography}},  \href{http://dx.doi.org/10.4310/ATMP.1998.v2.n2.a2}{{\sf Adv. Theor. Math. Phys.} {\sf {2} }{\sf (1998) }{\sf 253--291}}, \href{http://arxiv.org/abs/hep-th/9802150}{{\ttfamily arXiv:hep-th/9802150}}.

\bibitem{Banks:1998dd}
T.~Banks, M.~R. Douglas, G.~T. Horowitz, and E.~J. Martinec, {\it {AdS dynamics from conformal field theory}},  \href{http://arxiv.org/abs/hep-th/9808016}{{\ttfamily arXiv:hep-th/9808016}}.

\bibitem{Jain:2023fxc}
D.~Jain, S.~Kundu, S.~Minwalla, O.~Parrikar, S.~G. Prabhu, and P.~Shrivastava, {\it {The S-matrix and boundary correlators in flat space}},  \href{http://arxiv.org/abs/2311.03443}{{\ttfamily arXiv:2311.03443 [hep-th]}}.

\bibitem{Banerjee:2018gce}
S.~Banerjee, {\it {Null Infinity and Unitary Representation of The Poincare Group}},  \href{http://dx.doi.org/10.1007/JHEP01(2019)205}{{\sf JHEP} {\sf {01} }{\sf (2019) }{\sf 205}}, \href{http://arxiv.org/abs/1801.10171}{{\ttfamily arXiv:1801.10171 [hep-th]}}.

\bibitem{Lysov:2014csa}
V.~Lysov, S.~Pasterski, and A.~Strominger, {\it {Low\textquoteright{}s Subleading Soft Theorem as a Symmetry of QED}},  \href{http://dx.doi.org/10.1103/PhysRevLett.113.111601}{{\sf Phys. Rev. Lett.} {\sf {113} }{\sf no.~11, }{\sf (2014) }{\sf 111601}}, \href{http://arxiv.org/abs/1407.3814}{{\ttfamily arXiv:1407.3814 [hep-th]}}.

\bibitem{Polchinski:1999ry}
J.~Polchinski, {\it {S matrices from AdS space-time}},  \href{http://arxiv.org/abs/hep-th/9901076}{{\ttfamily arXiv:hep-th/9901076}}.

\bibitem{Susskind:1998vk}
L.~Susskind, {\it {Holography in the flat space limit}},  \href{http://dx.doi.org/10.1063/1.1301570}{{\sf AIP Conf. Proc.} {\sf {493} }{\sf no.~1, }{\sf (1999) }{\sf 98--112}}, \href{http://arxiv.org/abs/hep-th/9901079}{{\ttfamily arXiv:hep-th/9901079}}.

\bibitem{Gary:2009ae}
M.~Gary, S.~B. Giddings, and J.~Penedones, {\it {Local bulk S-matrix elements and CFT singularities}},  \href{http://dx.doi.org/10.1103/PhysRevD.80.085005}{{\sf Phys. Rev. D} {\sf {80} }{\sf (2009) }{\sf 085005}}, \href{http://arxiv.org/abs/0903.4437}{{\ttfamily arXiv:0903.4437 [hep-th]}}.

\bibitem{Gary:2009mi}
M.~Gary and S.~B. Giddings, {\it {The Flat space S-matrix from the AdS/CFT correspondence?}},  \href{http://dx.doi.org/10.1103/PhysRevD.80.046008}{{\sf Phys. Rev. D} {\sf {80} }{\sf (2009) }{\sf 046008}}, \href{http://arxiv.org/abs/0904.3544}{{\ttfamily arXiv:0904.3544 [hep-th]}}.

\bibitem{Giddings:1999jq}
S.~B. Giddings, {\it {Flat space scattering and bulk locality in the AdS / CFT correspondence}},  \href{http://dx.doi.org/10.1103/PhysRevD.61.106008}{{\sf Phys. Rev. D} {\sf {61} }{\sf (2000) }{\sf 106008}}, \href{http://arxiv.org/abs/hep-th/9907129}{{\ttfamily arXiv:hep-th/9907129}}.

\bibitem{Fitzpatrick:2011jn}
A.~L. Fitzpatrick and J.~Kaplan, {\it {Scattering States in AdS/CFT}},  \href{http://arxiv.org/abs/1104.2597}{{\ttfamily arXiv:1104.2597 [hep-th]}}.

\bibitem{Hijano:2019qmi}
E.~Hijano, {\it {Flat space physics from AdS/CFT}},  \href{http://dx.doi.org/10.1007/JHEP07(2019)132}{{\sf JHEP} {\sf {07} }{\sf (2019) }{\sf 132}}, \href{http://arxiv.org/abs/1905.02729}{{\ttfamily arXiv:1905.02729 [hep-th]}}.

\bibitem{Hijano:2020szl}
E.~Hijano and D.~Neuenfeld, {\it {Soft photon theorems from CFT Ward identites in the flat limit of AdS/CFT}},  \href{http://dx.doi.org/10.1007/JHEP11(2020)009}{{\sf JHEP} {\sf {11} }{\sf (2020) }{\sf 009}}, \href{http://arxiv.org/abs/2005.03667}{{\ttfamily arXiv:2005.03667 [hep-th]}}.

\bibitem{Komatsu:2020sag}
S.~Komatsu, M.~F. Paulos, B.~C. Van~Rees, and X.~Zhao, {\it {Landau diagrams in AdS and S-matrices from conformal correlators}},  \href{http://dx.doi.org/10.1007/JHEP11(2020)046}{{\sf JHEP} {\sf {11} }{\sf (2020) }{\sf 046}}, \href{http://arxiv.org/abs/2007.13745}{{\ttfamily arXiv:2007.13745 [hep-th]}}.

\bibitem{Li:2021snj}
Y.-Z. Li, {\it {Notes on flat-space limit of AdS/CFT}},  \href{http://dx.doi.org/10.1007/JHEP09(2021)027}{{\sf JHEP} {\sf {09} }{\sf (2021) }{\sf 027}}, \href{http://arxiv.org/abs/2106.04606}{{\ttfamily arXiv:2106.04606 [hep-th]}}.

\bibitem{Duary:2022pyv}
S.~Duary, E.~Hijano, and M.~Patra, {\it {Towards an IR finite S-matrix in the flat limit of AdS/CFT}},  \href{http://arxiv.org/abs/2211.13711}{{\ttfamily arXiv:2211.13711 [hep-th]}}.

\bibitem{Campoleoni:2023fug}
A.~Campoleoni, A.~Delfante, S.~Pekar, P.~M. Petropoulos, D.~Rivera-Betancour, and M.~Vilatte, {\it {Flat from anti de Sitter}},  \href{http://dx.doi.org/10.1007/JHEP12(2023)078}{{\sf JHEP} {\sf {12} }{\sf (2023) }{\sf 078}}, \href{http://arxiv.org/abs/2309.15182}{{\ttfamily arXiv:2309.15182 [hep-th]}}.

\bibitem{deGioia:2024yne}
L.~P. de~Gioia and A.-M. Raclariu, {\it {Celestial amplitudes from conformal correlators with bulk-point kinematics}},  \href{http://arxiv.org/abs/2405.07972}{{\ttfamily arXiv:2405.07972 [hep-th]}}.

\bibitem{Alday:2024yyj}
L.~F. Alday, M.~Nocchi, R.~Ruzziconi, and A.~Yelleshpur~Srikant, {\it {Carrollian Amplitudes from Holographic Correlators}},  \href{http://arxiv.org/abs/2406.19343}{{\ttfamily arXiv:2406.19343 [hep-th]}}.

\bibitem{Marotta:2024sce}
R.~Marotta, K.~Skenderis, and M.~Verma, {\it {Flat space spinning massive amplitudes from momentum space CFT}},  \href{http://arxiv.org/abs/2406.06447}{{\ttfamily arXiv:2406.06447 [hep-th]}}.

\bibitem{Maldacena:2015iua}
J.~Maldacena, D.~Simmons-Duffin, and A.~Zhiboedov, {\it {Looking for a bulk point}},  \href{http://dx.doi.org/10.1007/JHEP01(2017)013}{{\sf JHEP} {\sf {01} }{\sf (2017) }{\sf 013}}, \href{http://arxiv.org/abs/1509.03612}{{\ttfamily arXiv:1509.03612 [hep-th]}}.

\bibitem{deGioia:2022fcn}
L.~P. de~Gioia and A.-M. Raclariu, {\it {Eikonal approximation in celestial CFT}},  \href{http://dx.doi.org/10.1007/JHEP03(2023)030}{{\sf JHEP} {\sf {03} }{\sf (2023) }{\sf 030}}, \href{http://arxiv.org/abs/2206.10547}{{\ttfamily arXiv:2206.10547 [hep-th]}}.

\bibitem{Adamo:2017nia}
T.~Adamo, E.~Casali, L.~Mason, and S.~Nekovar, {\it {Scattering on plane waves and the double copy}},  \href{http://dx.doi.org/10.1088/1361-6382/aa9961}{{\sf Class. Quant. Grav.} {\sf {35} }{\sf no.~1, }{\sf (2018) }{\sf 015004}}, \href{http://arxiv.org/abs/1706.08925}{{\ttfamily arXiv:1706.08925 [hep-th]}}.

\bibitem{Adamo:2021rfq}
T.~Adamo, A.~Cristofoli, and P.~Tourkine, {\it {Eikonal amplitudes from curved backgrounds}},  \href{http://dx.doi.org/10.21468/SciPostPhys.13.2.032}{{\sf SciPost Phys.} {\sf {13} }{\sf no.~2, }{\sf (2022) }{\sf 032}}, \href{http://arxiv.org/abs/2112.09113}{{\ttfamily arXiv:2112.09113 [hep-th]}}.

\bibitem{Gonzo:2022tjm}
R.~Gonzo, T.~McLoughlin, and A.~Puhm, {\it {Celestial holography on Kerr-Schild backgrounds}},  \href{http://dx.doi.org/10.1007/JHEP10(2022)073}{{\sf JHEP} {\sf {10} }{\sf (2022) }{\sf 073}}, \href{http://arxiv.org/abs/2207.13719}{{\ttfamily arXiv:2207.13719 [hep-th]}}.

\bibitem{Balian:1976vq}
{\it {Methods in Field Theory. Les Houches Summer School in Theoretical Physics. Session 28, July 28-September 6, 1975}}, .
\newblock 1976.

\bibitem{Jevicki:1987ax}
A.~Jevicki and C.-k. Lee, {\it {The S Matrix Generating Functional and Effective Action}},  \href{http://dx.doi.org/10.1103/PhysRevD.37.1485}{{\sf Phys. Rev. D} {\sf {37} }{\sf (1988) }{\sf 1485}}.

\bibitem{Oblak:2015qia}
B.~Oblak, {\it {From the Lorentz Group to the Celestial Sphere}},
\newblock 8, 2015.
\newblock \href{http://arxiv.org/abs/1508.00920}{{\ttfamily arXiv:1508.00920 [math-ph]}}.

\bibitem{He:2014cra}
T.~He, P.~Mitra, A.~P. Porfyriadis, and A.~Strominger, {\it {New Symmetries of Massless QED}},  \href{http://dx.doi.org/10.1007/JHEP10(2014)112}{{\sf JHEP} {\sf {10} }{\sf (2014) }{\sf 112}}, \href{http://arxiv.org/abs/1407.3789}{{\ttfamily arXiv:1407.3789 [hep-th]}}.

\bibitem{Campiglia:2015qka}
M.~Campiglia and A.~Laddha, {\it {Asymptotic symmetries of QED and Weinberg\textquoteright{}s soft photon theorem}},  \href{http://dx.doi.org/10.1007/JHEP07(2015)115}{{\sf JHEP} {\sf {07} }{\sf (2015) }{\sf 115}}, \href{http://arxiv.org/abs/1505.05346}{{\ttfamily arXiv:1505.05346 [hep-th]}}.

\bibitem{Sahoo:2018lxl}
B.~Sahoo and A.~Sen, {\it {Classical and Quantum Results on Logarithmic Terms in the Soft Theorem in Four Dimensions}},  \href{http://dx.doi.org/10.1007/JHEP02(2019)086}{{\sf JHEP} {\sf {02} }{\sf (2019) }{\sf 086}}, \href{http://arxiv.org/abs/1808.03288}{{\ttfamily arXiv:1808.03288 [hep-th]}}.

\bibitem{Campiglia:2016hvg}
M.~Campiglia and A.~Laddha, {\it {Subleading soft photons and large gauge transformations}},  \href{http://dx.doi.org/10.1007/JHEP11(2016)012}{{\sf JHEP} {\sf {11} }{\sf (2016) }{\sf 012}}, \href{http://arxiv.org/abs/1605.09677}{{\ttfamily arXiv:1605.09677 [hep-th]}}.

\bibitem{AtulBhatkar:2019vcb}
S.~Atul~Bhatkar, {\it {Ward identity for loop level soft photon theorem for massless QED coupled to gravity}},  \href{http://dx.doi.org/10.1007/JHEP10(2020)110}{{\sf JHEP} {\sf {10} }{\sf (2020) }{\sf 110}}, \href{http://arxiv.org/abs/1912.10229}{{\ttfamily arXiv:1912.10229 [hep-th]}}.

\bibitem{Peraza:2023ivy}
J.~Peraza, {\it {Renormalized electric and magnetic charges for O(r$^{n}$) large gauge symmetries}},  \href{http://dx.doi.org/10.1007/JHEP01(2024)175}{{\sf JHEP} {\sf {01} }{\sf (2024) }{\sf 175}}, \href{http://arxiv.org/abs/2301.05671}{{\ttfamily arXiv:2301.05671 [hep-th]}}.

\bibitem{Choi:2024ygx}
S.~Choi, A.~Laddha, and A.~Puhm, {\it {Asymptotic Symmetries for Logarithmic Soft Theorems in Gauge Theory and Gravity}},  \href{http://arxiv.org/abs/2403.13053}{{\ttfamily arXiv:2403.13053 [hep-th]}}.

\bibitem{Nagy:2024dme}
S.~Nagy, J.~Peraza, and G.~Pizzolo, {\it {A General Hierarchy of Charges at Null Infinity via the Todd Polynomials}},  \href{http://arxiv.org/abs/2405.06629}{{\ttfamily arXiv:2405.06629 [hep-th]}}.

\bibitem{Skenderis:2008dh}
K.~Skenderis and B.~C. van Rees, {\it {Real-time gauge/gravity duality}},  \href{http://dx.doi.org/10.1103/PhysRevLett.101.081601}{{\sf Phys. Rev. Lett.} {\sf {101} }{\sf (2008) }{\sf 081601}}, \href{http://arxiv.org/abs/0805.0150}{{\ttfamily arXiv:0805.0150 [hep-th]}}.

\bibitem{DHoker:2010xwl}
E.~D'Hoker, P.~Kraus, and A.~Shah, {\it {RG Flow of Magnetic Brane Correlators}},  \href{http://dx.doi.org/10.1007/JHEP04(2011)039}{{\sf JHEP} {\sf {04} }{\sf (2011) }{\sf 039}}, \href{http://arxiv.org/abs/1012.5072}{{\ttfamily arXiv:1012.5072 [hep-th]}}.

\bibitem{Kim:2023vbj}
S.~Kim, P.~Kraus, and R.~M. Myers, {\it {Systematics of boundary actions in gauge theory and gravity}},  \href{http://dx.doi.org/10.1007/JHEP04(2023)121}{{\sf JHEP} {\sf {04} }{\sf (2023) }{\sf 121}}, \href{http://arxiv.org/abs/2301.02964}{{\ttfamily arXiv:2301.02964 [hep-th]}}.

\bibitem{Jorstad:2023ajr}
E.~J\o{}rstad, S.~Pasterski, and A.~Sharma, {\it {Equating Extrapolate Dictionaries for Massless Scattering}},  \href{http://arxiv.org/abs/2310.02186}{{\ttfamily arXiv:2310.02186 [hep-th]}}.

\bibitem{Himwich:2019dug}
E.~Himwich and A.~Strominger, {\it {Celestial current algebra from Low\textquoteright{}s subleading soft theorem}},  \href{http://dx.doi.org/10.1103/PhysRevD.100.065001}{{\sf Phys. Rev. D} {\sf {100} }{\sf no.~6, }{\sf (2019) }{\sf 065001}}, \href{http://arxiv.org/abs/1901.01622}{{\ttfamily arXiv:1901.01622 [hep-th]}}.

\end{thebibliography}\endgroup

\end{document}